\documentclass[manuscript,natbib=true,natbibopts={numbers,sort&compress}]{acmart}
\usepackage{soul}
\usepackage{hyperref}
\usepackage{comment}
\makeatletter
\providecommand{\ACM@linecountL}{}
\providecommand{\ACM@linecountR}{}
\expandafter\providecommand\csname ACM@linecountL\endcsname{}
\expandafter\providecommand\csname ACM@linecountR\endcsname{}
\makeatother
\usepackage{multirow}
\usepackage{listings}
\usepackage{fancybox}
\usepackage{enumitem}
\usepackage{graphicx}
\usepackage{color}
\usepackage{amsmath}
\usepackage{xspace}
\usepackage{url}
\usepackage{tikz}
\usepackage{caption}
\usepackage{pgfplots}
\usepackage{booktabs}
\usepackage{longtable}
\usepackage{tabularx}
\usepackage{array}
\usepackage{pdflscape} 
\usepackage{ragged2e}
\usepackage{balance}
\usepackage{subcaption}
\pgfplotsset{compat=1.16}
\usepackage{pgf-pie}
\usetikzlibrary{pgfplots.statistics,calc}
\usepackage{algorithm}
\usepackage{algorithmic}
\usepackage{adjustbox}
\usepackage{makecell}

\makeatletter
\newcommand{\mybox}[1]{%
	\setbox0=\hbox{#1}%
	\setlength{\@tempdima}{\dimexpr\wd0+13pt}%
	\begin{tcolorbox}[boxrule=0.5pt, colback=white, arc=4pt,
		left=6pt,right=6pt,top=6pt,bottom=6pt,boxsep=0pt]
		#1
	\end{tcolorbox}
}
\makeatother

\definecolor{songcolor}{RGB}{191,191,191}

\newcommand{\add}[1]{\textcolor{black}{#1}}

\newcommand{\respto}[1]{%
\leavevmode\hypertarget{response:#1}{}%
\label{response:#1}%
\fcolorbox{black}{black!15}{%
\bf
  \scriptsize Review {#1}}~%
}

\makeatother
\makeatother

\usepackage{xspace}

\usepackage{soul}
\usepackage{caption}
\usepackage{subcaption}
\usepackage{graphicx}
\usepackage{ifthen}
\usepackage[many]{tcolorbox}
\usepackage{enumitem}
\usepackage{wrapfig}
\usepackage{moresize}
\usepackage{fontawesome}
\usepackage{xstring}
\usepackage{booktabs}
\usepackage{tabularx}
\usepackage{tcolorbox}
\tcbuselibrary{listingsutf8}
\usepackage{enumitem}
\usepackage{url}
\usepackage{balance}

\AtBeginDocument{%
  }
    \newboolean{showcomments}
\setboolean{showcomments}{true}

\ifthenelse{\boolean{showcomments}}
{\newcommand{\nbc}[3]{
 {\colorbox{#3}{\bfseries\sffamily\scriptsize\textcolor{white}{#1}}}
 {\textcolor{#3}{\sf\small$\blacktriangleright$\textit{#2}$\blacktriangleleft$}}
 }
}
{\newcommand{\nbc}[3]{}
 }

\newcommand\gopi[1]{\nbc{Gopi}{#1}{red}}

\newcommand\tinyskip{\vspace{1.5pt}} 

\newtcolorbox{challengebox}[2][]{
top=0.15in,left=4pt,right=4pt,bottom=4pt,
fonttitle=\bfseries,
colbacktitle=gray,
colback=white,
colframe=gray!40!black,
enhanced,
breakable,
attach boxed title to top left={xshift=1.5em,yshift=-\tcboxedtitleheight/2},
boxed title style={size=small},
drop shadow={black!50!white},
parbox=false, 
title=#2,#1}

\newtcolorbox{sourcespotlight}[1][]{
  colback=gray!10,
  colframe=gray!55!black,
  boxrule=0.8pt,
  arc=1pt,
  left=6pt,
  right=6pt,
  top=6pt,
  bottom=4pt,
  enhanced,
  unbreakable,
  fonttitle=\bfseries,
  attach boxed title to top left={xshift=1.0em,yshift=-1mm},
  boxed title style={
    colback=teal!25,
    colframe=gray!55!black,
    sharp corners,
    boxrule=0.8pt,
    left=4pt,
    right=4pt,
    top=2pt,
    bottom=2pt,
  },
  title={\textcolor{orange!85!black}{\faLightbulbO}\ \textcolor{black}{Source Spotlight}},
  enlarge top by=2mm,
  #1
}

\newcommand{\assumpSep}{\par\smallskip\noindent\tikz{\draw[dashed,gray] (0,0) -- (\linewidth,0);}\smallskip\par}
\newtcolorbox{assumptiontradeoffbox}[1][]{
  colback=gray!08,
  colframe=gray!55!black,
  boxrule=0.8pt,
  arc=1pt,
  left=6pt,
  right=6pt,
  top=6pt,
  bottom=4pt,
  enhanced,
  unbreakable,
  fonttitle=\bfseries,
  attach boxed title to top left={xshift=1.0em,yshift=-1mm},
  boxed title style={
    colback=purple!18,
    colframe=gray!55!black,
    sharp corners,
    boxrule=0.8pt,
    left=4pt,
    right=4pt,
    top=2pt,
    bottom=2pt,
  },
  title={\textcolor{purple!75!black}{\faBalanceScale}\ \textcolor{black}{Assumptions \& Trade-offs}},
  enlarge top by=2mm,
  #1
}

\newtcolorbox{greenfieldbox}[1][]{
  colback=gray!08,
  colframe=gray!55!black,
  boxrule=0.8pt,
  arc=1pt,
  left=6pt,
  right=6pt,
  top=6pt,
  bottom=4pt,
  enhanced,
  unbreakable,
  fonttitle=\bfseries,
  attach boxed title to top left={xshift=1.0em,yshift=-1mm},
  boxed title style={
    colback=green!18,
    colframe=gray!55!black,
    sharp corners,
    boxrule=0.8pt,
    left=4pt,
    right=4pt,
    top=2pt,
    bottom=2pt,
  },
  title={\textcolor{green!55!black}{\faRoad}\ \textcolor{black}{Greenfield Road Ahead}},
  enlarge top by=2mm,
  #1
}

\begin{document}
\title{From Cool Demos to Production-Ready FMware: Core Challenges and a Technology Roadmap}

\author{Gopi Krishnan Rajbahadur}
\affiliation{%
  \institution{Centre for Software Excellence, Huawei Canada}
  \city{ON}
  \country{Canada}
}
\email{gopi.krishnan.rajbahadur1@huawei.com}

\author{Gustavo A. Oliva}
\affiliation{%
  \institution{Centre for Software Excellence, Huawei Canada}
  \city{ON}
  \country{Canada}
}
\email{gustavo.oliva@huawei.com}

\author{Dayi Lin}
\affiliation{%
  \institution{Centre for Software Excellence, Huawei Canada}
  \city{ON}
  \country{Canada}
}
\email{dayi.lin@huawei.com}

\author{Jiho Shin}
\affiliation{%
  \institution{Queen's University}
  \country{Canada}
}
\email{jiho.shin@queensu.ca}

\author{Ahmed E. Hassan}
\affiliation{%
  \institution{Queen's University}
  \country{Canada}
}
\email{ahmed@cs.queensu.ca}
\renewcommand{\shortauthors}{Rajbahadur et al.}

\begin{abstract}
%
The rapid expansion of foundation models (FMs), such as large language models (LLMs), has given rise to FMware, software systems that integrate FM(s) as core components. While building demonstration-level FMware is relatively straightforward, transitioning to production-ready systems presents numerous challenges, including reliability, high implementation costs, scalability, and compliance with privacy regulations. Our paper conducts a semi-structured thematic synthesis to identify key challenges in productionizing FMware across diverse data sources, including our industry experience developing FMArts, a FMware lifecycle engineering platform, and its integration into Huawei Cloud; grey literature; academic publications; hands-on involvement in the Open Platform for Enterprise AI (OPEA); organizing the AIware conference and bootcamp; and co-leading the ISO SPDX SBOM working group on AI and datasets. We identify critical issues in FM(s) selection, data and model alignment, prompt engineering, agent orchestration, system testing, and deployment, alongside cross-cutting concerns such as memory management, observability, and feedback integration. We discuss necessary technologies and strategies to address these challenges and offer guidance to enable the transition from demonstration systems to scalable, production-ready FMware solutions. Our findings underscore the importance of continued research and multi-industry collaboration to advance the development of production-ready FMware.
\end{abstract}

\RenewDocumentCommand{\respto}{m}{} 

\ccsdesc[500]{Software and its engineering~Software creation and management}
\ccsdesc[500]{Computing methodologies~Artificial intelligence}

\ccsdesc[500]{Software and its engineering}

\keywords{AI-powered Software, Production-ready FMware, FMware, Productionization}

\received{30 January 2025}

\maketitle
\section{Introduction}
\label{sec:intro}

FMware refers to software integrating foundation models (FMs), like large language models (LLMs), as core components~\cite{hassan2024rethinking}. Since ChatGPT's release in late 2022, the FMware landscape has exploded, with over 600,000 open-source models, including FM(s) and other AI models, now available on platforms like Hugging Face~\cite{factoredChoosingRight}. Goldman Sachs predicts FMware could boost global GDP by 7\%~\cite{goldmansachsGenerativeCould}.

Building impressive demos with FM(s) is relatively easy, but transitioning to production-ready FMware incurs significant challenges due to its inherent complexity as a compound system~\cite{zaharia2024shift,hassan2024rethinking}. Unlike static systems, FMware integrates multiple dynamic components for real-time updates, control, and adaptability~\cite{zaharia2024shift,ibmBuildingFoundation}. This complexity is reflected in industry adoption rates: a survey of 430 technology professionals revealed that only 10\% of organizations had launched FMware in production environments~\cite{intelOrganizationsSurveyed}. Respondents identified reliability, high implementation costs, latency, compliance, and privacy as the top challenges preventing them from moving from demos to production-ready FMware.

The transition to production-ready FMware is formidable due to its dynamic nature, which combines classical software engineering complexities with unique issues inherent to foundation models. Production-ready FMware must continuously evolve to meet customer expectations, requiring consistent performance, reliability, feature updates, and adherence to service level agreements (SLAs), all while managing operational costs. For instance, OpenAI's infrastructure costs for running ChatGPT in 2023 were estimated at \$700,000 per day~\cite{uniteFinancialChallenges}. Beyond these traditional challenges, FMware developers also face unique problems like handling hallucinations, increased inference costs, and orchestrating tasks across various AI components~\cite{opeaOpenPlatform,hassan2024rethinking}. LinkedIn's experience illustrates the effort required to reach production readiness: they achieved 80\% functionality in a month but spent four more months to complete the remaining 20\%, with diminishing returns on each additional 1\% improvement~\cite{linkedinMusingsBuilding}. Similarly, Microsoft and GitHub found that testing FMware becomes prohibitively expensive as complexity scales~\cite{parnin2023building}. These real-world examples underscore the need for robust, system-based approaches to build reliable, production-ready FMware solutions~\cite{zaharia2024shift,opeaOpenPlatform,hassan2024rethinking}.

In this paper, we employ a semi-structured thematic synthesis (more details in Section~\ref{sec:methodology}) to systematically identify key challenges in developing production-ready FMware. We draw on insights from multi-industry collaborations, conferences, customer meetings, hands-on development of the FMware lifecycle platform (FMArts)~\cite{hassan2024rethinking}, and literature surveys. In doing so, we provide the first comprehensive articulation of the recurrent issues across different stages of FMware development, consolidating them into challenges that hinder productionizing FMware. \respto{1-27a}
\add{While recent studies by Hassan~et~al.~\cite{hassan2024rethinking} and Chen~et~al.~\cite{chen2025empirical} primarily investigate the challenges of \textit{developing} FMware, our study adopts a distinct focus. We provide the first comprehensive synthesis of the challenges practitioners face across the full FMware lifecycle, specifically when transitioning systems from initial demonstrations and proofs-of-concept to robust and demanding production environments.}

This paper complements our vision for Software Engineering 3.0 (SE 3.0), where we advocate for an AI-native, intent-first approach where AI systems evolve from task-driven copilots to intelligent collaborators~\cite{hassan2024ainativesoftwareengineeringse}. Both SE 3.0 and our paper address the need to adapt software engineering practices to manage the complexity and dynamism of AI systems like FMware. By synthesizing hands-on knowledge from building production-ready FMware and interacting with industry experts, we underscore the urgency of addressing these challenges to move from ``cool'' demos to robust, production-ready solutions. Our study further highlights the importance of ongoing research and multi-industry collaboration as FMware continues to evolve.
\add{To support transparency and reuse, we provide a replication package containing the publicly
shareable artifacts and curated bibliography that underpin our synthesis \cite{replication_package}.}

This paper is organized as follows: Section~\ref{sec:background_scope} introduces the background and scope of our study, and Section~\ref{sec:related_work} reviews related work on readiness in software engineering, Neuralware, and FMware. Section~\ref{sec:methodology} describes our methodology. Section~\ref{sec:pipeline} outlines the FMware lifecycle for productionization, and Section~\ref{sec:issues} catalogs recurrent issues observed across the lifecycle stages. Section~\ref{sec:challenges} consolidates these issues into key cross-cutting challenges that hinder FMware production readiness and discusses future directions (including a roadmap in Section~\ref{sec:road_ahead}). Section~\ref{sec:limitations} discusses limitations. Finally, Section~\ref{sec:conclusion} concludes the study.




\add{\section{Background and Scope}
\label{sec:background_scope}}
\respto{2-4} \add{To provide a precise scope for this study and define the terminology used throughout the paper, we first distinguish between traditional machine learning models and FM(s), and then define the distinct classes of software systems built upon them.}

\add{
\subsection{Traditional Machine Learning Models vs. Foundation Models}
\label{sec:ml_vs_fm}
Traditional machine learning and deep learning models, such as convolutional neural networks (CNNs) or decision trees, are typically designed for single-purpose, deterministic tasks (e.g., image classification or structured prediction). These models operate on well-defined inputs and produce specific, bounded outputs. In contrast, \textit{FM(s)} are trained on vast, diverse datasets to support a broad range of downstream tasks. This category includes Large Language Models (LLMs) like GPT-4, as well as multimodal models like CLIP and Gemini~\cite{ibmFMs,mickinsey}. Unlike traditional models, FM(s) are general-purpose, adaptive, and prompt-driven, enabling flexible task specification at inference time without retraining.}

\add{
\subsection{AI-Powered Software Paradigms}
\label{sec:ai_paradigms}
\respto{2-3} Over the last two decades, the prevalent integration of traditional machine learning and, more recently, FM(s) into software has led to the emergence of distinct AI-powered software paradigms~\cite{hassan2024rethinking}. We briefly outline these paradigms below:}

\begin{itemize}
    \item \add{\textbf{Neuralware:} Software systems built around traditional machine learning or deep learning models. These systems follow well-defined engineering lifecycles that have been extensively studied in the literature~\cite{bastani2021applied, polyzotis2017data}. While Neuralware remains critical in practice, its deterministic lifecycle is outside the scope of this paper.}
    \item \add{\textbf{Promptware:} Software systems where workflows are primarily built around fixed prompts or templates that orchestrate calls to an FM(s) in a structured sequence. These systems are common in prompt-chaining and retrieval-augmented generation (RAG) pipelines, for example, document-grounded Q\&A and FAQ-style chatbots that retrieve context and then apply a prompt template to produce an answer~\cite{wu2022promptchainer, langchainRAGdoc, llamaIndexRag}. While Promptware is effective for applications with relatively static requirements, it is limited in adaptability when the workflow must change dynamically at runtime.}
    \item \add{\textbf{Agentware:} Software systems where autonomous agents, powered by FM(s), dynamically determine tasks, execution steps, and tool-use policies at runtime. Representative examples include tool-using agents following the Reason+Act paradigm (interleaving reasoning traces with actions), multi-agent conversation frameworks that coordinate specialized agents, and autonomous agent systems that iteratively plan and execute multi-step goals~\cite{yao2023react, wu2024autogen, yang2023auto}. Agentware enables emergent behaviors and multi-step orchestration beyond the capabilities of static Promptware.}
    \item \add{\textbf{FMware:} Following Hassan et al.~\cite{hassan2024rethinking}, we use the term \textit{FMware} to collectively refer to both Promptware and Agentware. While terms such as ``AI-powered software,'' ``LLM applications,'' ``GenAI applications,'' ``LLM-based agents,'' or ``LLM-powered systems'' are frequently used in prior work and practitioner discourse~\cite{wu2024autogen, langchainRAGdoc, huyen2023building}, they can be ambiguous about whether the system behavior is primarily prompt-chained or agentic. Furthermore, many production systems are hybrids that combine reliable Promptware flows with adaptive Agentware components. For example, a customer service platform that uses static prompts for FAQs but invokes agents for complex case resolution. FMware provides a useful abstraction to discuss cross-cutting production challenges that apply across systems employing both Promptware and Agentware components, and we use this term throughout our paper.}
\end{itemize}

\respto{1-26}\add{In our paper, we focus exclusively on the challenges concerning the production-readiness of \textit{FMware}, encompassing both Promptware and Agentware. We do not consider the challenges on \textit{Neuralware}, as prior work has extensively studied the engineering, testing, and production-readiness of traditional ML systems, including hidden technical debt and maintenance risks, process and best practices observed in industrial teams, and production-readiness rubrics and production-scale ML platforms~\cite{sculley2015hidden, amershi2019software, breck2016s, baylor2017tfx}.}

\add{Furthermore, our scope is limited to the production readiness and operation of FMware rather than the creation of FM(s). We do not cover challenges that model builders face when conducting novel RL techniques, new model architectures, or inference-time architectures at scale. Instead, we focus on three application areas that are critical for practitioners building production-ready FMware: (i) \textit{FM(s) adaptation} under Data and FM(s) Alignment (i.e., tailoring a selected FM(s) to a target domain/task through data preparation/curation and alignment techniques), including SFT (Supervised Fine Tuning), RFT (Reinforcement Learning based Fine Tuning), few-shot learning, and preference-based optimization; (ii) \textit{deployment and operations}, including hosting, monitoring, and continuous improvement pipelines; and (iii) \textit{agent development and orchestration}, focusing on tool use and multi-agent coordination. This scope enables us to study production-oriented challenges such as prompt engineering, cognitive observability, and controlled execution (see Sections~\ref{sec:issues} and~\ref{sec:challenges}).}

\section{Related Work}
\label{sec:related_work}
\respto{3-21}
\add{\subsection{Release and Production Readiness in Software Engineering}}
\label{sec:related_production_ready}


\add{Prior work in traditional software engineering has extensively studied \emph{release readiness}, i.e., whether a candidate version satisfies functional requirements and is fit to ship based on release-time quality signals and release processes~\cite{al2014monitoring,koprowski2014release,al2017two,port2013value,patel2024state}. In contrast, \emph{production readiness} concerns whether the deployed system can operate reliably in live environments, meeting operational constraints and service-level objectives, including latency, safety, cost, monitoring/observability, and on-call operability~\cite{asthana2009quantifying,cusick2013architecture}. Below, we first summarize readiness work in traditional software engineering and then discuss how readiness has been revisited in Neuralware and, more recently, in FMware.}

\add{\subsection{Traditional Release and Production Readiness.}
Early work by \citet{mockus2003analogy} modeled the workflow of work items (i.e., Modification Requests tracked in a change-management system) using historical analogs to forecast schedule risk and decide when a release is ready, showing how repository signals can predict slippage and throughput. Subsequent studies examined how organizations operationalize readiness gates and indicators, for example, readiness certification practices in high-assurance settings~\cite{port2013value} and project telemetry used to track readiness attributes such as defect find rate and bug fix rate~\cite{al2014monitoring}. Predictive approaches framed readiness as a classification problem using multi-project historical data~\cite{al2017two}, while other work contrasted competing measurement practices, e.g., defect-tracking, test-progress, and customer-centric indicators~\cite{koprowski2014release}. Complementing release-centric work, production readiness has been studied as ensuring safe and reliable operation post-deployment, including quantitative readiness perspectives over reliability and operational quality~\cite{asthana2009quantifying} and review practices that explicitly check deployment, capacity, and operability concerns~\cite{cusick2013architecture}.}


\add{\subsection{Production Readiness in Neuralware Systems}
\label{sec:related_neuralware_ready}
Production readiness considerations have also been specialized for Neuralware, where failures can arise from data and model behavior rather than only from code changes. Prior work highlights that ML systems introduce unique sources of technical debt and maintenance risk (e.g., hidden feedback loops and data dependencies)~\cite{sculley2015hidden} and face distinct production data-management challenges~\cite{polyzotis2017data}. For example, \citet{breck2017ml} introduced the ML Test Score, a rubric of actionable tests that span data, model, and operations, enabling teams to assess readiness beyond standard software practices, e.g., data invariants, model staleness checks, monitoring, and rollback safety. Complementary studies also document process and organizational practices for building and operating production ML systems in industry~\cite{amershi2019software} and describe production-scale platform support (e.g., TFX) to operationalize data validation, model analysis, and deployment pipelines~\cite{baylor2017tfx}. While our paper focuses on FMware rather than Neuralware, this body of work motivates why readiness criteria must evolve when AI components become first-class runtime dependencies.}

\add{\subsection{Emerging Work on Production-Ready FMware}
\label{sec:related_fmware_ready}
Recent work has begun to articulate readiness concerns specific to FMware. \citet{patel2024state} compiled a state-of-practice checklist for generative-AI products that extends traditional checks with FM-specific items, e.g., prompt and data governance, evaluation coverage for hallucination-prone tasks, privacy and compliance reviews, and operational guardrails. Complementing checklist-style guidance, \citet{parnin2023building} reported interview findings from teams building copilots, identifying pain points across the lifecycle, e.g., expensive prompt and evaluation loops, data curation overheads, and gaps in tooling for testing and monitoring. The O’Reilly series~\cite{oreillyWhatLearned} distilled lessons from a year of building with LLMs, emphasizing evaluation discipline, guardrails, data quality, and operational hygiene as prerequisites for robust production systems. Nahar et al.~\cite{nahar2024beyond} conducted a mixed-methods study, 26 interviews, and a 332-participant survey, surfacing emerging solutions focused on quality assurance in products that integrate foundation models.}

\add{Beyond readiness checklists, a growing body of work has started to propose evaluation and diagnostics techniques for production FMware subsystems, particularly grounding and RAG pipelines. Representative examples include reference-free RAG evaluation frameworks such as RAGAS~\cite{es2023ragas}, ARES~\cite{saadfalcon2024ares}, and RAGChecker~\cite{ru2024ragchecker}, which aim to separately assess retrieval quality, faithfulness, and answer relevance and thus better support iterative improvement.}

\add{For Agentware, recent work has argued that outcome-only benchmarks are insufficient and has proposed richer observability- and process-centric evaluation. For example, Moshkovich et al.~\cite{moshkovich2025beyondblackbox} discuss how agentic system evaluation should leverage runtime logs and observability signals, while Liu et al.~\cite{liu2025processcentric} propose process-centric analyses (Graphectory) that characterize agent trajectories beyond end outcomes.}

\add{In parallel, early work has started to develop FMware-specific perspectives on operability, including cognitive observability for FM-powered agents~\cite{rombaut2024watson} and software performance engineering considerations for FMware systems~\cite{zhang2024software}.}

\add{While these studies provide valuable readiness signals and subsystem-specific methods, they do not offer a lifecycle-wide synthesis of recurrent, practitioner-observed issues and their consolidation into a compact set of cross-cutting challenges. Our study differs by synthesizing diverse grey literature and practitioner sources with academic work to chart recurrent issues across the full FMware lifecycle, then organizing these into practitioner-oriented challenges that inform a technology roadmap for engineering production-ready FMware.}

\renewcommand{\arraystretch}{1.1}
\begin{table*}[t]
\caption{\respto{3-14a}\add{\textbf{Comparison with prior surveys on foundation-model-powered software (FMware)}. Columns indicate each study’s scope, whether the evidence base includes grey literature (Grey lit.), end-to-end lifecycle coverage (E2E lifecycle), explicit focus on production readiness (Prod. readiness), coverage of agentic systems (Agentic), provision of a research/practice roadmap (Roadmap), and discussion of FMware operation concerns (FMware ops.). Scope abbreviations: FM4SE = FMs for software engineering; SE lifecycle = Software Engineering lifecycle; CodeGen = Code Generation; TestGen = Test Generation; Agent eval. = Agent Evaluation; SDLC benchmarks = benchmarks mapped to Software Development Lifecycle (SDLC) phases; FM Dev. challenges = FM application development challenges; Prod-ready FMware = production-ready FMware. Symbols: $\bigcirc$ indicates covered, $\bullet$ indicates partially covered, and $-$ indicates not covered.}}
\label{tab:fmware-survey-comparison}
\setlength{\tabcolsep}{5pt}
\resizebox{\textwidth}{!}{
\begin{tabular}{lccccccc}
\toprule
Study & Scope & Grey lit. & E2E lifecycle & Prod. readiness & Agentic & Roadmap & FMware ops. \\
\midrule
Jin et al. \cite{jin2024llms} & FM4SE & – & $\bigcirc$ & – & $\bullet$ & – & – \\
Fan et al. \cite{fan2023large} & FM4SE & – & $\bigcirc$ & – & – & – & – \\
Zhang et al. \cite{zhang2023survey} & SE lifecycle & – & $\bigcirc$ & – & – & – & – \\
Hou et al. \cite{hou2024large} & FM4SE & – & $\bigcirc$ & – & – & – & – \\
Jiang et al. \cite{jiang2024survey} & CodeGen & – & – & – & – & – & – \\
Wang et al. \cite{wang2024software} & TestGen & – & – & $\bullet$ & – & – & – \\
Yehudai et al. \cite{yehudai2025survey_eval_agents} & Agent eval. & – & – & – & $\bigcirc$ & $\bullet$ & – \\
Wang et al. \cite{wang2025sdlc_benchmarks} & SDLC benchmarks & – & $\bigcirc$ & – & $\bullet$ & $\bullet$ & – \\
Chen et al. \cite{chen2025empirical} & FM Dev. challenges & $\bigcirc$ & $\bigcirc$ & $\bullet$ & – & – & $\bigcirc$ \\
Hassan et al. \cite{hassan2024rethinking} & FMArts/FMware & – & $\bigcirc$ & $\bullet$ & $\bigcirc$ & $\bullet$ & $\bigcirc$ \\
\textbf{This paper} & \textbf{Prod-ready FMware} & \textbf{$\bigcirc$} & \textbf{$\bigcirc$} & \textbf{$\bigcirc$} & \textbf{$\bigcirc$} & \textbf{$\bigcirc$} & \textbf{$\bigcirc$} \\
\bottomrule
\end{tabular}}
\end{table*}

\subsection{Surveys in FMware Research}
\label{sec:related_surveys}
\respto{3-14a} \add{Table~\ref{tab:fmware-survey-comparison} positions our study relative to prior surveys and empirical works along two dimensions: (i) \emph{scope} (end-to-end lifecycle coverage, production readiness, agentic systems, roadmap, and FMware operations) and (ii) \emph{evidence base} (whether the synthesis relies primarily on academic studies or also incorporates grey literature and practitioner artifacts).}

While several surveys have examined FMware and the role of LLMs in software engineering, they predominantly focus on academic publications or practitioner interviews. Jin et al.~\cite{jin2024llms} provide a comprehensive exploration of LLMs and their application as autonomous agents in software engineering. Similarly, Fan et al.~\cite{fan2023large} survey LLMs' potential in software development activities, including debugging and refactoring, emphasizing hybrid approaches that combine classical software engineering techniques with LLMs.

Zhang et al.~\cite{zhang2023survey} investigate the impact of LLMs across five phases of the software engineering lifecycle, leveraging academic datasets to highlight technical challenges such as model tuning and evaluation. Hou et al.~\cite{hou2024large} curated a systematic review of 395 research papers to understand how LLMs are used in various software engineering tasks and on which tasks LLMs have shown success. Jiang et al.~\cite{jiang2024survey} focus on LLMs for code generation, presenting a taxonomy of existing approaches based on structured academic benchmarks such as HumanEval and MBPP. 

Wang et al.~\cite{wang2024software} and others emphasize the use of LLMs in software testing, particularly for unit test generation and debugging, using standard datasets like Defects4J. \add{Complementary to these task-centric surveys, recent work has started to survey evaluation methodologies for LLM-based agents~\cite{yehudai2025survey_eval_agents} and benchmarks for CodeLLMs and agents mapped across software development life cycle phases~\cite{wang2025sdlc_benchmarks}. However, all of these studies share a common limitation: an overreliance on academic datasets and benchmarks without addressing the contributions of open-source projects, forums, and informal knowledge exchange.}
\respto{3-14a} \add{As summarized in Table~\ref{tab:fmware-survey-comparison}, these studies largely center on LLM4SE task capabilities and development-time considerations, and they only partially cover production readiness and FMware-native operability concerns. Moreover, many rely primarily on academic datasets and benchmarks, with limited incorporation of practitioner artifacts, open-source working-group materials, and informal knowledge exchange that often surface production failures and mitigations.}

Different from the aforementioned studies, Chen et al.~\cite{chen2025empirical} provide the most relevant foundation for our work by focusing on the challenges faced by FM(s) application developers. They present a comprehensive empirical study, mining 29,057 posts from the OpenAI developer forum to construct a detailed taxonomy of challenges. Their taxonomy highlights critical issues such as prompt engineering, API limitations, rate constraints, and hallucination management. To validate their findings, they extend their analysis to GitHub issues for LLaMa and Gemini, demonstrating the generalizability of their taxonomy across platforms. 
\respto{1-27b}\add{However, Table~\ref{tab:fmware-survey-comparison} highlights that their study remains centered on API-driven development, does not cover the full FMware lifecycle, and does not derive a production-readiness roadmap that addresses operability at scale.}

\respto{2-2}\add{Hassan et al.~\cite{hassan2024rethinking} examines FMware across the lifecycle through the lens of their FMArts platform, primarily targeting challenges in \emph{building} FMware (e.g., prompt engineering, alignment data management, and workflow design). In contrast, our study foregrounds the end-to-end journey from demo to production and the associated production-readiness barriers, including controlled execution for predictable behavior, observability for multi-agent workflows, lifecycle-wide testing under non-determinism, and performance engineering under strict Service Level Objectives (SLOs). Consistent with Table~\ref{tab:fmware-survey-comparison}, our contributions differ in two ways: (1) we map recurrent issues that impact productionization across the full FMware lifecycle (Section~\ref{sec:pipeline}), and (2) we consolidate them into actionable technology challenges and solution directions that inform a roadmap for production-grade FMware. Together, these works are complementary: Hassan et al. focus on developing trustworthy FMware, whereas our study focuses on making FMware production-ready.}

\respto{2-2}\add{While recent studies by Hassan~et~al.~\cite{hassan2024rethinking} and Chen~et~al.~\cite{chen2025empirical} primarily investigate the challenges of \textit{developing} FMware, our study focuses on the demo-to-production transition and the associated production-readiness barriers. In summary, Table~\ref{tab:fmware-survey-comparison} highlights that our study's scope uniquely combines end-to-end lifecycle coverage with an explicit production-readiness focus, including FMware operations and a challenge-driven roadmap, grounded in practitioner-facing evidence beyond academic benchmarks.}

\section{Methodology}
\label{sec:methodology}

\respto{1-25a} \respto{3-7a} 
\add{We conducted a semi-structured thematic synthesis, inspired by thematic analysis~\cite{butler2024objectives,cruzes2010synthesizing}, to identify and organize key challenges in developing production-ready FMware. Our approach collected relevant sources and grouped issues into themes through expert-driven discussions. To mitigate bias, we triangulated insights from industry discussions, conference reports, and academic studies, emphasizing actionable insights and practical relevance.}

\tinyskip \noindent \textbf{Step 1: Data Collection.} We collected data from various sources to gain insights into the challenges that researchers and practitioners encounter when productionizing FMware.

\respto{3-17a}\add{We follow a share-when-possible policy: we include public sources in the replication package \cite{replication_package} and summarize non-public materials only through aggregated recurrent issues and themes to preserve confidentiality.
Table~\ref{tab:data_provenance} summarizes our data sources and sharing policy \cite{replication_package}.}

\begin{table*}[t]
\centering
\small
\caption{\respto{3-17a} \add{Overview of the data sources used in our synthesis and the associated sharing policy for the replication package \cite{replication_package} (public artifacts are provided with full citations and URLs; confidential inputs are represented only via metadata and aggregated issues/themes).}}
\label{tab:data_provenance}
\begin{tabularx}{\textwidth}{p{0.35\textwidth} p{0.09\textwidth} X}
\toprule
\textbf{Source category} & \textbf{Availability} & \textbf{What we share} \\
\midrule
Community notes and working-group minutes & Public & Included in the replication package with full citations and URLs.  \\ \hline
Summit, conference, and workshop artifacts & Public & Included in the replication package with full citations and URLs. \\ \hline
Practitioner blogs and academic papers & Public & Included in the replication package, or organized in lifecycle stage(s). \\ \hline
Internal meeting notes and post-event reports & Confidential & Metadata only (e.g., event, date, lifecycle stage tags, derived issues/themes), no document contents. \\ \hline
Experiential knowledge from collaborations and product work & Confidential & Aggregated themes and anonymized examples, no proprietary or identifying details. \\
\bottomrule
\end{tabularx}
\end{table*}

\begin{itemize}[wide = 0pt, itemsep = 1.5pt, topsep = 1.5pt]
    \item \textit{OPEA Project Participation}: As active participants, we played a significant role in the OPEA initiative (Open Platform for Enterprise AI (OPEA)~\cite{opeaOpenPlatform}). In particular, the last author leads OPEA's research working group. OPEA is a global multi-company collaborative initiative focused on addressing the challenges of making FMware enterprise-ready. OPEA provides detailed frameworks, architectural blueprints, and a four-step assessment for evaluating FMware in terms of enterprise readiness. We used meeting notes (some of which are publicly available~\cite{atlassianOPEACommunity}) from OPEA discussions, which involved 43 organizations (e.g., Intel, AMD, RedHat, Docker, JFrog, Anyscale, LlamaIndex, and SAP). \respto{3-17a}\add{Publicly archived OPEA notes are included in the replication package, while internal notes that contain confidential organizational or participant details are incorporated only through aggregated issues and themes.}
    \item \textit{SPDX Working Group}: First author co-led SPDX community meetings (an ISO/IEC 5692:2021 Software Bill of Materials standard~\cite{spdx}) over the past 4 years, contributing to discussions on the AI Bill of Materials (AI BOM), including challenges in building production-ready FMware. The meeting notes, available publicly~\cite{spdxminutes}, were incorporated into our analysis. \respto{3-17a}\add{These minutes are included in the replication package with URLs for verification.}
    \item \textit{Inception and Launch of the AIware Conferences, summit and Bootcamp}: Authors of this paper spearheaded the first ACM International Conference on AI-Powered Software (AIware)~\cite{aiware}, co-located with FSE 2024, the FM+SE summit series~\cite{fmseFMSESummit} and the AIware Bootcamp~\cite{aiwarebootcampAIwareLeadership}. Engaging with participants from Google, Microsoft, GitHub, and other industry and academic professionals provided valuable insights into the challenges of productionizing FMware, which we summarized in a report used in our analysis~\cite{hassan2024fmse}. \respto{3-18}\add{These internal post-conference reports informed our thematic synthesis, but they are used only to derive aggregated, non-identifying artifacts (e.g., summaries and consolidated issues/themes), and the underlying reports cannot be released because they contain sensitive information.}
    \item \textit{Conference and Workshop Attendance and Reports}: Since late 2022, all authors have actively participated in several top conferences, workshops, and developer meetings relevant to FMware, such as ICSE, FSE, FM+SE 2030, FM+SE Tokyo, SEMLA, OSS Summits, Ray Summits, and ai\_dev Summits. Although these events covered a range of topics beyond the challenges of productionizing FMware, we gained valuable insights by listening to and interacting with researchers and practitioners involved in productionizing their own FMware. After each conference, the attending authors compiled detailed reports summarizing key discussions and relevant research on FMware. These reports formed the foundation for our thematic analysis. \respto{3-18}\add{As these reports contain sensitive company information and participant details, we do not release them; instead, we only incorporate aggregated insights in the paper.}
    \item \textit{Internal Industry Experience}: Our collaborations with customers and development teams to understand FMware's functional and non-functional requirements. Additionally, all authors contributed to developing and productionizing FMArts, a comprehensive lifecycle engineering platform for FMware~\cite{hassan2024rethinking}. Our practical experience in creating FMware with FMArts and making it production-ready~\cite{hassan2024rethinking} and integrating it into Huawei products (e.g., Huawei Cloud) informed the thematic analysis. \respto{3-17a}\add{Due to confidentiality constraints, we incorporate this input through aggregated themes and anonymized examples rather than sharing proprietary details.}
    \item \textit{In-depth Literature Review}: We began with a review of grey literature, including blog posts, whitepapers, and developer forums, and noticed recurring discussions about the production readiness of FMware. Building on these insights, we conducted a focused review of academic papers to find supporting evidence. Through this activity, we provide a comprehensive analysis of the challenges in deploying FMware, grounding all issues in Section~\ref{sec:issues} with citations and minimizing subjective bias. \respto{3-17a}\add{We include a consolidated, stage-organized bibliography of these public sources in the replication package, with citations and URLs.}
\end{itemize}
The collected materials were reviewed to identify recurring issues relevant to FMware production readiness. These insights formed the basis for subsequent grouping and theme development.

\tinyskip\noindent\textbf{Step 2: Data Analysis and Grouping.}
\respto{3-17a}\add{Our synthesis draws on public sources (including grey literature) and academic work, and incorporates non-public inputs only through aggregated issues and themes. Where possible, we cite public evidence directly (e.g., via Source Spotlight callouts in Section~\ref{sec:issues}) and triangulate across sources to reduce the influence of any single stream.}
\respto{1-7}\add{In this step, we extract \emph{recurrent issues} as concrete, evidenced problem instances from the collected materials and tag each issue with the relevant FMware lifecycle stage(s) (Figure~\ref{fig:lifecycle}) to preserve context. We then group closely related recurrent issues into \emph{groupings} as an intermediate organizational layer before higher-level abstraction.}
After data collection, three of the four authors reviewed the materials to identify recurring issues. Instead of formal iterative coding, we relied on a collaborative and expert-driven approach to extract key challenges and categorize them. We grouped issues into preliminary themes based on shared characteristics or relevance to specific stages of the FMware lifecycle (Figure~\ref{fig:lifecycle}). For example, issues related to \textit{Data and FM(s) Alignment} included challenges such as \textit{low data quality} and \textit{insufficient data diversity}. This approach emphasized practical categorization rather than exhaustive formal coding.

\respto{1-7}\add{\textbf{\textit{Running Example}}: within \textit{Data and FM(s) Alignment}, we treat ``low data quality'' as a recurrent issue and group it with related issues such as ``low domain coverage'' under \textit{Alignment Data Quality}.}

\tinyskip\noindent\textbf{Step 3: Collaborative Discussion to Identify Themes.}
\respto{1-7}\add{In Step 3, we synthesize \emph{themes} by abstracting over one or more groupings to capture broader, potentially cross-cutting patterns that may span lifecycle stages.} The authors collaboratively refined the initial groupings into overarching themes (which we present as challenges) through discussions informed by their domain expertise. These themes spanned multiple FMware lifecycle stages and were grounded in the collected data, but did not follow the strict iterative refinement protocols of thematic analysis. Instead, the grouping process was guided by the practical relevance of the challenges and their applicability to real-world FMware development.

\respto{1-7}\add{\textbf{\textit{Running Example}}: we synthesize \textit{Alignment Data Quality} together with grounding-related groupings (e.g., \textit{Insufficient Grounding Quality}) into a theme such as \textit{Insufficient Data and Grounding Quality Across the FMware Stack}.}

\tinyskip\noindent\textbf{Step 4: Thematic Consolidation.}
\respto{1-7}\add{In Step 4, we consolidate one or more themes into practitioner-facing \emph{challenges}, formulated as actionable barriers to making FMware production-ready.} In the final step, the authors revisited and refined the overarching themes to ensure they comprehensively represented the identified challenges. The process prioritized clarity and relevance for practitioners while acknowledging the absence of exhaustive formal coding. \respto{1-7}\add{Section~\ref{sec:challenges} presents these themes, offering a systematic view of critical challenges in transitioning FMware from demos to production-ready systems.}

\add{\textbf{\textit{Running Example}}: we combine a data-and-grounding theme with a quality-assurance and governance theme to form the \textit{Built-in Quality} challenge.}

\section{The Lifecycle in Productionizing FMware}
\label{sec:pipeline}
\begin{figure*}
  \centering
  \includegraphics[width=1.02\textwidth]{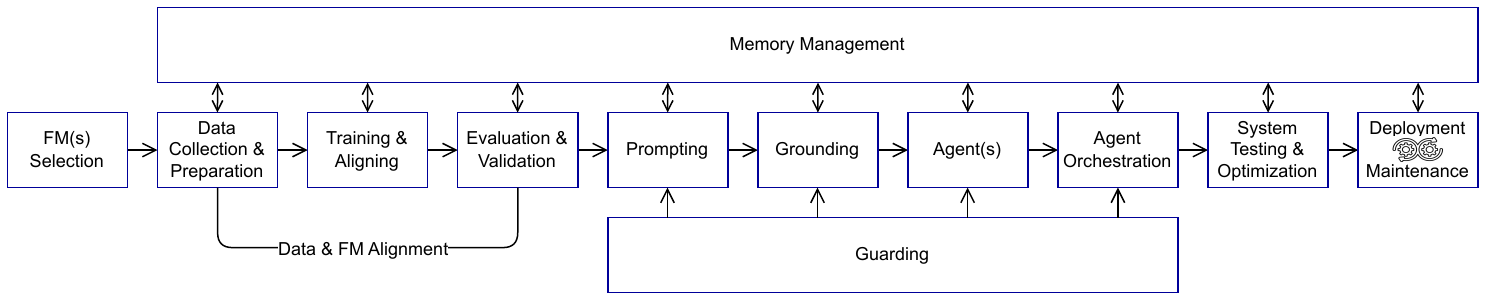}
  \caption{\respto{1-6b}\add{FMware lifecycle with procedural stages from FM(s) selection to deployment, with memory management and guarding shown as cross-cutting concerns for context preservation, safety, and compliance.}}
  \label{fig:lifecycle}
\end{figure*}

In this section, we outline the stages of the FMware engineering lifecycle, as depicted in Figure~\ref{fig:lifecycle}, and explain each stage.
\respto{1-6a}\add{
The FMware lifecycle begins with FM(s) selection, where teams choose externally hosted FM(s) or locally hosted ones that satisfy key constraints such as task performance, latency, privacy, and cost. Next, teams perform Data and FM(s) Alignment, where the selected FM(s) are adapted for the target task through data alignment (e.g., data preparation and curation) and FM(s) alignment (e.g., SFT, RLHF) on the prepared datasets. Prompting specifies the task interface and defines structured expectations for model behavior, such as input/output formats, constraints, and behavioral rules the model is expected to follow. Grounding connects the system to external knowledge through methods like Retrieval-Augmented Generation (RAG) or structured data sources to keep responses accurate, current, and reliable. Agent construction and orchestration integrate tools, memory, and policies to execute multi-step tasks. After this, System Testing and Optimization validate the quality, safety, and performance of the constructed FMware as a whole under realistic load and operational budgets. Deployment and Maintenance involve continuous monitoring and feedback loops that drive iterative updates to model selection, alignment, and prompt or policy designs. Memory management and Guarding are cross-cutting concerns that span multiple stages: Memory preserves relevant state for long-horizon tasks and efficient context usage in most stages, while Guarding enforces safety, compliance, and schema correctness across inputs (prompts), retrieved evidence (grounding), tool calls, and outputs during the agent's orchestration.} 

\respto{1-19} \respto{2-7} \add{We derive this lifecycle from our survey and industry experience, and present it as a descriptive baseline rather than a prescriptive process model (e.g., CRISP-DM~\cite{wirth2000crisp} or CPMAI~\cite{cpmai}). While not every FMware system traverses every stage (e.g., lightweight applications may require minimal alignment), the stages capture a common progression observed in practice. Most stages (\textit{FM(s) Selection}, \textit{Data and FM(s) Alignment}, \textit{Prompting}, \textit{Grounding}, \textit{Testing}, \textit{Deployment}, \textit{Memory Management}, and \textit{Guarding}) apply to both Promptware and Agentware; \textit{Agent Construction} and \textit{Orchestration} are primarily associated with Agentware.}

\respto{1-24} \add{This lifecycle view is consistent with practitioner LLMOps guidance that frames productionization as an end-to-end, continuously operated pipeline rather than a one-off integration~\cite{brousseau2025llms}.}

\subsection{FM(s) Selection}
The first step in developing FMware is to select the FM(s) to be used. There exist over 600,000 open-source models available (among FM(s) and the more traditional machine learning models)~\cite{factoredChoosingRight} on HuggingFace. Typically, the number of FM(s) to be selected depends on the complexity and functionality required by the FMware. For example, a simple chatbot may only need one FM, while complex systems might require multiple FMs for their different functionalities.

\subsection{Data and FM(s) Alignment}
\label{sec:dataFMalignment}
After selecting FM(s), the next crucial step is data and model alignment. 
\respto{1-2}\add{Data and model alignment refers to the end-to-end process of preparing high-quality datasets, adapting the FM(s) to domain-specific requirements and preferences, and systematically validating their behavior to ensure reliable, safe, and context-appropriate performance throughout its lifecycle~\cite{xia2024understanding}.
Following the structure proposed by \citet{wang2023aligning}, we divide this stage into three core alignment categories as follows.}


\add{\textbf{1) Data Collection and Preparation:} This sub-stage focuses on both acquisition and curation of diverse, high-quality, task-specific data. \textbf{a) Acquisition} involves gathering data from multiple sources, such as human-labeled datasets, publicly available corpora, enterprise data pipelines, and synthetic datasets generated through controlled processes or other FM(s). \textbf{b) Curation} is a critical and distinct step that goes beyond raw collection, ensuring that the data is representative, ethical, compliant, and tailored to FMware's target domain. This includes quality assessment (e.g., detecting noise, bias, or duplication), selection and filtering to balance domains and reduce overrepresentation, augmentation to address gaps or edge cases, and annotation refinement to ensure clear, unambiguous labels. The goal of this sub-stage is to build datasets that improve both input diversity and output quality, enabling FM(s) to operate effectively in real-world production settings while maintaining fairness, compliance, and reliability.}

\add{\textbf{2) Alignment Methods:} This sub-stage focuses on alignment, which we define as the process of adapting an FM(s) so that its behaviors and outputs are consistent with human intentions, safety requirements, and domain-specific constraints. While alignment enables the chosen FM(s) to perform well in the target domain, they also ensure that the FM(s) do so in a reliable and trustworthy manner. Teams can accomplish alignment through several key approaches: \textbf{a) Supervised Fine-Tuning (SFT)} involves carefully curated, task-specific datasets to directly guide the model toward desired behaviors. Parameter-efficient methods make SFT more scalable and cost-effective, enabling rapid iteration without extensive computational resources, e.g., LoRA and QLoRA. \textbf{b) Reinforcement Learning (RL) with Feedback Signals} uses a reward signal to refine model behavior. In addition to human- or AI-preference feedback (e.g., RLHF/RLAIF), teams also use task- and execution-derived signals, such as success/failure of tool calls, constraint violations, unit-test outcomes, or other automated evaluators, to shape behavior toward safety, compliance, and task success. \textbf{c) Preference-based and Inference-Time Alignment} includes Direct Preference Optimization (DPO) and related methods that optimize directly from preference data without explicit reward modeling. More broadly, inference-time alignment controls generation behavior through decoding and policy constraints, improving reliability and safety without updating model parameters.} \respto{1-24} \add{Also, model alignment extends beyond training to include inference-time behavior shaping. Parameters such as temperature, top-\textit{k} sampling, nucleus (\textit{p}) sampling, and repetition penalties critically affect the reliability and reproducibility of FM(s) outputs. Integrating these decoding controls within the alignment framework ensures that generation settings remain consistent across runs and that evaluation outcomes are reproducible \cite{bouchard2024building}.}

\add{\textbf{3) Evaluation and Validation:} The final sub-stage of the FMware lifecycle focuses on continuous and adaptive evaluation to ensure that alignment efforts remain effective over time. We structure this evaluation along three interconnected dimensions. \textbf{a) User-centric evaluation} involves direct feedback from human stakeholders, emphasizing whether the FMware aligns with operational goals such as safety, compliance, and reliability in real-world contexts. This includes expert reviews, usability studies, and operational audits to capture nuanced judgments that benchmarks alone cannot. \textbf{b) Benchmark-centric evaluation} provides scalable and repeatable quantitative assessments using static and dynamic benchmarks. These benchmarks measure reasoning, robustness, safety, and domain-specific performance while adapting to evolving operational demands to prevent overfitting to fixed datasets. \textbf{c) FM-as-judge evaluation} leverages FM(s) themselves as evaluators to handle large-scale, continuous monitoring that would be infeasible for humans alone. This includes meta-evaluation models, self-reflection loops, and hybrid pipelines that combine automated evaluation with selective human oversight. Together, these three dimensions form a closed-loop feedback system in which evaluation results are continuously fed back into data curation and training processes. This ensures that FMware not only performs well at deployment but also adapts dynamically to new conditions, maintaining trustworthiness and operational effectiveness throughout its lifecycle.}

\subsection{Prompting} 
\label{sec:prompting}
Prompting involves designing and refining prompts to guide FM(s) toward desired outputs using precise instructions and contextual cues. \respto{2-18}\add{One common mechanism is \emph{in-context learning (ICL)}, which conditions the model on examples placed inside the prompt rather than updating model parameters. These examples typically show input and output pairs, task constraints, and acceptance criteria, which serve as a soft specification for the desired behavior. In production, examples must be representative and sufficiently rich, which includes common cases and edge cases, so the model generalizes reliably under real workloads.} While demos can get away with simple or non-optimized prompts, production-ready FMware demands meticulous engineering for consistent and reliable performance~\cite{parthasarathy2024ultimate}.

\subsection{Grounding}
\label{sec:grounding}
\respto{1-4} \add{Grounding is the process of providing relevant, accurate, and appropriate context at inference time to mitigate hallucinations and link FM(s) responses to verifiable external data.} While techniques like In-Context Learning (ICL) and Retrieval-Augmented Generation (RAG) are the primary mechanisms for this, production-ready grounding requires moving beyond simple retrieval. \respto{1-24} \add{Advanced operations such as query expansion, sub-question decomposition, re-ranking, and recursive retrieval are often necessary to handle complex queries reliably. Furthermore, recent practitioner guidelines emphasize that grounding must be engineered as a durable infrastructure rather than ad hoc prompt tuning~\cite{kouri2025langchain}. This involves building centralized knowledge hubs that ingest heterogeneous sources into shared indices, utilizing hybrid retrieval (combining dense and lexical search), and enforcing citations to evidence as default behaviors. To ensure production reliability, these pipelines are often coupled with rigorous quantitative evaluators, such as recall@k, faithfulness, and latency profiling~\cite{bouchard2024building}.} \respto{1-4} \add{It is important to distinguish grounding from training-time alignment techniques like Supervised Fine-Tuning (SFT) or Reinforcement Learning (RL) (detailed in Section \ref{sec:dataFMalignment}). While SFT and RL adapt the model's internal behaviors and preferences, Grounding focuses exclusively on providing real-time, domain-specific context to ensure outputs remain current and verifiable in dynamic environments.}

\subsection{Agent(s)} Agents in FMware are autonomous components that leverage the decision-making and reasoning abilities of FM(s) to achieve tasks with minimal human input. They dynamically execute actions, interact with environments, and collaborate with other agents. \respto{1-5} \add{These agents span diverse domains, from software engineering tools like SWE-agents~\cite{yang2024swe}, to open-ended embodied agents like Voyager~\cite{wang2023voyager}, customer support systems~\cite{koualty2024generative}, sales and marketing applications~\cite{shareef2024retailgpt}, computer use agents \cite{computeruse} that interact directly with software interfaces, and deep research agents \cite{deepresearch} that conduct multi-step information synthesis.} By combining capabilities from previous software generations \add{with autonomous learning and adaptation, this} new generation \add{of agents enhances} the functionality of FMware \add{across multiple application domains}.

\subsection{Agent Orchestration}
\label{sec:agent_orchestration}
\add{Orchestrating multiple agents in FMware (both FM-based and traditional) ensures the handling of complex tasks by assigning sub-tasks to specialized agents, thereby improving scalability and reliability~\cite{rasal2024navigating}. This approach mirrors microservices architecture, where modular components collaborate toward a common goal to integrate flexibility and maintainability into production systems.}

\respto{1-24} \add{In practice, frameworks such as LangChain and LlamaIndex are used to implement these multi-step, tool-using agents on top of Prompting and Grounding stages~\cite{bouchard2024building}. Whether utilizing pre-determined flows or dynamic FM-generated workflows, production agentic systems are typically built as engineered compositions of the same core levers (e.g., prompt templates, retrieval components, and tool-call policies). Consequently, Agent Orchestration requires rigorous observability and guarding to be production-ready.}


\subsection{Guarding}
\label{sec:guarding}
Guarding ensures that FM(s) and Agents using FM(s) outputs are safe, accurate, and policy-compliant. Guardrails, e.g., implement ``rails'' for input moderation, fact-checking, and hallucination control, maximizing reliability of outputs in production FMware~\cite{rebedea2023nemo,dong2024framework}. \respto{1-3}\add{These protections are applicable for both Promptware and Agentware. In Promptware, output level guardrails operate directly on model inputs and responses; in Agentware, guarding extends to prompts, retrievals, tool calls, intermediate outputs, and memory interactions. Consistent with this scope, Figure~\ref{fig:lifecycle} depicts Guarding as a cross-cutting concern spanning Prompting and Grounding in addition to Agent and Agent Orchestration.} While demos may ignore these precautions, production systems must prioritize guardrails to ensure compliance, trust, and error prevention~\cite{rebedea2023nemo, dong2024framework}.

\subsection{FMware System Testing and Optimization}
\label{sec:testNoptimize}
FMware system testing and optimization are more complex than traditional software due to the non-deterministic nature of FM(s). Unlike conventional systems, where the behavior is more predictable, FMware involves continuously evolving FM(s) (particularly when using third-party FM(s)) that can silently degrade system performance without explicit changes to the underlying code.

Moreover, in production-ready FMware, system-wide testing must account for the dynamic interactions between agents, scalability concerns, and maintaining robustness under varying workloads. These challenges demand rigorous optimization, as the FMware needs to handle real-world complexities like fluctuating performance, inconsistent results, and unexpected edge cases. This goes beyond the controlled environments seen in demos, where simplicity and consistency are more achievable~\cite{ma2024my}. \respto{2-11}\add{Cross-cutting testing concerns, such as the lack of assertion-based unit tests and overreliance on text-based evaluation, are further discussed in Sections~\ref{sec:lacking_automated_tests}--\ref{sec:textbased_eval}}

\subsection{Deployment and Maintenance}
\label{sec:deploymentNmaintenance}
FMware deployment requires ongoing monitoring, maintenance, and adaptation. Unlike traditional software, FMware deployment must account for non-deterministic outputs and continuous learning. In production, the focus is on ensuring reliability, efficiency, and scalability under real-world conditions.

\subsection{Memory Management}
\label{sec:memory_management}
\respto{1-9} \add{Memory management refers to the processes and techniques by which an FMware system captures, organizes, serves, and maintains memory to support long-horizon reasoning and reliable operations. Concretely, memory management addresses four questions: \emph{what to add} (which new artifacts or signals to persist), \emph{what to update} (how to revise or replace stale knowledge as products, policies, or user needs evolve), \emph{what to serve} (how to prioritize and retrieve the most relevant fragments into prompts or tool calls at inference time), and \emph{what to delete} (how to prune obsolete or low-signal data to avoid bloat and context dilution). To operationalize these decisions at scale, FMware employs \emph{retrieval policies}, \emph{indexing strategies}, and \emph{memory-augmented generation} mechanisms to enable precise, cost-aware memory operations~\cite{borgeaud2022improving,jin2024ragcache,packer2023memgpt}. These mechanisms are central to long-horizon reasoning under limited context windows.} Memory management is a cross-cutting concern that affects multiple stages of the FMware lifecycle, including data alignment, prompting, grounding, agent orchestration, system optimization, and deployment.

\respto{1-9} \add{Effective memory management supplies FM(s) and agent components with relevant context, and supports storing and updating knowledge over time. These choices affect operational properties such as inference latency, retry behavior, and scalability~\cite{kwon2023efficient,li2024llm,llmpilot}. We distinguish two key memory types: (i) \emph{Episodic memory} (short-term, task-specific context within an ongoing session, e.g., prompts, intermediate reasoning steps, temporary results), which is widely used in interactive systems and single-session agents~\cite{fountas2024human,pink2025position}; and (ii) \emph{Long-term memory} (persistent knowledge across sessions, e.g., historical interactions, compliance rules, organizational policies, evolving domain knowledge), which is essential for personalization, compliance auditing, and multi-session workflows~\cite{chhikara2025mem0,tan2025prospect}.}

During deployment, memory management supports model swapping, where efficient context handling can prevent performance degradation from latency issues~\cite{llmpilot}. Memory is also important for managing retries by preserving context, preventing costly retry storms, and maintaining performance under load~\cite{microsoftRetryStorm,luo2024arena}. In addition, memory management supports multi-agent coordination by enabling controlled sharing of state and context. Robust memory management contributes to optimized dynamic workloads and regulatory compliance~\cite{hassan2024rethinking,wang2023survey}.

\respto{1-9} \add{Well-designed memory management has direct operational implications: (i) \emph{retry behaviors} (reusing stored state rather than re-executing full pipelines), (ii) \emph{multi-agent coordination} (safe, selective context sharing to maintain coherence across agents), and (iii) \emph{compliance auditing} (preserving decision trails for governance, safety, and regulatory review). As a cross-cutting concern, memory management underpins reliability, scalability, and organizational trust across the FMware lifecycle.}

\section{Recurrent Issues}
\label{sec:issues}

\respto{2-6} \add{This section catalogs recurrent, stage-specific problems practitioners encounter when productionizing FMware. Each issue is grounded in empirical observations from our survey and practice and mapped to the lifecycle stages in Figure~\ref{fig:lifecycle}. For ease of navigation, we organize issues into lifecycle-aligned subsections (e.g., FM(s) selection, prompting, grounding), with a final subsection for cross-cutting issues that span multiple stages, enabling direct mapping between recurrent issues and their lifecycle stages.}
\add{Within each stage, we report concrete practitioner pain points rather than abstract themes, keeping issues and evidence close to their operational context and avoiding conflation with the cross-cutting challenges synthesized later in Section~\ref{sec:challenges}.}

\respto{3-17b} \add{To support quick comparison across issues, we summarize each issue’s \emph{Symptom}, \emph{Conditions}, \emph{Consequences}, and \emph{Prototypical Signals} in Tables~\ref{tab:issues_fm_selection}--\ref{tab:issues_crosscutting} (grouped per lifecycle stage). To make the provenance of practitioner-facing evidence visible while preserving confidentiality, we also include \emph{Source Spotlight} callouts in the text where we can directly cite non-academic sources (e.g., field deployment notes, working-group discussions, or practitioner blog posts).} \respto{1-21} \add{Finally, Table~\ref{tab:traceability} provides a traceability map from lifecycle stages (Section~\ref{sec:pipeline}) to recurrent issues (Section~\ref{sec:issues}) and the cross-cutting challenges they motivate (Section~\ref{sec:challenges}), with representative supporting evidence.}

\subsection{FM(s) Selection}

\subsubsection{Difficulty Balancing Functional Requirements With Performance and Costs}
\label{sec:balancing_perf_cost}
\add{Selecting FM(s) for FMware requires balancing functionality against performance, infrastructure, and costs~\cite{kamath2024llms}. Demos often default to powerful general-purpose models like GPT-5.2 or Gemini-3 to maximize capability, typically neglecting the latency and resource constraints of high-volume usage. For instance, hosting a model like LLaMa 70B locally demands approximately 140GB of VRAM~\cite{chavan2024faster}, a prohibitive requirement for many production environments.}

\respto{2-9} \add{In production, the optimal choice depends heavily on the deployment context. Owner-managed environments, such as healthcare or on-device assistants~\cite{gunter2024apple}, prioritize privacy and latency. Teams in these settings often migrate to smaller or compressed models, utilizing techniques like quantization and pruning to reduce infrastructure overhead~\cite{lin2024awq, hohman2024model}. Conversely, cloud-hosted applications like coding assistants (e.g., Lovable, ReplIt) often rely on large external FM(s) where the provider manages inference scaling. Even in these cases, the model providers themselves are increasingly utilizing Mixture-of-Experts (MoE) architectures to balance capability with efficiency~\cite{moe_in_LLM}. The recurrent issue here is that neither path is free: smaller models risk degrading answer quality, while larger models risk violating cost or latency Service Level Objectives (SLOs)~\cite{chen2024role,irugalbandara2024scaling}.}


\begin{table}[t]
\centering
\begingroup
\scriptsize
\setlength{\tabcolsep}{1.5pt}
\renewcommand{\arraystretch}{0.92}
\setlength{\emergencystretch}{1em}
\newcommand{\riSep}{\newline--\ }
\newcommand{\riRowSep}{\addlinespace\midrule\addlinespace}

\caption{Recurrent issues in FM(s) Selection stage with symptoms, conditions, consequences, and prototypical signals.}
\label{tab:issues_fm_selection}

\begin{tabular}{@{}
  >{\raggedright\arraybackslash}p{0.16\textwidth}
  >{\raggedright\arraybackslash}p{0.21\textwidth}
  >{\raggedright\arraybackslash}p{0.21\textwidth}
  >{\raggedright\arraybackslash}p{0.19\textwidth}
  >{\raggedright\arraybackslash}p{0.23\textwidth}
  >{\raggedright\arraybackslash}p{0.001\textwidth}
@{}}
\toprule
\textbf{Recurrent Issues} &
\textbf{Symptoms} &
\textbf{Conditions} &
\textbf{Consequences} &
\textbf{\shortstack[l]{Prototypical\\Signals}} &
{} \\
\midrule

\ref{sec:balancing_perf_cost} Difficulty Balancing Functional Requirements With Performance and Costs &
-- Struggle to pick one FM that meets capability plus latency, cost, privacy constraints &
-- Demo to sustained traffic\riSep Edge/on-prem memory limits\riSep Multi-turn interactions inflate context &
-- Over-sized: SLO misses, cost overruns\riSep Under-sized: capability gaps, degraded quality, brittle prompt workarounds, retries\riSep Ownership costs still high (autoscaling, upgrade churn) &
-- Tail latency (p95/p99) up\riSep Token cost per successful task up\riSep GPU VRAM pressure (often tens to hundreds of GB depending on precision and serving setup)\riSep Declining accepted responses\riSep Retry rate tracks traffic bursts &
\\ 

\bottomrule
\end{tabular}
\endgroup
\end{table}

\subsection{Data and FM Alignment}
\subsubsection{Low Data Quality}
\label{sec:low_data_qual}
\add{Data quality is a foundational bottleneck in FMware production~\cite{albalak2024survey}. As noted by \citet{djuhera2025fixing}, successful alignment depends heavily on specific dataset attributes, including the clarity of prompts, the diversity of responses, and the complexity of represented tasks. However, curating data that meets these standards is difficult because the specific quality dimensions required for a target domain are often unknown until after alignment fails. Consequently, practitioners frequently struggle to acquire ``gold-standard'' data that adequately represents the nuances of the production environment~\cite{albalak2024survey}.}





\subsubsection{Low Domain Coverage}
\label{sec:low_domain_cov}
\add{\emph{Domain coverage} denotes how well the data, and consequently the model, represent the target domain, including typical use cases, edge cases, compliance requirements, and rare but critical scenarios. When coverage is poor, gaps in data representation surface as unpredictable behaviors in production, which undermines reliability and increases the risk of biased or non-compliant outcomes~\cite{li2023ood_eval_survey,yang2023ood_generalization_nlp}.} Ensuring broad domain coverage is key to trustworthy FMware. Insufficient domain representation can result in biased and unreliable outputs.

\subsubsection{ Low Data Efficiency}
\label{sec:low_data_eff}
\add{Achieving data efficiency is a dual challenge that impacts both model training and production inference. Practitioners frequently struggle to determine the precise ``sufficiency threshold'' for their tasks, as the number of data points required for effective fine-tuning or the optimal number of examples for in-context learning remains an open research question~\cite{chen2023maybe}. This uncertainty often leads to defensive engineering: teams may curate unnecessarily large alignment datasets or over-stuff context windows with redundant examples. Furthermore, the increasing use of verbose Chain-of-Thought (CoT) reasoning strategies significantly inflates output token counts \cite{Aytes2025SketchofThoughtELA}. While these strategies improve reasoning, they impose a heavy tax on the context window, directly degrading latency and increasing inference costs without always yielding proportional value \cite{su2025between, wu2025more}.}


\begin{table}[t]
\centering
\begingroup
\scriptsize
\setlength{\tabcolsep}{1.5pt}
\renewcommand{\arraystretch}{0.92}
\setlength{\emergencystretch}{1em}
\newcommand{\riSep}{\newline--\ }
\newcommand{\riRowSep}{\addlinespace\midrule\addlinespace}

\caption{Recurrent issues in Data and FM Alignment stage with symptoms, conditions, consequences, and prototypical signals.}
\label{tab:recurrent_issues_matrix}

\begin{tabular}{@{}
  >{\raggedright\arraybackslash}p{0.16\textwidth}
  >{\raggedright\arraybackslash}p{0.21\textwidth}
  >{\raggedright\arraybackslash}p{0.21\textwidth}
  >{\raggedright\arraybackslash}p{0.19\textwidth}
  >{\raggedright\arraybackslash}p{0.23\textwidth}
  >{\raggedright\arraybackslash}p{0.001\textwidth}
@{}}
\toprule
\textbf{Recurrent Issues} &
\textbf{Symptoms} &
\textbf{Conditions} &
\textbf{Consequences} &
\textbf{\shortstack[l]{Prototypical\\Signals}} &
{} \\
\midrule

\ref{sec:low_data_qual} Low Data Quality &
-- Brittle/inconsistent/wrong outputs due to noisy labels\riSep Stale facts\riSep Spurious correlations &
-- Rapid heterogeneous datasets\riSep Limited provenance\riSep Web-scale scraping without post-hoc curation\riSep Fast-changing domains &
-- Lower accuracy and trust\riSep Higher cost (retries, human overrides)\riSep Safety and compliance risk &
-- Human-output disagreement up\riSep Metric drift after data refresh\riSep Frequent prompt hot-fixes masking data issues &
\\ \riRowSep

\ref{sec:low_domain_cov} Low Domain Coverage &
-- Fails on edge cases or specialised subdomains underrepresented in alignment data or retrieval indices &
-- Long-tail domains\riSep Evolving rules\riSep Multilingual or multi-regional requirements\riSep Sparse or proprietary knowledge &
-- Systematic capability gaps\riSep Biased or non-inclusive behaviour\riSep Higher human fallback\riSep Reduced trust and adoption &
-- Errors concentrated in specific intents/entities\riSep High null retrieval for niche queries\riSep Frequent escalation on specialised topics &
\\ \riRowSep

\ref{sec:low_data_eff} Low Data Efficiency &
-- Token and bandwidth grow faster than traffic\riSep Context saturated with low-value content\riSep Storage/retrieval costs dominate &
-- Multi-turn workflows\riSep Verbose intermediate reasoning (CoT)\riSep Broad retrieval without filtering\riSep Generous logging retention &
-- SLO violations from high generation latency\riSep Cost overruns from unnecessary output tokens\riSep Context exhaustion and truncation &
-- Average tokens per task up (especially output tokens)\riSep Bytes per request up\riSep High cache miss with low retrieval relevance\riSep Irrelevant context share in audits increases &
\\

\bottomrule
\end{tabular}
\endgroup
\end{table}





\subsection{Prompting}
\subsubsection{God Prompts}
\label{sec:god_prompts}
\respto{2-8}\add{\emph{God prompt} is a large and monolithic instruction block that mixes task directives, examples, rules, error handling, and even tool usage descriptions into one place, which effectively encodes a fixed workflow in natural language (similar to God classes in traditional programming). ``God prompts'' typically attempts to handle multiple tasks at once, leading to unpredictable outputs and complicated debugging and maintenance~\cite{oreillyWhatLearned,parthasarathy2024ultimate}. Different prompting strategies like \emph{Chain-of-thought} or \emph{checklist prompting}, which may increase the size of the prompt does not contribute to this recurrent issue. A God prompt may include chain-of-thought-style phrasing, but it typically also accumulates additional directives (e.g., multiple tasks, rules, and edge-case handling) that make the prompt multi-purpose and tightly coupled.}

\respto{2-8}\add{God prompts impact both non-reasoning and reasoning models (i.e., models that perform an implicit chain-of-thought \cite{wei2022chain} generating a response, e.g., GPT-4o, Claude-Opus). In fact, several frontier reasoning FM providers recommend using modular and task-specific prompts that preserve a clear alignment between instructions and objectives~\cite{prompt_guide}. In production, tasks should be split into smaller, focused prompts and be decoupled for better accuracy, debuggability, and maintainability~\cite{oreillyWhatLearned}. Modular prompts reduce \emph{lost in the middle} effects~\cite{liu2023lost} and \emph{needle in a haystack} failures~\cite{wang2024multimodal}, which are common when instructions are overly long or mix unrelated goals. Industry guides likewise recommend simple, direct, single-purpose prompts rather than mixed-purpose blocks~\cite{prompt_guide,reasoning_best_practices}.}





\begin{sourcespotlight}
\add{\citet{oreillyWhatLearned} suggest: ``Have small prompts that do one thing, and only one thing, well'' and note that the ``God Object'' anti-pattern applies to prompts as well.}
\end{sourcespotlight}

\subsubsection{Lack of Built in QA Checks for Prompts}
\label{sec:prompt_QA}
Production-ready FMware demands robust QA for prompts, including semantic and structural validation. Without built-in QA checks, prompts may produce erroneous or inappropriate responses, leading to failures and loss of user trust~\cite{kamoi2024evaluating,ahmed2023better,wang2024minstrel}. \add{In practice, prompt QA treats prompts as versioned artefacts and uses automated validation and regression tests to check output format, instruction adherence, and safety constraints on representative and adversarial inputs.} 





\subsubsection{Lack of Prompt Portability Across FM(s)}
\label{sec:prompt_portability}
Prompts optimized for one FM(s) typically fail with others due to differing architectures and training data~\cite{chang2024efficient}, which limits portability and complicates software evolution.  





\subsubsection{Absence of FM(s) Specific Optimizations}
\label{sec:fm_optimizations}
FM-specific optimizations can boost performance by up to 10\% by tailoring prompts to each model's architecture~\cite{chen2024mapo}. These enhancements create dependencies that complicate updates and maintenance~\cite{sabbatella2024prompt}.  





\subsubsection{Complexity in Determining Failure Rationale}
\label{sec:complexity_failure_rationale}
Diagnosing whether failures result from prompt issues or FM(s) shortcomings is difficult, which hinders root cause identification and delays resolution~\cite{ashtari2023discovery,amujo2024good}. \add{For example, a customer-support agent which decides refunds and outputs a JSON rationale. After adding a new policy rule and a few in-context examples, the agent starts denying eligible refunds. The regression is hard to attribute: it may stem from prompt interactions (conflicting rules, a misleading example, or changed instruction priority), or from an FM limitation (instability under long, multi-part prompts). Practitioners typically triage this by replaying a fixed test set, ablating prompt components (rules, examples, ordering), and cross-checking across models to localize the cause~\cite{ashtari2023discovery,amujo2024good}.}





\subsubsection{Non Representative or Insufficient In-Context Learning Examples}
\label{sec:no_good_icl_ex}
Providing insufficient or non-representative examples for in-context learning degrades FM(s) performance. For instance, \citet{wang2023learning} shows that random in-context examples yield suboptimal outputs. Production systems require comprehensive and relevant examples for reliability.  


\begin{table}[t]
\centering
\begingroup
\scriptsize
\setlength{\tabcolsep}{1.5pt}
\renewcommand{\arraystretch}{0.92}
\setlength{\emergencystretch}{1em}
\newcommand{\riSep}{\newline--\ }
\newcommand{\riRowSep}{\addlinespace\midrule\addlinespace}

\caption{Recurrent issues in Prompting stage with symptoms, conditions, consequences, and prototypical signals.}
\label{tab:recurrent_issues_matrix}

\begin{tabular}{@{}
  >{\raggedright\arraybackslash}p{0.16\textwidth}
  >{\raggedright\arraybackslash}p{0.21\textwidth}
  >{\raggedright\arraybackslash}p{0.21\textwidth}
  >{\raggedright\arraybackslash}p{0.19\textwidth}
  >{\raggedright\arraybackslash}p{0.23\textwidth}
  >{\raggedright\arraybackslash}p{0.001\textwidth}
@{}}
\toprule
\textbf{Recurrent Issues} &
\textbf{Symptoms} &
\textbf{Conditions} &
\textbf{Consequences} &
\textbf{\shortstack[l]{Prototypical\\Signals}} &
{} \\
\midrule

\ref{sec:god_prompts} God Prompts &
-- Monolithic prompts cause surprising behaviours\riSep Hidden coupling across tasks\riSep Small edits trigger regressions &
-- Incremental prompt accretion\riSep Mixed intents in one workflow\riSep No modular decomposition or regression tests\riSep Weak observability of intermediates &
-- Hard root cause analysis and slow mitigation\riSep Higher token usage\riSep Poor transfer across models &
-- Output variance rises for same intent\riSep Frequent prompt hotfixes with collateral regressions\riSep Prompt length and unrelated instructions grow &
\\ \riRowSep

\ref{sec:prompt_QA} Lack of Built in QA Checks for Prompts &
-- Silent prompt drift, schema violations, policy breaches reach production outputs &
-- Manual edits without review gates\riSep No schema/contract checks\riSep Limited red teaming or adversarial evaluation &
-- More incidents and moderation escalations\riSep Compliance exposure\riSep Costly human-in-the-loop overrides &
-- Spikes in moderation flags\riSep Downstream schema parsing errors\riSep Rising share of guardrail interventions &
\\ \riRowSep

\ref{sec:prompt_portability} Lack of Prompt Portability Across FM(s) &
-- Prompt works on one FM but underperforms or fails on another with similar capability &
-- Model/vendor switches for cost, latency, privacy\riSep Mixed-model stacks\riSep Staged side-by-side rollouts &
-- Migration delays\riSep Duplicated maintenance\riSep Non-comparable evaluations\riSep Increased lock-in &
-- Large accuracy/safety deltas on swap\riSep Model-specific prompt forks proliferate\riSep Canary skip rate grows &
\\ \riRowSep

\ref{sec:fm_optimizations} Absence of FM(s) Specific Optimizations &
-- Generic prompts underperform tuned variants\riSep Tuned prompts hard to reuse across versions &
-- Heterogeneous model fleet with periodic updates\riSep Tight SLOs\riSep Limited automation for prompt variant management &
-- Regressions during upgrades\riSep Prompt sprawl and configuration drift\riSep Brittle release pipelines &
-- Step drops in win rate after FM updates\riSep Many near-duplicate prompts keyed to model IDs\riSep Rollback frequency rises after refreshes &
\\ \riRowSep

\ref{sec:complexity_failure_rationale} Complexity in Determining Failure Rationale &
-- Intermittent failures with unclear attribution to prompt phrasing, retrieval context, or model behaviour &
-- Limited intermediate logging\riSep No ablation-style prompt tests\riSep Shared prompts across multiple tasks confound attribution &
-- Complex debugging\riSep Costly overcorrection (prompt and model changed together)\riSep Repeated incidents on same intents &
-- Postmortems often cite unclear root causes\riSep Frequent parallel prompt and model changes in one release\riSep Reproduction rate of issue reports is often low &
\\ \riRowSep

\ref{sec:no_good_icl_ex} Non Representative or Insufficient In-Context Learning Examples &
-- Overfits to trivial patterns\riSep Fails on realistic edge cases\riSep Bias toward frequent templates &
-- Minimal example curation\riSep Random sampling\riSep Domain shift between examples and live traffic\riSep Tasks evolve without refreshing examples &
-- Lower accuracy and consistency\riSep More verbose instructions inflate tokens\riSep Increased fallback to human review &
-- Curated eval vs production gap\riSep Token per task rises due to compensating instructions\riSep Errors concentrate on intents missing from examples &
\\

\bottomrule
\end{tabular}
\endgroup
\end{table}





\subsection{Grounding}
\subsubsection{Low Information Density in Grounding Data}
\label{sec:low_density}

\respto{2-19}\add{Low-density grounding information refers to situations where the available grounding sources (e.g., documents, logs, policies, or retrieval corpora) provide insufficient, fragmented, or stale evidence to support accurate responses, leading to higher hallucination risk and unstable behaviour \cite{laban2024summary,wang2024adapting}.}
In production, rich and well-structured data is essential for accurate outputs. \citet{laban2024summary} shows that poorly structured and vast information leads to failures in long context tasks, while \citet{wang2024adapting} emphasizes efficient data compression and selection to maintain performance and relevance in complex queries.






\subsubsection{Irrelevant Grounding Data}
\label{sec:irrelevant_grounding}
Incorporating irrelevant grounding data confuses FM(s) and reduces performance. \citet{liu2024lost} shows that models struggle with long context information, especially when crucial data is buried, while \citet{cuconasu2024power} notes how noisy data worsens the focus of retrieval systems. For production environments, filtering and curating data are crucial to ensure only pertinent information is used.





\subsubsection{Over Complicating Solutions With Advanced Techniques}
\label{sec:over_complicating_solutions}
Using complex methods when simpler ones suffice adds unnecessary overhead. For instance, neural retrieval can raise computational costs without added benefit when a keyword search would be enough~\cite{oreillyWhatLearned,sawarkar2024blended}. In production, simplicity promotes scalability and maintainability.


\begin{table}[t]
\centering
\begingroup
\scriptsize
\setlength{\tabcolsep}{1.5pt}
\renewcommand{\arraystretch}{0.92}
\setlength{\emergencystretch}{1em}
\newcommand{\riSep}{\newline--\ }
\newcommand{\riRowSep}{\addlinespace\midrule\addlinespace}

\caption{Recurrent issues in Grounding stage with symptoms, conditions, consequences, and prototypical signals.}
\label{tab:recurrent_issues_matrix}

\begin{tabular}{@{}
  >{\raggedright\arraybackslash}p{0.16\textwidth}
  >{\raggedright\arraybackslash}p{0.21\textwidth}
  >{\raggedright\arraybackslash}p{0.21\textwidth}
  >{\raggedright\arraybackslash}p{0.19\textwidth}
  >{\raggedright\arraybackslash}p{0.23\textwidth}
  >{\raggedright\arraybackslash}p{0.001\textwidth}
@{}}
\toprule
\textbf{Recurrent Issues} &
\textbf{Symptoms} &
\textbf{Conditions} &
\textbf{Consequences} &
\textbf{\shortstack[l]{Prototypical\\Signals}} &
{} \\
\midrule

\ref{sec:low_density} Low Information Density in Grounding Data &
-- Vague/repetitive/off-target responses when retrieval returns long passages with little salient content per token &
-- Unnormalised document dumps\riSep Long unstructured pages\riSep Verbose logs/transcripts\riSep No summarisation or chunk-level relevance scoring &
-- Lower signal-to-noise prompts\riSep Higher latency and cost\riSep Reduced answer fidelity due to diluted key facts &
-- Tokens per retrieved passage rise with flat/falling accuracy\riSep Frequent truncation at context limits\riSep Users report hedging or missed facts present in corpus &
\\ \riRowSep

\ref{sec:irrelevant_grounding} Irrelevant Grounding Data &
-- Cites sources that do not answer the question\riSep Contradicts trusted references\riSep Hallucinates despite retrieval &
-- Recall-heavy retrieval without precision filters\riSep Weak query reformulation\riSep Stale/mismatched indices\riSep Mixed-quality corpora without tiers &
-- Accuracy drops despite higher retrieval volume\riSep Moderation risk\riSep Wasted tokens\riSep User frustration &
-- Low top-$k$ precision in offline evals\riSep High irrelevant chunk share in audits\riSep Answer quality may improve when retrieval is disabled (a signal that retrieval is adding noise) &
\\ \riRowSep

\ref{sec:over_complicating_solutions} Over Complicating Solutions With Advanced Techniques &
-- Complex pipelines match or underperform simpler baselines while consuming more compute and engineering effort &
-- Optimising for novelty early\riSep Default dense retrieval/reranking without baseline checks\riSep Limited ablation vs lexical or rule-based approaches &
-- Higher latency and spend\riSep Fragile dependencies slow incident response\riSep Harder onboarding and maintenance &
-- Lexical baseline may match or beat dense stack on key tasks\riSep Cost per successful task rises after adding complexity\riSep Regressions tied to rerankers/rewriters &
\\ 
\bottomrule
\end{tabular}
\endgroup
\end{table}





\begin{sourcespotlight}
\add{\citet{oreillyWhatLearned} suggest ``Don’t forget keyword search; use it as a baseline and in hybrid search.'' when using RAG. They mention that "keyword search is usually more computationally efficient" and they also borrow words from famous people like Aravind Srinivas, CEO Perplexity.ai, and Beyang Liu, CTO Sourcegraph, saying that dense embedding search alone do not guarantee good search results; in many cases, keyword-based methods like BM25 are more reliable and efficient, so starting with keyword search is recommended before adding semantic or embedding-based retrieval.}
\end{sourcespotlight}

\subsection{Agent(s)}
\subsubsection{God Agents}
\label{sec:god_agents}
Monolithic ``God agents'' (similar to God prompts) that handle many tasks create complexity and maintenance challenges in production. \add{Whether in software development, customer service, or research contexts,} smaller specialized agents are preferred for modularity and reliability, offering better scalability and maintainability~\cite{rasal2024navigating,chen2024coder}.





\subsubsection{Too Low Level Tool Usage Abstraction}
\label{sec:low_level_tool}
Agents that require developers to interact with low-level tools increase system complexity and hinder scalability in production~\cite{yang2024swe}. When agents demand manual management of API calls or intricate state transitions, they become error-prone and harder to maintain. \add{This challenge affects agents across domains, from software engineering agents managing code repositories to customer service agents handling natural language understanding and dialogue management systems~\cite{koualty2024generative}.} High-level abstractions, which hide tool interactions, are crucial for reducing bugs and allowing developers to focus on business logic rather than technical details. \add{Recent practice is converging on standardized agent-tool interfaces such as the Model Context Protocol (MCP) \cite{hasan2025model}, which externalizes tools behind a client-server abstraction and reduces bespoke, low-level integrations. Related ecosystem efforts also promote shared conventions for agent configuration and guidance (e.g., AGENTS.md) and first-class tool-calling and hosted connector support, which collectively shift developers away from manual API orchestration toward reusable, portable tool abstractions.}





\subsubsection{Capability Centric Instead of Use Case Centric Documentation for Tools}
\label{sec:use_case_centric}
Documentation that focuses on tool capabilities rather than specific use cases makes it harder for FM(s) to leverage these tools. \add{Beyond missing use cases, unoptimized tool descriptions are themselves a frequent source of tool-use failures: vague or incomplete descriptions, unclear parameter semantics, and uninformative success or error messages can mislead the model about when and how to invoke a tool, since these artifacts are injected into the model context and shape its behavior \cite{roig2025tool_descriptions}.} FM(s) need clear and practical examples to operate effectively, as seen in adaptive systems like Voyager, where agents excel at tool use by leveraging concrete use cases~\cite{wang2023voyager}. \add{This pattern extends across domains: sales agents benefit from example customer interaction flows~\cite{shareef2024retailgpt}, customer service agents benefit from example conversation flows and FAQ structures~\cite{koualty2024generative}, computer use agents \cite{computeruse} require demonstrated interface interaction patterns, and research agents \cite{deepresearch} need exemplar multi-step synthesis workflows.} Without this, the learning curve steepens, increasing the risk of errors and slowing agent evolution~\cite{wang2023voyager,yang2024swe,chen2024coder}. In production, use case-driven guidance is crucial for building reliable systems across these application domains~\cite{wang2023voyager}.





\begin{sourcespotlight}
\add{Anthropic~\cite{computeruse} mentions, ``Instead of making specific tools to help Claude complete individual tasks, we're teaching it general computer skills, allowing it to use a wide range of standard tools and software programs designed for people.'' OpenAI~\cite{deepresearch} mentions, ``Deep research independently discovers, reasons about, and consolidates insights from across the web. To accomplish this, it was trained on real-world tasks requiring browser and Python tool use...''}
\end{sourcespotlight}


\begin{table}[t]
\centering
\begingroup
\scriptsize
\setlength{\tabcolsep}{1.5pt}
\renewcommand{\arraystretch}{0.92}
\setlength{\emergencystretch}{1em}
\newcommand{\riSep}{\newline--\ }
\newcommand{\riRowSep}{\addlinespace\midrule\addlinespace}

\caption{Recurrent issues in Agent(s) stage with symptoms, conditions, consequences, and prototypical signals.}
\label{tab:recurrent_issues_matrix}

\begin{tabular}{@{}
  >{\raggedright\arraybackslash}p{0.16\textwidth}
  >{\raggedright\arraybackslash}p{0.21\textwidth}
  >{\raggedright\arraybackslash}p{0.21\textwidth}
  >{\raggedright\arraybackslash}p{0.19\textwidth}
  >{\raggedright\arraybackslash}p{0.23\textwidth}
  >{\raggedright\arraybackslash}p{0.001\textwidth}
@{}}
\toprule
\textbf{Recurrent Issues} &
\textbf{Symptoms} &
\textbf{Conditions} &
\textbf{Consequences} &
\textbf{\shortstack[l]{Prototypical\\Signals}} &
{} \\
\midrule

\ref{sec:god_agents} God Agents &
-- One agent accumulates many roles and tools\riSep Decision traces opaque\riSep Fixes in one area affect others &
-- PoC growth without clear boundaries\riSep Shared memory/tools across unrelated skills\riSep Missing contracts for handoffs &
-- Slow incident response and fragile deployments\riSep Limited parallelism\riSep Costly retraining or re-prompting when requirements change &
-- Rising span of control\riSep Long multi-purpose policies\riSep Frequent regressions after minor edits\riSep Declining success on previously stable tasks &
\\ \riRowSep

\ref{sec:low_level_tool} Too Low Level Tool Usage Abstraction &
-- Business logic tangled with tool protocol details\riSep Small API changes cascade\riSep Onboarding needs deep internals &
-- Direct invocation of heterogeneous tools without adapters\riSep Ad hoc state machines\riSep Inconsistent error handling and retries &
-- Higher defect rates, slower feature velocity\riSep Duplicated integration code\riSep Hard to scale to new tools/vendors &
-- High fraction of changes touch tool bindings\riSep Frequent hot-fixes after upstream API updates\riSep Divergent patterns for same tool across teams &
\\ \riRowSep

\ref{sec:use_case_centric} Capability Centric Instead of Use Case Centric Documentation for Tools &
-- Agents misuse tools or fail to invoke them\riSep Prompts reference tool abilities without task scaffolds &
-- Capability lists without end-to-end examples\riSep Missing input-output schemas\riSep Lack negative examples and guardrails &
-- Lower task success\riSep Higher token use from exploratory calls\riSep Inconsistent behaviour\riSep Increased human escalation &
-- High no-op/error-return rates\riSep Repetitive clarification turns pre-tool\riSep Performance improves when adding task-oriented examples &
\\

\bottomrule
\end{tabular}
\endgroup
\end{table}

\subsection{Agent Orchestration}
\subsubsection{Over Reliance on FM(s) Planning Capability}
\label{sec:over_planning}
Relying solely on an FM's planning capabilities in production FMware introduces risks like unpredictability and lack of transparency, often leading to errors and unreliable outcomes due to issues such as hallucinations or failure to align with business logic~\cite{guo2024large,huang2024understanding}. In contrast, demos may succeed with FM-based planning due to their controlled environment and limited scope. \respto{2-10}\add{In production systems, planning and acting are often separated into distinct phases, where the model first analyzes and decomposes the task, then proceeds to execution and code generation. Systems such as Claude Code and Replit exemplify this pattern, with a planning or architect mode distinct from a writer or executor mode \cite{claude_tips}. This separation also encourages prompt modularity by assigning each phase a narrow objective, which is consistent with guidance that recommends simple, single-purpose prompts rather than mixed-purpose instruction blocks~\cite{prompt_guide, reasoning_best_practices}.}

\respto{2-10}\add{Experience with early agentic frameworks shows the pitfalls of over-reliance on FM planning capability. AutoGPT and BabyAGI frequently exhibited recursive loops, drifting goals, and incomplete tasks when high-level planning and execution were combined without constraints \cite{yang2023auto, babyagi}. Even when planning is separated from acting, very large or abstract plans produced in a single step tend to drift, accumulate errors, and complicate debugging when reused or scaled. Practical recommendations are therefore to use planning selectively and within bounded scopes, to iterate plans with feedback loops, and to instrument each step with observability hooks and control points so deviations can be intercepted early.}





\begin{sourcespotlight}
 \add{\citet{claude_tips} describes how Claude has modes beyond just code generation, specifically noting: "Try the enable-architect mode for more complex tasks," which highlights a distinct planning/architect mode that is separate from the writing/execution phase. Similar to Section \ref{sec:use_case_centric}, OpenAI \cite{prompt_guide, reasoning_best_practices} suggests: ``Keep prompts simple and direct: The models excel at understanding and responding to brief, clear instructions.''}
\end{sourcespotlight}


\begin{table}[t]
\centering
\begingroup
\scriptsize
\setlength{\tabcolsep}{1.5pt}
\renewcommand{\arraystretch}{0.92}
\setlength{\emergencystretch}{1em}
\newcommand{\riSep}{\newline--\ }
\newcommand{\riRowSep}{\addlinespace\midrule\addlinespace}

\caption{Recurrent issues in Agent Orchestration stage with symptoms, conditions, consequences, and prototypical signals.}
\label{tab:recurrent_issues_matrix}

\begin{tabular}{@{}
  >{\raggedright\arraybackslash}p{0.16\textwidth}
  >{\raggedright\arraybackslash}p{0.21\textwidth}
  >{\raggedright\arraybackslash}p{0.21\textwidth}
  >{\raggedright\arraybackslash}p{0.19\textwidth}
  >{\raggedright\arraybackslash}p{0.23\textwidth}
  >{\raggedright\arraybackslash}p{0.001\textwidth}
@{}}
\toprule
\textbf{Recurrent Issues} &
\textbf{Symptoms} &
\textbf{Conditions} &
\textbf{Consequences} &
\textbf{\shortstack[l]{Prototypical\\Signals}} &
{} \\
\midrule

\ref{sec:over_planning} Over Reliance on FM(s) Planning Capability &
-- Plans vague or too high-level\riSep Decomposition opaque\riSep Execution drifts without checkpoints or stop conditions &
-- End-to-end planning delegated to one FM call\riSep No explicit task graphs/policies\riSep Weak verification of intermediates\riSep Limited guardrails/tool preconditions &
-- Flaky production behaviour\riSep Higher cost from unnecessary tool calls and retries\riSep Harder compliance and audit (missing rationale and step evidence) &
-- High variance in step counts for same intent\riSep Frequent mid-flight loops/reversions\riSep Improved stability with schema-constrained subtasks or simpler planners &
\\

\bottomrule
\end{tabular}
\endgroup
\end{table}

\subsection{Guarding}
\subsubsection{Simple Keyword Based Guarding is Ineffective}
\label{sec:keyword_guarding_ineffective}
Relying solely on keyword-based guarding is ineffective in production environments. Simple keyword filters can be easily bypassed or can mistakenly block legitimate content due to language ambiguity. In production-ready FMware, sophisticated guardrails such as fact checking, semantic filtering, or policy-guided decoding are essential for discerning intent and context, reducing both false positives and false negatives~\cite{rebedea2023nemo,dong2024framework}.





\subsubsection{Lack of Hallucination Guardrails}
\label{sec:lack_hall_guardrails}
Hallucinations pose significant risks in production systems, especially in domains such as healthcare or law, where factual accuracy is crucial. Fabricated outputs can spread misinformation, undermine user trust, and create legal liabilities~\cite{ayyamperumal2024current,dahl2024hallucinating}. In addition, FM(s) often lack calibrated confidence estimation, which makes it difficult to flag uncertain outputs for human review before harm occurs. User feedback mechanisms are frequently underutilized, which prevents continuous correction of inaccuracies~\cite{ayyamperumal2024current,dahl2024hallucinating,spiess2024quality}.


\begin{table}[t]
\centering
\begingroup
\scriptsize
\setlength{\tabcolsep}{1.5pt}
\renewcommand{\arraystretch}{0.92}
\setlength{\emergencystretch}{1em}
\newcommand{\riSep}{\newline--\ }
\newcommand{\riRowSep}{\addlinespace\midrule\addlinespace}

\caption{Recurrent issues in Guarding stage with symptoms, conditions, consequences, and prototypical signals.}
\label{tab:recurrent_issues_matrix}

\begin{tabular}{@{}
  >{\raggedright\arraybackslash}p{0.16\textwidth}
  >{\raggedright\arraybackslash}p{0.21\textwidth}
  >{\raggedright\arraybackslash}p{0.21\textwidth}
  >{\raggedright\arraybackslash}p{0.19\textwidth}
  >{\raggedright\arraybackslash}p{0.23\textwidth}
  >{\raggedright\arraybackslash}p{0.001\textwidth}
@{}}
\toprule
\textbf{Recurrent Issues} &
\textbf{Symptoms} &
\textbf{Conditions} &
\textbf{Consequences} &
\textbf{\shortstack[l]{Prototypical\\Signals}} &
{} \\
\midrule

\ref{sec:keyword_guarding_ineffective} Simple Keyword Based Guarding is Ineffective &
-- Harmful outputs slip through word lists\riSep Benign content over-blocked\riSep Noisy UX and high moderator load &
-- Static allow/block lists without semantic checks\riSep Limited context awareness in multi-turn\riSep No structure/schema validation before delivery &
-- Incidents and user friction\riSep Brittle rules require constant manual updates\riSep Compliance exposure &
-- High false-positive/false-negative rates in moderation logs\riSep Repeated rule additions for same behaviour\riSep Improvement with semantic/policy-aware filters &
\\ \riRowSep

\ref{sec:lack_hall_guardrails} Lack of Hallucination Guardrails &
-- Confident but incorrect answers\riSep Fabricated citations\riSep Weak calibration of confidence vs correctness &
-- No retrieval verification or grounding checks\riSep Missing fallbacks for human review\riSep No uncertainty thresholds\riSep Limited post-generation validation &
-- Reputational and regulatory risk\riSep Costly remediation\riSep Increased support load\riSep Reduced adoption due to loss of trust &
-- Spike in factual-error issues\riSep Frequent user reports despite high internal scores\riSep Quality often improves when enabling fact-checking or evidence policies &
\\ 

\bottomrule
\end{tabular}
\endgroup
\end{table}





\subsection{FMware System Testing and Optimization}
\subsubsection{Lack of Latency Handling Mechanisms}
\label{sec:lack_latency_handling}
Latency is a major challenge for production-ready FMware, leading to slow responses or timeouts under real-world conditions. \citet{kwon2023efficient} show that serving modern large language models is both memory- and compute-intensive, motivating specialized serving techniques just to sustain acceptable latency and throughput. \citet{santilli2023accelerating} shows that the sequential nature of model inference introduces delays, which makes real-time performance difficult~\cite{santilli2023accelerating}. Variable loads further complicate the issue, since latency spikes during high demand can render systems unresponsive~\cite{microsoftLatencyGuidebook}.





\begin{sourcespotlight}
\add{Microsoft \cite{microsoftLatencyGuidebook} explicitly lists "the overall load on the deployment \& system" as one of the four primary factors that influence FM(s) response time. This factor refers to how system performance can degrade under unpredictable or high demand, leading to latency spikes and possible unresponsiveness, especially when the load is variable and not well-managed.}
\end{sourcespotlight}

\subsubsection{Lack of Retry Optimizations}
\label{sec:lack_retry_opt}
Failure to implement intelligent retry strategies reduces FMware reliability, especially with cloud-hosted FM(s). Transient errors are common, and retry storms where repeated requests overwhelm a system can be costly, increasing error rates and decreasing availability~\cite{scaleRateLimits,microsoftRetryStorm}. FMware also benefits from intelligent retries for strategies such as re-prompting for self-consistency, where multiple attempts may be needed to ensure the correct output~\cite{signozOptimizingOpenAI}.





\begin{sourcespotlight}
\add{Scale AI~\cite{scaleRateLimits} mentions, ``Adding random jitter to the delay helps retries from all hitting at the same time.''; indicating that without proper retry strategies, repeated requests can concentrate and cause issues (``retry storms''). Microsoft Ignite \cite{microsoftRetryStorm} explains that when a cloud service becomes unavailable or imposes throttling/rate limits, frequent client retries ``can prevent the service from recovering and worsen the problem." It notes that ``excessive connection attempts during recovery can overwhelm the service and intensify the original problem," explicitly describing how retry storms occur and their costly impact: increased error rates and reduced availability. The article even uses terms like ``a thundering herd" to characterize these storms. SigNoz \cite{signozOptimizingOpenAI} covers implementing retries and error handling as best practices for OpenAI API integration. Specifically, it explains the use of exponential backoff for retries to avoid overwhelming the API and provides a sample code that retries API calls if they hit rate limits.}
\end{sourcespotlight}

\subsubsection{High Cost of Regression Testing}
\label{sec:high_cost_regres_test}
Regression testing in FMware is expensive and time-consuming due to non-determinism and complex interactions. As FM(s) APIs evolve, performance can degrade silently, which requires frequent re-evaluation across many scenarios to ensure reliability~\cite{ma2024my,scottlogicTestingLLMBased,humanloop}. Exact match baselines are inadequate; similarity and correctness testing are needed to detect subtle behavior changes, which increases complexity and cost, especially at scale~\cite{ma2024my,scottlogicTestingLLMBased}.





\begin{sourcespotlight}
\add{Hayes \cite{humanloop} explains that FM(s) apps involve many components (e.g., prompt templates, data sources, memory, agent control flow) that interact in unpredictable ways, making regression testing challenging and expensive because changes in one part (like a model or API) can have unexpected system-wide effects. Scott Logic Blog\cite{scottlogicTestingLLMBased} discusses the challenges of regression testing for LLM-based applications and explicitly notes that testing costs and complexity are far higher than for traditional apps, primarily due to non-determinism and unpredictable output. It states that as AI APIs evolve, performance can silently degrade, so frequent re-evaluation across scenarios is required to ensure reliability.}
\end{sourcespotlight}

\subsubsection{Absence of Software Performance Engineering Practices}
\label{sec:abs_perf_engg}
The absence of established performance engineering practices, such as model swapping and capacity planning, complicates optimization and scaling. In production, model swaps require careful planning and compatibility testing to secure performance gains. Without these measures, FMware risks suboptimal resource use and higher costs~\cite{li2024llm,huang2024new}. 
\respto{2-20}\add{Inadequate optimization strategies, i.e., system-level Software Performance Engineering (SPE) practices that target production latency, throughput, and cost SLOs for running FMware (rather than optimizing or training the FM(s) itself), can bottleneck FM(s) inference~\cite{llmpilot}, which reinforces the need for robust practices \add{(e.g., missing performance budgets or explicit SLOs, lack of autoscaling policies tuned to FMware workloads, and absence of shadow or A/B testing when comparing alternative models or pipelines)}.}






\subsubsection{Lack of Controlled Execution Mechanisms}
\label{sec:lack_control_exec}
\respto{1-10a} \add{Controlled execution is the disciplined management of FMware system behaviors to enable repeatable runs and systematic exploration of non-deterministic executions for reliability and insight. (details in Section \ref{sec:controlled_exec}).}
The lack of controlled execution mechanisms complicates the verification of fixes and tracking the impact of updates. Without feature flags, staging environments, or canary releases, testing changes before full deployment becomes risky, since updates can introduce new issues or break functionality~\cite{wu2024secgpt}. Systems such as SecGPT~\cite{wu2024secgpt} emphasize isolating updates and managing risks in complex environments where many components interact.


\begin{table}[t]
\centering
\begingroup
\scriptsize
\setlength{\tabcolsep}{1.5pt}
\renewcommand{\arraystretch}{0.92}
\setlength{\emergencystretch}{1em}
\newcommand{\riSep}{\newline--\ }
\newcommand{\riRowSep}{\addlinespace\midrule\addlinespace}

\caption{Recurrent issues in FMware Testing and Optimization stage with symptoms, conditions, consequences, and prototypical signals.}
\label{tab:recurrent_issues_matrix}

\begin{tabular}{@{}
  >{\raggedright\arraybackslash}p{0.16\textwidth}
  >{\raggedright\arraybackslash}p{0.21\textwidth}
  >{\raggedright\arraybackslash}p{0.21\textwidth}
  >{\raggedright\arraybackslash}p{0.19\textwidth}
  >{\raggedright\arraybackslash}p{0.23\textwidth}
  >{\raggedright\arraybackslash}p{0.001\textwidth}
@{}}
\toprule
\textbf{Recurrent Issues} &
\textbf{Symptoms} &
\textbf{Conditions} &
\textbf{Consequences} &
\textbf{\shortstack[l]{Prototypical\\Signals}} &
{} \\
\midrule

\ref{sec:lack_latency_handling} Lack of Latency Handling Mechanisms &
-- P95/p99 tail latency rises under load\riSep Timeouts\riSep Wide variability for similar tasks &
-- No parallelism/pipelining in tool calls\riSep Oversized prompts/contexts\riSep Synchronous external calls on hot path\riSep Slow autoscaling under bursty traffic &
-- SLO misses and abandoned sessions\riSep Upstream throttling\riSep Higher cost from retries and over-provisioning\riSep Reduced user trust &
-- Tail latency rises with modest QPS increases\riSep Queue depth spikes\riSep Token count correlates with response time\riSep High variance across identical intents &
\\ \riRowSep

\ref{sec:lack_retry_opt} Lack of Retry Optimizations &
-- Bursts of 429/5xx trigger cascades of automatic retries that amplify load &
-- Uniform backoff without jitter\riSep No circuit breakers or budgets\riSep Fan-out multiplies retries across steps\riSep Missing idempotency or deduplication &
-- Availability dips and cost spikes\riSep Duplicated work\riSep Inconsistent state\riSep Noisy alerts slow root cause analysis &
-- Retry-to-success ratio worsens\riSep Correlated spikes in tokens and errors\riSep Downstream saturation without proportional traffic growth &
\\ \riRowSep

\ref{sec:high_cost_regres_test} High Cost of Regression Testing &
-- Regressions slip past exact-match tests\riSep Large suites costly\riSep Quality drift after provider updates &
-- Non-deterministic outputs without seeds/controls\riSep Mixed prompts across tasks\riSep Incomplete golden sets\riSep No canary evaluation on live-like traffic &
-- Slower release cycles and higher engineering load\riSep Undetected quality drops\riSep Rising evaluation spend with unclear signal &
-- Cost per test run rises\riSep Flaky reruns\riSep Offline-production divergence\riSep Step changes after FM version bumps &
\\ \riRowSep

\ref{sec:abs_perf_engg} Absence of Software Performance Engineering Practices &
-- Static deployments and fixed model choices\riSep Limited headroom\riSep No systematic cost-latency-quality experiments &
-- No performance budgets or SLOs\riSep Missing autoscaling tuned to FM workloads\riSep No shadow or A/B tests for alternative models &
-- Overspending for marginal gains\riSep Inability to sustain spikes\riSep Prolonged incidents due to rigid capacity &
-- Throughput flat despite more hardware\riSep Inconsistent latency after scale-up\riSep Quality-cost curves not characterised across model options &
\\ \riRowSep

\ref{sec:lack_control_exec} Lack of Controlled Execution Mechanisms &
-- Fixes not reliably reproducible\riSep Experiments hard to constrain\riSep Incident timelines mix concurrent changes &
-- No deterministic replay of inputs and contexts\riSep Missing feature flags and routing\riSep Low-fidelity staging\riSep Entangled prompt and model changes &
-- Slow and risky releases\riSep Difficult regression attribution\riSep Repeated rollbacks\riSep Prolonged incidents &
-- Low reproduction rate in postmortems\riSep Frequent hotfix rollbacks\riSep Quality swings during releases\riSep Stability improves with canary/shadow &
\\ 

\bottomrule
\end{tabular}
\endgroup
\end{table}





\subsection{Deployment and Maintenance}
\subsubsection{Lack of Efficient Feedback Technology}
\label{sec:lack_eff_feedback}
Collecting and integrating feedback seamlessly is essential for refining models, addressing issues, and aligning with user needs. \respto{1-11}\add{In production contexts, this feedback is not limited to explicit signals (e.g., thumbs up or thumbs down) but also includes automated, implicit, and passive signals such as user behavior patterns, downstream corrections, and system-level performance indicators. The emphasis is on scalable and continuous collection that minimizes disruption while still surfacing meaningful signals.} Without these mechanisms, developers miss critical insights into user behavior, which leads to stagnation in performance and adaptability. \citet{luo2024arena} demonstrates the importance of a robust feedback loop in the presence of continuous FM(s) and FMware evolution, showing how the absence of efficient data collection delays improvements by failing to identify and address FM(s) weaknesses. Production environments, unlike demos, require \add{automated, iterative, and multi-channel feedback systems} to enable continuous learning, performance optimization, and user engagement~\cite{wang2023survey,luo2024arena,awscloudAmazonServices}.





\begin{sourcespotlight}
 \add{AWS \cite{awscloudAmazonServices} describes the general idea of the ``Data Flywheel'' and continuous data momentum, mentioning components like multi-phase data movement, building data-driven apps, and using feedback loops for business acceleration.}
\end{sourcespotlight}

\subsubsection{Lack of FMware Native Observability}
\label{sec:lack_native_obs}
The lack of FMware native observability makes it difficult to diagnose issues, optimize performance, and ensure compliance in production-ready FMware~\cite{hassan2024rethinking}. Traditional tools focus on functional metrics such as latency, execution traces, and throughput, which are insufficient for capturing internal reasoning and decision making of FM(s)~\cite{hassan2024rethinking}. This opacity hinders root cause analysis, especially when errors arise from complex reasoning or coordination between agents. Regulatory compliance often requires detailed logs and auditing, which are difficult to maintain without comprehensive observability. \respto{1-24} \add{The LangChain playbook’s emphasis on AI-specific testing, traceability, and platform-level telemetry \cite{kouri2025langchain} underscores that many current deployments lack FMware-native QA and observability, not just general monitoring. Kouri~\cite{kouri2025langchain} further documents patterns where missing semantic assertions, absent trace correlation between prompts, tools, and outputs, and lack of business-aligned metrics lead directly to silent regressions in multi-agent systems. This practitioner evidence aligns with our observations that evaluation and observability must be redesigned for FMware semantics, strengthening the empirical basis for the recurrent issues we describe.}





\subsubsection{Ineffective FM(s) Update Mechanisms}
\label{ineffective_fm_update}
Maintaining and updating FM(s) in production FMware presents significant issues compared to traditional software~\cite{hassan2024rethinking}. Fixes can take months, and updates are non-deterministic; providing more training data does not ensure specific issues are resolved. \respto{2-21}\add{Improvements may not work as expected, and previously stable features can break. In addition, improvements noted in release notes (typically a high level capability statement from a provider (e.g. “designed to spend more time thinking,” “improved reasoning and math,” or “improved image understanding”~\cite{gpt_release_notes, claude_release_notes, gemini_release_notes})) do not directly translate to exposed features (i.e., a concrete and testable behavior inside the FMware system (e.g. a higher win rate on a defined task set, support for a specific function calling schema, or meeting a compliance rubric on targeted intents)). This phenomenon makes it difficult for developers to test what changed or to verify improved capability. The gap between high-level claims and system features creates ambiguity about what to verify after an update.} Furthermore, such unpredictability complicates testing and maintenance and makes it difficult to prioritize efforts. In production FMware, these issues undermine reliability, performance, and user trust.





\begin{sourcespotlight}
\add{Public release notes from OpenAI~\cite{gpt_release_notes}, Anthropic~\cite{claude_release_notes}, and Google~\cite{gemini_release_notes} consistently announce broad “reasoning” or “understanding” upgrades, which do not reflect concrete testable features.}
\end{sourcespotlight}


\begin{table}[t]
\centering
\begingroup
\scriptsize
\setlength{\tabcolsep}{1.5pt}
\renewcommand{\arraystretch}{0.92}
\setlength{\emergencystretch}{1em}
\newcommand{\riSep}{\newline--\ }
\newcommand{\riRowSep}{\addlinespace\midrule\addlinespace}

\caption{Recurrent issues in Deployment and Maintenance stage with symptoms, conditions, consequences, and prototypical signals.}
\label{tab:recurrent_issues_matrix}

\begin{tabular}{@{}
  >{\raggedright\arraybackslash}p{0.16\textwidth}
  >{\raggedright\arraybackslash}p{0.21\textwidth}
  >{\raggedright\arraybackslash}p{0.21\textwidth}
  >{\raggedright\arraybackslash}p{0.19\textwidth}
  >{\raggedright\arraybackslash}p{0.23\textwidth}
  >{\raggedright\arraybackslash}p{0.001\textwidth}
@{}}
\toprule
\textbf{Recurrent Issues} &
\textbf{Symptoms} &
\textbf{Conditions} &
\textbf{Consequences} &
\textbf{\shortstack[l]{Prototypical\\Signals}} &
{} \\
\midrule

\ref{sec:lack_eff_feedback} Lack of Efficient Feedback Technology &
-- Feedback sparse/delayed/noisy, trends hard to detect, improvements hard to attribute &
-- Reliance on explicit ratings only\riSep Missing passive telemetry and downstream corrections\riSep No feedback budgets/sampling\riSep Limited privacy-aware instrumentation &
-- Slow iteration\riSep Blind spots in safety and quality\riSep Regressions persist\riSep Wasted spend on low-signal experiments &
-- Flat/inconsistent win rates despite changes\riSep Low task coverage by feedback events\riSep Gains appear offline but not in production telemetry &
\\ \riRowSep

\ref{sec:lack_native_obs} Lack of FMware Native Observability &
-- Incidents hard to reproduce and explain\riSep Traces show I/O but not intermediate reasoning\riSep Incomplete audit trails &
-- No capture of intermediate steps/tool calls\riSep Missing schema/contract logs\riSep Weak linkage between retrieval evidence and generated content\riSep Fragmented dashboards &
-- Longer time to mitigation\riSep Recurring failures on same intents\riSep Inability to demonstrate compliance\riSep Reduced stakeholder trust &
-- Postmortems cite insufficient data\riSep High share of unknown root causes\riSep Quality swings correlate with hidden model or prompt changes &
\\ \riRowSep

\ref{ineffective_fm_update} Ineffective FM(s) Update Mechanisms &
-- Silent behaviour shifts after version bumps\riSep Regressions on stable tasks\riSep Mismatch between provider notes and observed changes &
-- Tight coupling between prompts and one FM\riSep Lack canary/shadow rollouts\riSep Incomplete golden sets\riSep No fallbacks/traffic routing during upgrades &
-- Repeated rollbacks and hotfixes\riSep Slower release cadence\riSep Higher evaluation and incident response costs\riSep Reduced user confidence &
-- Step changes in production metrics around updates\riSep Offline-online divergence\riSep Increased variance for identical intents post update &
\\ 

\bottomrule
\end{tabular}
\endgroup
\end{table}

\subsection{Memory Management}
\subsubsection{Inefficient Knowledge Representation}
\label{sec:ineff_knldg_repr}
Inefficient knowledge representation in FMware leads to storing redundant or irrelevant data, which inflates computational costs and response times. For instance, \citet{guo2024knowledgenavigator} highlights that poor knowledge representation can cause inconsistent reasoning and incorrect outputs when retrieval systems are not optimized. In addition, irrelevant information retrieved due to inefficient knowledge representation can skew FM(s) outputs~\cite{wu2024easily}.





\subsubsection{Cumbersome and Error-Prone in Memory and Across Memory Knowledge Management}
\label{sec:cumbersome_memory}
Managing memory across multiple systems in FMware is complex and prone to errors such as synchronization issues and data corruption. \citet{xie2023adaptive} and \citet{chen2022rich} emphasize that conflicting knowledge within memory systems can lead to inconsistencies and errors when FM(s) must navigate between contradictory pieces of information. \citet{packer2023memgpt} highlights that production-ready FMware requires robust memory systems, since context loss or memory mismanagement can degrade user experience or cause failures~\cite{packer2023memgpt}.


\begin{table}[t]
\centering
\begingroup
\scriptsize
\setlength{\tabcolsep}{1.5pt}
\renewcommand{\arraystretch}{0.92}
\setlength{\emergencystretch}{1em}
\newcommand{\riSep}{\newline--\ }
\newcommand{\riRowSep}{\addlinespace\midrule\addlinespace}

\caption{Recurrent issues in Memory Management stage with symptoms, conditions, consequences, and prototypical signals.}
\label{tab:recurrent_issues_matrix}

\begin{tabular}{@{}
  >{\raggedright\arraybackslash}p{0.16\textwidth}
  >{\raggedright\arraybackslash}p{0.21\textwidth}
  >{\raggedright\arraybackslash}p{0.21\textwidth}
  >{\raggedright\arraybackslash}p{0.19\textwidth}
  >{\raggedright\arraybackslash}p{0.23\textwidth}
  >{\raggedright\arraybackslash}p{0.001\textwidth}
@{}}
\toprule
\textbf{Recurrent Issues} &
\textbf{Symptoms} &
\textbf{Conditions} &
\textbf{Consequences} &
\textbf{\shortstack[l]{Prototypical\\Signals}} &
{} \\
\midrule

\ref{sec:ineff_knldg_repr} Inefficient Knowledge Representation &
-- Retrieval returns long low-salience passages\riSep Repeated facts across chunks\riSep Conflicting snippets confuse downstream reasoning &
-- Flat stores without schemas\riSep Coarse chunking\riSep Missing entity linking/normalisation\riSep Indices from raw dumps without curation &
-- Higher token usage and latency\riSep Lower answer fidelity due to diluted context\riSep Brittle prompts to compensate for representation gaps &
-- Tokens per retrieved passage rise while accuracy stalls\riSep Frequent truncation at context limits\riSep Audits show redundant or contradictory evidence &
\\ \riRowSep

\ref{sec:cumbersome_memory} Cumbersome and Error-Prone in Memory and Across Memory Knowledge Management &
-- Stale/conflicting entries across sessions\riSep Episodic memory disagrees with long-term stores\riSep Updates fail to propagate consistently &
-- Multiple memory backends with ad hoc schemas\riSep Weak write/update policies\riSep No dedup/conflict resolution\riSep Async pipelines without reconciliation &
-- User-facing inconsistency\riSep Unexpected behaviour during retries/rollbacks\riSep Hard-to-debug incidents tied to hidden memory state\riSep Wasted storage/compute &
-- Same query differs across sessions\riSep Duplicate memory records\riSep Cache invalidation spikes\riSep Fixes require manual memory surgery &
\\ 

\bottomrule
\end{tabular}
\endgroup
\end{table}





%
\subsection{Cross-Cutting Challenges}
\subsubsection{\add{[FMware System Testing and Optimization, Deployment and Maintenance, Guarding]} Lack of Automated Testing Capabilities}
\label{sec:lacking_automated_tests}
The lack of automated testing in FMware hinders efficient issue detection and continuous integration. Existing automated tools often miss the nuanced complexities in FMware, which makes them inadequate for production systems where manual testing is impractical at scale~\cite{abeysinghe2024challenges,hassan2024rethinking}. Without automation, development slows, costs increase, and regressions go undetected.





\subsubsection{\add{[FMware System Testing and Optimization, Prompting, Agent(s), Memory Management]} Lack of Assertion Based Unit Tests}
\label{sec:lack_assertion}
Writing assertion-based unit tests for FMware is difficult due to the non-deterministic nature of FM(s), which makes expected outputs hard to define~\cite{abeysinghe2024challenges,scottlogicTestingLLMBased}. Traditional tests rely on fixed outcomes, but FMware varies across runs, especially in agent construction and memory management~\cite{scottlogicTestingLLMBased}. Despite these challenges, such tests remain critical to detect regressions during refactoring or updates~\cite{abeysinghe2024challenges}.





\begin{sourcespotlight}
\add{Scott Logic Blog\cite{scottlogicTestingLLMBased} mentions, ``The most significant challenge in testing LLM-based applications is the non-deterministic output result... LLM-based applications can provide different responses, even with the same input. The unpredictable outcomes make traditional testing approaches, especially test automation, extremely difficult.''}
\end{sourcespotlight}

\subsubsection{\add{[FMware System Testing and Optimization, Data and FM(s) Alignment, Grounding]} Text Based Evaluation Leads to Overestimation of Quality}
\label{sec:textbased_eval}
Text-based evaluation in FMware often overestimates quality by missing deeper issues such as relevance, factual accuracy, or alignment with business rules. \citet{gao2024llm} highlights that current text-based Natural Language Generation (NLG) evaluations miss context and factual correctness, while \citet{hasanbeig2023allure} shows that over-reliance on surface-level outputs can mask underlying performance issues.





\subsubsection{\add{[Data and FM(s) Alignment, Deployment and Maintenance]} Ineffective and Inefficient AI-as-Judge Technologies}
\label{sec:ai_as_judge}
\respto{2-22}\add{An \emph{AI-as-judge} system uses one or more models to automatically evaluate the outputs of another model or of an FMware pipeline, rather than relying only on human reviewers. A separate model scores properties such as correctness, safety, or style and serves as an automated oracle~\cite{zheng2023judging, zhuge2024agent}. This approach is widely used in large-scale settings for RLHF-style evaluation, automated testing, and continuous monitoring, where human-only review cannot match scale and speed.} AI as a judge system can be inefficient and ineffective due to cost and bias. \citet{thakur2024judging} note that models hallucinate and err on complex decisions, and running advanced models such as GPT-4 for ongoing evaluations is expensive and often misaligned with specific business needs~\cite{thakur2024judging,newsYcombinator}. Custom evaluation pipelines are frequently required.





\begin{sourcespotlight}
\add{A post in Hacker News \cite{newsYcombinator} mentions, ``GPT-4 is crazy expensive, paying \$20/mo only gets you 25 messages/3 hours and it's crazy slow. The API is rather expensive too.'' A comment agrees mentioning, ``5 cents. Per resume. \$500 per 10k. ... you have to convince my boss to pay for something that otherwise would have been free... My boss wants to know why we would pay any money for a less reliable solution (GPT serialization is not nearly as reliable as a standard django form).'' Regarding reliability and verifying correctness, they also mention, ``I have tried GPT-3.5 and GPT-4 for this type of task - the 'near perfect results' is really problematic because you need to verify that it's likely correct, notify you if there's issues, and even then you aren't 100\% sure that it selected the correct first/last name. This is compared to a standard html form. Which is.... very reliable and (for us) automatically has error handling built in...''}
\end{sourcespotlight}

\subsubsection{\add{[Data and FM(s) Alignment, Deployment and Maintenance, Memory Management]} Difficulty Navigating the Regulatory and Legal Compliance Minefield}
\label{sec:regulatory_legal}
Legal compliance and licensing complexities permeate all stages of the FMware lifecycle. FM(s) must align with the legal requirements of deployment regions, e.g., some models and providers cannot be used in certain jurisdictions~\cite{hacker2023regulating,cnnApplesChina}. 
\add{For example, following restrictions in the British Columbia public sector, the University of British Columbia (UBC) restricted use of the DeepSeek application (mobile, desktop, and browser access) on devices that access university systems, citing privacy and security risks~\cite{ubc2025deepseek}. Similar restrictions have been applied in the United States, where Commerce Department bureaus prohibited DeepSeek on government-furnished equipment, and Microsoft reported that it does not allow employees to use the DeepSeek application due to data-security concerns~\cite{reuters2025deepseek_commerce,reuters2025deepseek_microsoft}.} Licensing of FM(s) data, including synthetic data, preference data, and user feedback, further complicates compliance. If a model generates synthetic data, licensing of the source data may constrain downstream use. When user feedback is incorporated, later opt-out requests create challenges for removing their data from models and related components. Traditional software bills of materials cover code and dependencies, but FMware’s scope spans data, models, and generated artifacts, which exceed what conventional SBOMs track. FM(s) must align with the legal requirements of deployment regions, e.g., some models and providers cannot be used in certain jurisdictions~\cite{hacker2023regulating,cnnApplesChina}.





\begin{sourcespotlight}
\add{CNN \cite{cnnApplesChina} reports, ``ChatGPT (soon to be integrated into Siri) is banned in China... China is one of the first countries in the world to regulate the generative AI technology that powers these popular services... requiring companies to seek approval before deployment.''}
\end{sourcespotlight}


\begin{table}[t]
\centering
\begingroup
\scriptsize
\setlength{\tabcolsep}{1.5pt}
\renewcommand{\arraystretch}{0.92}
\setlength{\emergencystretch}{1em}
\newcommand{\riSep}{\newline--\ }
\newcommand{\riRowSep}{\addlinespace\midrule\addlinespace}

\caption{Recurrent issues in Cross-Cutting stages with symptoms, conditions, consequences, and prototypical signals.}
\label{tab:issues_crosscutting}

\begin{tabular}{@{}
  >{\raggedright\arraybackslash}p{0.16\textwidth}
  >{\raggedright\arraybackslash}p{0.21\textwidth}
  >{\raggedright\arraybackslash}p{0.21\textwidth}
  >{\raggedright\arraybackslash}p{0.19\textwidth}
  >{\raggedright\arraybackslash}p{0.23\textwidth}
  >{\raggedright\arraybackslash}p{0.001\textwidth}
@{}}
\toprule
\textbf{Recurrent Issues} &
\textbf{Symptoms} &
\textbf{Conditions} &
\textbf{Consequences} &
\textbf{\shortstack[l]{Prototypical\\Signals}} &
{} \\
\midrule

\ref{sec:lacking_automated_tests} Lack of Automated Testing Capabilities &
-- Limited/brittle suites miss behaviour drift\riSep Evaluations run ad hoc\riSep Issues found late in production &
-- Non-deterministic outputs without controls\riSep Heterogeneous tasks lacking shared oracles\riSep Sparse golden sets\riSep Limited synthetic edge-case generation &
-- Slower release cadence\riSep Higher incident frequency\riSep Rising evaluation spend for limited signal\riSep Hard to demonstrate reliability &
-- High flake rate and reruns\riSep Offline vs production divergence\riSep Repeated regressions on previously fixed intents &
\\ \riRowSep

\ref{sec:lack_assertion} Lack of Assertion Based Unit Tests &
-- Tests overfit exact text or become too permissive, real faults slip through and create false confidence &
-- No schema/contract-based assertions\riSep Missing invariants on intermediate steps\riSep No seeded or controlled runs for reproducibility &
-- Hidden behaviour drift and brittle releases\riSep Hard to localise failures to prompts vs tools/models\riSep Increased manual review burden &
-- High exact-match failure with low correlation to user-visible quality\riSep Tests rewritten after minor prompt changes\riSep Stability improves with structured assertions &
\\ \riRowSep

\ref{sec:textbased_eval} Text Based Evaluation Leads to Overestimation of Quality &
-- High scores on text similarity/preference benchmarks but poor task completion, compliance, or user satisfaction &
-- Generic metrics without domain grounding\riSep Limited counterfactual tests\riSep Evaluations ignore retrieval evidence and policy constraints &
-- Premature deployment\riSep Wasted iteration on prompt polish instead of data/tooling\riSep Loss of stakeholder trust when real-world results lag &
-- Strong offline scores with weak production KPIs\riSep Gaps between rubric human review and automatic metrics\riSep Quality improves with evidence checks or policy-aware scoring &
\\ \riRowSep

\ref{sec:ai_as_judge} Ineffective and Inefficient AI-as-Judge Technologies &
-- Judge disagrees with domain experts on hard cases\riSep Costs scale linearly with volume\riSep Verdicts lack stable rationale &
-- Single-model judges without calibration/diversity\riSep No human spot checks\riSep Generic rubrics ignoring domain rules\riSep Limited sampling/adjudication &
-- Biased or noisy quality signals\riSep Misdirected optimisation\riSep Escalating evaluation spend without proportional gains &
-- Low inter-rater agreement (judge vs humans)\riSep High variance across runs\riSep Cost per evaluated sample dominates experimentation budgets &
\\ \riRowSep
\ref{sec:regulatory_legal} Difficulty Navigating the Regulatory and Legal Compliance Minefield &
-- Ambiguity about permissible data/model use\riSep Incomplete lineage in audits\riSep Rollouts stall due to regional or licensing constraints &
-- Mixed provenance corpora\riSep Opaque vendor terms\riSep Cross-border deployments\riSep Feedback integration without granular consent tracking\riSep No model/data SBOM equivalents &
-- Delays or forced de-scopes\riSep Takedown or retraining\riSep Contractual exposure and reputational risk\riSep Fragmented deployments by region &
-- Legal reviews block releases\riSep Repeated lineage requests teams cannot satisfy\riSep Region-specific configs proliferate without clear rationale &
\\
\bottomrule
\end{tabular}
\endgroup
\end{table}

\subsection{From Demos to Production: Practice-Driven Findings}
\label{sec:reflective_new_findings}
\respto{3-14b}
\add{Having cataloged recurrent issues across the FMware lifecycle (including cross-cutting concerns), we close this section by distilling a small set of \emph{operationally dominant} findings that emerged most strongly from our practitioner-grounded and grey-source synthesis. Many underlying model limitations are already widely recognized in prior work (e.g., non-determinism, prompt sensitivity, hallucination); in contrast, the points below emphasize how those limitations become acute when FMware transitions from \emph{demos} to \emph{production} (where traffic scale, operational governance, and downstream side effects make failures visible and costly). This synthesis clarifies what our evidence adds beyond known FM properties and provides a bridge from stage-specific issues (Section~\ref{sec:issues}) to the cross-cutting challenges in Section~\ref{sec:challenges}.}

\add{
\begin{itemize}
  \item \textbf{FM(s) update unpredictability in multi-provider ecosystems.}
  Teams report that provider-driven updates can introduce silent capability drift, behavior regressions, or breaking changes that invalidate prior validation, forcing governance mechanisms (release monitoring, canaries, rollback plans, and version pinning) to become first-class engineering concerns.
  \item \textbf{Lifecycle-wide testing for agentic FMware, not prompt-only evaluation.}
  Practitioner evidence repeatedly emphasizes that reliability hinges on end-to-end tests that span prompts, retrieval, tool calls, memory writes, and downstream side effects, because failures often emerge only through multi-stage interactions.
  \item \textbf{FMware-native observability, including decision-level traces.}
  Beyond standard logging, practitioners call for observability that captures why the system acted as it did (reasoning-path and decision evidence across tool-use and retrieval), enabling actionable debugging rather than post-hoc guessing.
  \item \textbf{Controlled execution under stochasticity and changing external context.}
  A recurring practice-driven need is repeatability via snapshotting (prompts, retrieval corpora, model/config versions, sampling parameters, and tool-call context) to support regression analysis and incident forensics despite non-deterministic behavior.
  \item \textbf{Resource-aware QA as a binding constraint on what “rigour” is feasible.}
  At production scale, the cost and latency of evaluation pushes teams toward caching, reuse, and budgeted test scheduling, shifting QA from “run everything” to “prioritize evidence under constraints,” which is rarely treated as central in prior FMware discussions.
\end{itemize}}

\add{\noindent
Together, these findings highlight where the hardest production problems arise (updates, end-to-end interactions, trace fidelity, reproducibility, and evaluation cost). They directly motivate the challenge framing and solution directions synthesized next in Section~\ref{sec:challenges}.}

\section{Challenges}
\label{sec:challenges}

In this section, we present the results of our semi-structured thematic synthesis and outline the key challenges in productionizing FMware. It is important to note that the challenges presented in this version of the paper address only a subset of the identified recurrent issues, as shown in Table~\ref{tab:traceability}. Future revisions will expand this coverage as the field evolves, since the set of recurrent issues reflects the current state of production FMware and may change over time as new challenges emerge and others are mitigated through improved tools, practices, and standards. 
\begin{table*}[t]
\centering

\definecolor{chBlue}{HTML}{0072B2}      
\definecolor{chSky}{HTML}{56B4E9}       
\definecolor{chGreen}{HTML}{009E73}     
\definecolor{chOrange}{HTML}{E69F00}    
\definecolor{chVermillion}{HTML}{D55E00}
\definecolor{chPurple}{HTML}{CC79A7}    

\newcommand{\chtag}[2]{\begingroup\setlength{\fboxsep}{1.0pt}\colorbox{#1!18}{\textcolor{#1!85!black}{\textsf{#2}}}\endgroup}
\newcommand{\chTesting}{\chtag{chBlue}{Testing}}
\newcommand{\chObservability}{\chtag{chGreen}{Observability}}
\newcommand{\chControlledExec}{\chtag{chPurple}{Controlled Exec.}}
\newcommand{\chResourceQA}{\chtag{chOrange}{Resource-Aware QA}}
\newcommand{\chFeedback}{\chtag{chVermillion}{Feedback}}
\newcommand{\chBuiltIn}{\chtag{chSky}{Built-in Quality}}

\newcommand{\issueTag}[2]{\begingroup\setlength{\fboxsep}{0.6pt}\colorbox{#1!18}{\textcolor{#1!85!black}{\scriptsize\textsf{#2}}}\endgroup}
\newcommand{\itTesting}{\issueTag{chBlue}{T}}
\newcommand{\itObservability}{\issueTag{chGreen}{O}}
\newcommand{\itControlledExec}{\issueTag{chPurple}{C}}
\newcommand{\itResourceQA}{\issueTag{chOrange}{R}}
\newcommand{\itFeedback}{\issueTag{chVermillion}{F}}
\newcommand{\itBuiltIn}{\issueTag{chSky}{B}}
\newcommand{\issue}[2]{#1\,\,#2}

\caption{\respto{1-21} \add{Traceability from lifecycle stages (Section~\ref{sec:pipeline}) to recurrent issues (Section~\ref{sec:issues}), cross-cutting challenges (Section \ref{sec:challenges}), and representative evidence. \emph{Legend:} \chTesting{}~Testing; \chObservability{}~Observability; \chControlledExec{}~Controlled Execution; \chResourceQA{}~Resource-Aware QA; \chFeedback{}~Feedback Integration; \chBuiltIn{}~Built-in Quality. Issue markers in the \textbf{Recurrent issues} column use the same colors (T/O/C/R/F/B); multiple markers indicate overlap across challenges. Recurrent issues shown without markers are not used to define the challenges in this paper version; see Section~\ref{sec:challenges} for details.}}
\label{tab:traceability}

\begingroup
\scriptsize            
\setlength{\tabcolsep}{3pt} 
\renewcommand{\arraystretch}{1.05} 

\begin{tabular}{p{1.5cm}p{7cm}p{2.9cm}p{2.3cm}}
\toprule
\textbf{Lifecycle stage} & \textbf{Recurrent issues} & \textbf{Related challenges } & \textbf{Evidences} \\ \hline
\midrule
FM(s) Selection
& \issue{\itBuiltIn\itResourceQA}{\ref{sec:balancing_perf_cost} Difficulty Balancing Functional Requirements With Performance and Costs}
& \chBuiltIn{}, \chResourceQA{}
& \cite{kamath2024llms,chavan2024faster,chen2024role,irugalbandara2024scaling,gunter2024apple,moe_in_LLM,lin2024awq,hohman2024model} \\ \hline

Data \& FM Alignment
& \issue{\itBuiltIn}{\ref{sec:low_data_qual} Low Data Quality}; \issue{\itBuiltIn}{\ref{sec:low_domain_cov} Low Domain Coverage}; \issue{\itResourceQA}{\ref{sec:low_data_eff} Low Data Efficiency}
& \chBuiltIn{}, \chResourceQA{}
& \cite{albalak2024survey,djuhera2025fixing,li2023ood_eval_survey,yang2023ood_generalization_nlp,chen2023maybe,Aytes2025SketchofThoughtELA,su2025between,wu2025more} \\ \hline

Prompting
& \issue{\itBuiltIn}{\ref{sec:god_prompts} God Prompts}; \issue{\itBuiltIn}{\ref{sec:prompt_QA} Lack of Built in QA Checks for Prompts}; \ref{sec:prompt_portability} Lack of Prompt Portability Across FMs; \ref{sec:fm_optimizations} Absence of FM Specific Optimizations; \issue{\itObservability}{\ref{sec:complexity_failure_rationale} Complexity in Determining Failure Rationale}; \issue{\itBuiltIn}{\ref{sec:no_good_icl_ex} Non Representative or Insufficient In-Context Learning Examples}
& \chBuiltIn{}, \chObservability{}
& \cite{oreillyWhatLearned, parthasarathy2024ultimate, wei2022chain, liu2023lost, wang2024multimodal, prompt_guide, reasoning_best_practices, kamoi2024evaluating, ahmed2023better, wang2024minstrel, chang2024efficient, chen2024mapo, sabbatella2024prompt, ashtari2023discovery, amujo2024good, wang2023learning} \\ \hline

Grounding
& \issue{\itBuiltIn}{\ref{sec:low_density} Low Information Density in Grounding Data}; \issue{\itBuiltIn}{\ref{sec:irrelevant_grounding} Irrelevant Grounding Data}; \ref{sec:over_complicating_solutions} Over Complicating Solutions With Advanced Techniques
& \chBuiltIn{}
& \cite{laban2024summary, wang2024adapting, liu2024lost, cuconasu2024power, oreillyWhatLearned, sawarkar2024blended} \\ \hline

Agent(s) \& Orchestration
& \issue{\itBuiltIn}{\ref{sec:god_agents} God Agents}; \issue{\itBuiltIn}\ref{sec:low_level_tool} Too Low Level Tool Usage Abstraction; \ref{sec:use_case_centric} Capability Centric Instead of Use Case Centric Documentation for Tools; \issue{\itBuiltIn}{\ref{sec:over_planning} Over Reliance on FM Planning Capability}
& \chBuiltIn{}
& \cite{rasal2024navigating, chen2024coder, yang2024swe, koualty2024generative, wang2023voyager, shareef2024retailgpt, computeruse, deepresearch, guo2024large, huang2024understanding, claude_tips, prompt_guide, reasoning_best_practices, yang2023auto, babyagi, hasan2025model, roig2025tool_descriptions} \\ \hline

FMware System Testing \& Optimization
& \issue{\itResourceQA}{\ref{sec:lack_latency_handling} Lack of Latency Handling Mechanisms}; \issue{\itResourceQA}{\ref{sec:lack_retry_opt} Lack of Retry Optimizations}; \issue{\itResourceQA}{\ref{sec:high_cost_regres_test} High Cost of Regression Testing}; \issue{\itResourceQA}\ref{sec:abs_perf_engg} Absence of Software Performance Engineering Practices; \issue{\itControlledExec}{\ref{sec:lack_control_exec} Lack of Controlled Execution Mechanisms}
& \chResourceQA{}, \chControlledExec{}
& \cite{kwon2023efficient, santilli2023accelerating, microsoftLatencyGuidebook, scaleRateLimits, microsoftRetryStorm, signozOptimizingOpenAI, ma2024my, scottlogicTestingLLMBased, humanloop, li2024llm, huang2024new, llmpilot, wu2024secgpt} \\ \hline

Deployment \& Maintenance
& \issue{\itFeedback}{\ref{sec:lack_eff_feedback} Lack of Efficient Feedback Technology}; \issue{\itObservability}{\ref{sec:lack_native_obs} Lack of FMware Native Observability}; \issue{\itControlledExec\itTesting}{\ref{ineffective_fm_update} Ineffective FM Update Mechanisms}
& \chFeedback{}, \chObservability{}, \chControlledExec{}, \chTesting{}
& \cite{wang2023survey, luo2024arena, awscloudAmazonServices, hassan2024rethinking, gpt_release_notes, claude_release_notes, gemini_release_notes, kouri2025langchain} \\ \hline

Memory Management
& \issue{\itBuiltIn}\ref{sec:ineff_knldg_repr} Inefficient Knowledge Representation; \issue{\itFeedback}{\ref{sec:cumbersome_memory} Cumbersome and Error-Prone in Memory and Across Memory Knowledge Management}
& \chFeedback{} \chBuiltIn{}
& \cite{guo2024knowledgenavigator, wu2024easily, xie2023adaptive, chen2022rich, packer2023memgpt} \\ \hline

Guarding
& \issue{\itBuiltIn}\ref{sec:keyword_guarding_ineffective} Simple Keyword Based Guarding is Ineffective; \issue{\itBuiltIn}{\ref{sec:lack_hall_guardrails} Lack of Hallucination Guardrails}
& \chBuiltIn{}
& \cite{rebedea2023nemo, dong2024framework, ayyamperumal2024current, dahl2024hallucinating, spiess2024quality} \\ \hline

Cross-cutting / Multi-stage
& \issue{\itTesting}{\ref{sec:lacking_automated_tests} Lack of Automated Testing Capabilities}; \issue{\itTesting}{\ref{sec:lack_assertion} Lack of Assertion Based Unit Tests}; \issue{\itTesting}{\ref{sec:textbased_eval} Text Based Evaluation Leads to Overestimation of Quality}; \issue{\itResourceQA}\issue{\itTesting}{\ref{sec:ai_as_judge} Ineffective and Inefficient AI-as-Judge Technologies}; \issue{\itBuiltIn}{\ref{sec:regulatory_legal} Difficulty Navigating the Regulatory and Legal Compliance Minefield}
& \chTesting{}, \chResourceQA{}, \chBuiltIn{}
& \cite{abeysinghe2024challenges, hassan2024rethinking, gao2024llm, hasanbeig2023allure, zheng2023judging, zhuge2024agent, reuters2025deepseek_microsoft, thakur2024judging, newsYcombinator, hacker2023regulating, cnnApplesChina, reuters2025deepseek_commerce, ubc2025deepseek, scottlogicTestingLLMBased} \\
\bottomrule
\end{tabular}

\endgroup
\end{table*}


For each challenge, we provide an overview, a critical analysis of the state-of-the-practice, and our vision for addressing it. \respto{2-5}\add{Importantly, these challenges are not uniform in their origin: some are primarily symptoms of the field’s nascency, such as missing standards, immature tooling, and limited operational practices, while others appear more fundamental to FMware systems, including underspecified requirements, non-deterministic behavior, and hard-to-fully-verify interactions with open-world data and tools.} Tackling these challenges requires innovative technologies driven by research at the intersection of software engineering and AI. Our vision highlights these technologies, their functional requirements, and points to promising research directions.


\subsection{Testing}
\label{sec:testing}

\noindent \textbf{Overview.} Poor testing practices present significant roadblocks in the transition from demo FMware to production-ready systems. As identified in Section~\ref{sec:pipeline}, the lack of robust, assertion-based unit tests, the absence of automated testing, and the over-reliance on surface-level text-based evaluation~\cite{abeysinghe2024challenges,scottlogicTestingLLMBased,hassan2024rethinking} often lead to overestimation of FMware quality. In real-world production settings, they result in unreliable outputs, latent defects, and increased regression costs. To address these issues, FMware requires innovative testing mechanisms that account for the unique complexities of FM(s) and their non-deterministic behavior. \respto{2-12}\add{At its core, FMware testing validates the entire production system, where failures can arise from the interaction between the model and prompts, retrieval and grounding, memory, guardrails, tools, external APIs, and orchestration.}


\add{System testing remains necessary even with a well-aligned frontier FM. A high-quality FM(s) can still fail once integrated into a pipeline because defects often originate from interfaces and context rather than from the model alone. For example:}
\add{
\begin{enumerate}
    \item A model that is safe in isolation emits unsafe SQL when a downstream agent provides ambiguous or adversarial instructions.
    \item With RAG, the model misinterprets outdated or poorly structured entries, yielding incorrect answers despite being capable on clean corpora.
    \item In multi-agent flows, coordination errors lead to loops or conflicting outputs even when each agent passes its unit checks.
    \item Memory or grounding subsystems supply stale or incomplete context, producing silent reasoning errors that alignment cannot anticipate.
\end{enumerate}
}
\add{In practice, production teams therefore layer integration harnesses, regression suites, canaries, and user-in-the-loop validation to test prompts, agent actions, grounding data, and downstream effects end-to-end. Industry reports note that alignment alone is insufficient and emphasize multi-layer testing pipelines and CI-based evaluations~\cite{testing_ai_gen, copilot_best_practice}. In fact, even a perfectly aligned model or a frontier model must be tested within the surrounding FMware to guarantee safe, predictable, and supportable behavior in production.}

\respto{2-5}\add{This challenge is primarily a \textsc{Fundamental limitation}, since many FMware behaviors lack a fully specified, deterministic oracle (especially for open-ended text, tool choices, and multi-step agent traces). As a result, correctness must often be checked via approximate properties rather than exact equality (though some narrow, well-specified tasks can still use deterministic oracles). The recommendations below assume (i) the team can define testable properties or invariants (schemas, contracts, metamorphic relations, regression expectations), (ii) representative evaluation data can be curated or sampled from production traces, and (iii) any AI-as-judge component is calibrated (spot-checked by humans, drift-monitored) to bound bias and cost. In regulated or safety-critical domains, the guidance further assumes stricter auditability requirements, favoring deterministic validators and conservative pass criteria.}




\smallskip \noindent \textbf{Critical Analysis of the State-of-the-Practice.} FMware testing remains an immature area, particularly in comparison to well-established practices in traditional software systems. The current testing pipeline for FMware relies heavily on manual processes for \textit{test generation}. The manual identification of metamorphic relations (MRs), which are dependencies between different inputs and expected outputs, is a key bottleneck in the process. This task, while feasible for small-scale demos, becomes impractical and error-prone in production-ready FMware systems. The lack of automated test generation severely hampers the speed and coverage of testing in production systems.


Moreover, the role of \textit{AI-as-judge} technologies in testing pipelines adds another layer of complexity. Current AI judges, particularly those using large FM(s) like GPT-4, exhibit several critical flaws in evaluating correctness. These technologies often favor outputs based on superficial attributes, such as formatting or the length of responses, rather than factual accuracy or alignment with real-world constraints~\cite{thakur2024judging,chen2024humans}. For instance, research shows that simply reversing the order of candidate responses can dramatically alter the AI's judgment, resulting in false positives or negatives~\cite{thakur2024judging,chen2024humans,koo2024benchmarkingcognitivebiaseslarge}. The failure to maintain consistency in judging introduces significant risks when scaling FMware for production environments. Research has also pointed to a misalignment between AI decisions and human expectations. For example, Koo\textit{~et~al.}~\cite{koo2024benchmarkingcognitivebiaseslarge} calculate the Rank-Biased Overlap (RBO) score to measure the agreement between human preferences and model evaluations in ranking-generated texts across 16 different FM(s) (RBO varies from 0 to 1, and higher values indicate higher agreement). The average RBO was 0.44, with GPT-4 scoring 0.47. Another major issue arises from the associated cost and latency with AI-as-judge systems. Relying on large models like GPT-4 is computationally expensive and not viable for long-term or large-scale production-grade testing~\cite{parnin2023building}. 

\add{Formal verification techniques, such as SMT or symbolic reasoning, offer partial relief but only for bounded, schema-defined components of FMware, for instance, validating tool contracts, prompt grammars, or API parameter constraints. They cannot yet capture open-ended semantic drift or non-deterministic natural-language outputs. Consequently, formal reasoning provides provable compliance for well-specified artifacts, while stochastic behaviors must still be validated empirically or via AI-based judges.}

\smallskip \noindent \textbf{Our Vision.} To address these limitations, we envision an approach that combines \textit{automated test generation}, a \textit{next-generation AI-judge creator}, and \add{bounded formal verification for structured components. Together, these form a hybrid testing stack where deterministic checks and probabilistic evaluation coexist.}

\noindent \textit{-- Automated test generation.} The process begins with gathering user feedback in real-world scenarios, such as thumbs-up or thumbs-down data, which provides rich, actionable insights. By integrating this feedback with pre-existing domain knowledge, \textit{automated} metamorphic relations (MRs) are generated for specific attributes of interest. These MRs are then applied to conduct metamorphic testing (MT) on FMware, ensuring that different properties and behaviors of the system are adequately tested. Following the MT, human experts evaluate the results to identify any discrepancies or potential improvements, further refining the MRs. This iterative cycle of automated generation, testing, and human-driven feedback continuously enhances the accuracy and relevance of the MRs, ultimately leading to more robust and reliable FMware.


\noindent \textit{-- Next-generation AI-judge framework.} The development of such a framework is crucial for improving the reliability of evaluations. Unlike existing solutions, this framework would guide developers in crafting tailored prompts that align with specific business logic and domain constraints. These judges would be trained to focus not just on the superficial aspects of FM(s) outputs (e.g., format, length) but on deeper attributes like factual accuracy, consistency with previous outputs, and compliance with application-specific requirements. To train the judges in a more structured, guided, flexible, evolvable, and cost-effective manner, we foresee the use of \textit{curriculum engineering} (more details in Section~\ref{sec:built-in-quality}). As a result, we would also obtain a more lightweight and efficient model, further contributing to cost-effectiveness.
\add{The judge itself can act as an adaptive verifier whose scoring rules evolve under human oversight, reducing drift over time. By pairing these lightweight judges with symbolic validators, teams achieve tiered assurance, symbolic for deterministic artifacts, heuristic for open-ended content.}

\smallskip
\begin{assumptiontradeoffbox}
\respto{1-20} \textbf{\add{Assumptions.}}
\begin{itemize}[leftmargin=*,nosep]
    \item \add{The FMware under test has \emph{at least partially structured artifacts} (e.g., tool contracts/schemas, API parameter constraints, or structured prompts) that can be validated deterministically.}
    \item \add{Teams can maintain a \emph{reference workload} (golden sets, canary traffic, or representative scenarios) to continuously assess regressions as prompts, tools, and models evolve.}
    \item \add{Human oversight is available for calibration (spot checks / adjudication), because fully automated judging remains brittle on hard, domain-specific cases.}
\end{itemize}

\assumpSep

\textbf{\add{Trade-offs.}}
\begin{itemize}[leftmargin=*,nosep]
    \item \add{\emph{Assurance depth vs. throughput:} deeper verification (formal checks, stricter judges, larger test suites) improves confidence but slows CI cycles and increases evaluation cost.}
    \item \add{\emph{Determinism vs. coverage:} symbolic/static checks provide strong guarantees for structured components, but open-ended natural-language behavior still requires empirical evaluation (humans or AI-as-judge), which introduces variance.}
    \item \add{\emph{Cost vs. fidelity:} larger judges (e.g., frontier FM(s)) tend to be more capable but are expensive for continuous use; smaller judges are cheaper but require tighter scoping and more calibration to avoid drift.}
\end{itemize}

\tinyskip
\add{Overall, these assumptions and trade-offs are realistic for many production settings, but they should be stated explicitly because they do not hold for fully open-ended systems without schemas, stable reference workloads, or access to human adjudication.}
\end{assumptiontradeoffbox}




\subsection{Observability}
\label{sec:observability}
\noindent \textbf{Overview.} Observability in FMware systems is crucial to ensure transparency, traceability, and reliability throughout their lifecycle. However, as outlined in Section~\ref{sec:pipeline}, FMware presents unique challenges in observability due to its reliance on dynamic, non-deterministic components like FM(s) and autonomous agents. Recurrent issues, such as the complexity in determining the rationale behind system failures (i.e., whether stemming from limitations in prompts or FM(s) themselves) and the lack of FMware-native observability mechanisms~\cite{hassan2024rethinking}, highlight the need for novel solutions. Unlike traditional software, FMware systems demand observability approaches that can capture both \textit{functional} and \textit{cognitive} aspects of AI agents and models.

\respto{2-5}\add{This challenge is primarily an \textsc{Evolving practice}, since today’s observability stacks for FMware remain immature (missing shared standards for traces, prompts, retrieval context, and agent state), even though the underlying need for end-to-end visibility is long-term. The recommendations below assume (i) teams can log or sample higher-fidelity artifacts (prompt context, retrieved passages, tool I/O, intermediate states) with appropriate privacy controls, (ii) traces can be linked to versions of models, prompts, tools, and data, and (iii) the organization can tolerate some latency and storage overhead for telemetry and governance. In privacy-constrained deployments, the guidance assumes additional redaction, hashing, or aggregation, which may reduce diagnostic utility.}

\respto{2-13}\add{Two complementary layers of observability exist. At the \emph{FM(s) level}, interpretability work studies internal neurons, attention heads, or attribution maps, useful for model introspection but limited for system diagnosis. At the \emph{FMware level}, the goal is end-to-end visibility across prompts, retrieval, grounding, memory, guardrails, and tools. Most real incidents originate here. Fast-path production models optimized for latency rarely emit chain-of-thought traces, and even when present, these traces can be noisy or fabricated~\cite{chen2025reasoning}. Hence, FM-level observability provides semantic introspection but not actionable debugging.}

\add{Recent frameworks such as Watson~\cite{rombaut2024watson} introduce \emph{surrogate agents} that reconstruct causal reasoning traces without instrumenting production runs. We adopt this principle in proposing \emph{fidelity-verified decision traces} that connect low-level events (FM(s) calls, retrievals, and tool executions) to the high-level rationale inferred by the surrogate, yielding a consistent “what–why” map akin to the What-If Tool for interactive probing of ML models~\cite{whatiftool2019}, for debugging and audit. This makes observability both explanatory and verifiable without perturbing live systems.}

\smallskip \noindent \textbf{Critical Analysis of the State-of-the-Practice.} Current observability tools in FMware, while evolving, remain grounded in classical software observability practices. Tools like \textit{OpenLLMetry}~\cite{traceloopOpensourceObservability} and LangSmith~\cite{langchainLangSmith} focus on low-level resource monitoring, capturing traces of FM(s) calls, vector DB interactions, and user prompts, along with metrics such as latency and resource utilization~\cite{kouri2025langchain}. \respto{1-12}\add{In practice, these signals are largely surface-level (e.g., latency, token counts, and request traces), which can indicate \emph{what} failed but often provide limited evidence for \emph{why} it failed.} However, these frameworks overlook the \textit{cognitive} processes driving FM-based decisions, particularly in complex multi-agent FMware systems where outcomes can be non-deterministic or emergent (e.g., through agent interactions or prompt misinterpretation). \respto{1-12}\add{This gap contrasts with the training layer, where the Ultra-Scale Playbook demonstrates fine-grained instrumentation of throughput, utilization, and communication behavior at scale~\cite{tazi2025ultra}; comparable discipline at the application layer, such as systematic tracing of model invocations, retrieval steps, tool calls, routing decisions, and cache behaviors and correlating them with quality and cost outcomes, is still uncommon. As a result, observability remains closer to basic logging than an empirical basis for diagnosing incidents, validating changes, and enforcing SLAs in complex FMware pipelines.} 

Workflow-level tooling can also provide high-level overviews of agent workflows, enabling developers to trace where agents get stuck or fail (e.g., via visual run histories and step-by-step execution views in workflow platforms such as Microsoft Power Automate). While helpful, such tooling remains focused on functional observability and typically does not capture how agents reason, plan, or coordinate their decisions. As FMware systems scale and become more complex, the gap between functional monitoring and decision-level observability widens.

\smallskip \noindent \textbf{Our Vision.} Addressing the aforementioned limitations in FMware observability requires a shift in both methodology and technology. One of the key innovations needed is a \textit{general observability framework} for FMware that can capture both functional performance metrics and the internal cognitive processes of FM(s) and agents. This framework must allow developers to probe into multiple depths of abstraction, providing traceability not only at the system level but also across the decision-making stages of autonomous agents. \add{Concretely, we extend FMware observability along five axes: (1) \textit{Output Integrity Monitoring}, to detect hallucinations, factual drift, and correctness regressions; (2) \textit{Semantic Feedback Integration}, capturing explicit ratings and implicit user corrections to improve alignment; (3) \textit{Reasoning Path Observability}, to trace intermediate reasoning or coordination between agents; (4) \textit{Decision-Level Traces}, linking outcomes to prompt spans, retrieved evidence, and invoked tools; and (5) \textit{Privacy-Preserving Failure Clustering}, aggregating anomalies without leaking user data. Each axis introduces distinct feasibility constraints, while (1--2) are already implementable via telemetry hooks, (3--5) require architectural changes to agents and runtime environments.}

We envision an \textit{observability analytics engine} that visualizes and analyzes events at higher abstraction levels, helping developers quickly pinpoint root causes in complex multi-agent workflows. \add{These enhancements redefine observability from passive logging to active sense-making: integrity monitors and feedback signals highlight emerging failure modes, while decision-level and reasoning-path traces enable developers to explain \emph{why} outcomes occurred rather than only \emph{what} failed.} To achieve this, we propose a ``plane flight recorder'' for agents, inspired by aviation black boxes. In FMware, this recorder would selectively capture an agent's internal reasoning steps and communications, allowing developers to trace decision-making pathways and understand how agents reach conclusions.

However, recording an agent's thoughts can introduce the \textit{observer effect}, where observation alters behavior (e.g., adding ``think step-by-step'' to a prompt changes the output). To mitigate this, we propose a \textit{surrogate agent} that enables debugging without directly interfering with the system. The original problem and result are sent to the surrogate agent (whose goal is to reason rather than solve the problem), which reasons verbosely to infer the original agent's thought process. While the surrogate's reasoning might differ, it still provides transparency and insight into decision-making. \add{At the same time, cognitive traces and surrogate-style analyses incur measurable token and latency overhead, so systems must make these trade-offs explicit and configurable.} We further envision that research should explore techniques to enhance the trustworthiness of the surrogate agent's output.

\smallskip
\begin{assumptiontradeoffbox}
\respto{1-20} \textbf{\add{Assumptions.}}
\begin{itemize}[leftmargin=*,nosep]
    \item \add{At least partial cognitive metadata can be extracted (or reconstructed via surrogates) without violating privacy constraints.}
    \item \add{The system can support a configurable observability mode (e.g., sampling, on-demand deep traces, or shadow/diagnostic runs) so richer traces are not required on every request.}
    \item \add{Teams have the operational maturity to act on observability signals (triage workflows, ownership, and remediation loops), otherwise richer traces become expensive logging with limited impact.}
\end{itemize}

\assumpSep

\textbf{\add{Trade-offs.}}
\begin{itemize}[leftmargin=*,nosep]
    \item \add{\emph{Observability depth vs. operational efficiency:} rich traces improve explainability and diagnosis but increase token/latency overhead, storage, and review burden; minimal traces preserve throughput but reduce diagnostic precision.}
    \item \add{\emph{Instrumentation control vs. behavioral realism:} tightly controlled diagnostic runs simplify attribution, but rare, stochastic, and interaction-driven failures may only surface under real traffic and open-world tool responses.}
    \item \add{\emph{Privacy vs. fidelity:} higher-fidelity traces (prompts, retrieved passages, tool outputs) are often the most useful for debugging, but they are also the most sensitive and may require redaction, hashing, or access controls that reduce utility.}
\end{itemize}

\tinyskip
\add{These assumptions and trade-offs are plausible for production teams that can budget for telemetry and governance. However, they may not hold for privacy-constrained deployments (e.g., regulated domains) or for systems that cannot tolerate any latency overhead, in which case the approach must rely more heavily on sampling, aggregation, and surrogate/offline analysis.}
\end{assumptiontradeoffbox}

\subsection{Controlled Execution}
\label{sec:controlled_exec}

\noindent \textbf{Overview.} Controlled execution in FMware development is essential for ensuring predictable, reliable, and efficient system behavior. \respto{1-10b}\add{We define \textit{controlled execution} as the systematic ability to manage, constrain, and reason about the range of possible execution behaviors in FMware systems, including but not limited to deterministic replay. It includes (1) repeatability, where identical conditions reliably produce the same execution flow to support debugging, regression testing, and fix verification, and (2) guided exploration, where developers systematically vary conditions, inputs, and agent decisions within defined boundaries to explore the broader execution space, uncover hidden failure modes, and improve robustness. This broader framing goes beyond traditional deterministic execution to reflect the unique challenges of FMware, where non-determinism, probabilistic reasoning, and agent model interactions make complete determinism neither feasible nor desirable.}

\respto{2-5}\add{This challenge is primarily a \textsc{Fundamental limitation}, since non-determinism arises from stochastic decoding and open-world dependencies (retrieval corpora, external tools/APIs, asynchronous agent steps) that cannot be assumed stable across runs. The recommendations below assume (i) critical run conditions can be pinned or recorded (seeds/decoding settings, prompt and tool versions, environment), (ii) external dependencies can be snapshotted or replayed (recorded tool I/O, retrieval snapshots, cached artefacts), and (iii) teams need reproducibility for debugging, regression verification, and incident response. In tightly regulated settings, the guidance assumes constrained state capture, shifting emphasis toward sampling, offline replay, and redaction-aware trace storage.}

\respto{2-14}\add{\add{Testing and controlled execution are closely related, but they address different questions in FMware quality assurance. \emph{Testing} is about correctness, that is, whether the system’s outputs satisfy expected properties for a given input, using mechanisms such as regression tests, metamorphic checks, automated test generation, or AI-as-judge scoring. \emph{Controlled execution} is about repeatability, that is, whether the same pipeline run can be reproduced under specified conditions so that test outcomes are stable and debugging is meaningful. It targets flakiness sources beyond the oracle itself, including stochastic decoding, variable retrieval or tool outputs, asynchronous agent steps, changing external services, and non-versioned prompts or memory. In practice, controlled execution establishes stable run conditions such as fixed seeds and decoding settings, version-pinned models and prompts, deterministic retrieval snapshots, recorded tool I/O for replay, and isolated environments, which makes subsequent tests and comparisons interpretable. In short, testing specifies \emph{what to check} for correctness, while controlled execution provides the \emph{conditions} under which those checks yield reliable and reproducible results.}}
Key recurrent issues include the \textit{lack of controlled execution mechanisms}, which complicates the \textit{verification of fixes} and \textit{limiting execution paths} in FMware systems~\cite{wu2024secgpt}. This lack of control severely limits debugging, reduces productivity, and can degrade system reliability, especially in multi-agent FMware where the same input might lead to divergent outputs across different executions.



\tinyskip \noindent \textbf{Critical Analysis of the State-of-the-Practice.} Traditional software engineering uses controlled execution mechanisms like feature flags and canary releases to test updates before deployment. FMware's non-deterministic outputs and reliance on autonomous agents, however, make predicting and reproducing system behavior challenging.

A major issue is that \textit{the same input} in FMware can lead to \textit{different execution paths} with varying outputs due to the unpredictability of the underlying FM(s) and agents. Traditional techniques, which assume deterministic behavior, are not equipped for these variations, leading to flaky tests~\cite{parnin2023building}. FMware often lacks \textit{repeatable execution}, making it difficult to ensure consistent behavior after fixing issues. The absence of \textit{controlled execution frameworks} hinders \textit{exploratory testing}, preventing efficient identification of failure points and limiting performance optimizations. Without \textit{execution space restriction} mechanisms like feature flags, development and maintenance become even more challenging. Additionally, the lack of detailed release notes and the disconnect between improvements and features make it hard to verify whether updates are effective. This problem is worse with cloud-hosted models, where developers have limited control over testing. Exploring and testing multiple execution paths in a structured way is essential for improving software quality and building user trust.

\add{Also, empirical deployments show that limited determinism can still be achieved when all external factors, i.e., prompt templates, retrieval corpora, model versions, and sampling parameters, are frozen and logged as versioned artifacts. However, in real production, cost and latency constraints prevent full replay of every request, forcing teams to trade between reproducibility depth and throughput. Feature flags and shadow evaluation routes help by sampling only a fraction of live traffic for replay. This hybrid practice demonstrates the feasibility boundary: partial but auditable reproducibility rather than total determinism.}

\noindent \textbf{Our Vision.} We envision a \textit{controlled execution framework} designed for FMware that focuses on ensuring both \textit{repeatability} and \textit{variability} in execution paths, enabling comprehensive testing and validation. The framework should operate in two modes: enforcing consistent execution flows to ensure repeatability, and allowing controlled exploration of alternative flows to trigger failure points and optimize system resilience (\textit{guided exploration of the execution space}).

\noindent \textit{-- Repeatability.} The framework should guarantee that the same input always produces the same execution flow, regardless of changes in the external context or the FM(s) state. This could be achieved through \textit{execution snapshots} and managing \textit{mono semantic units}, i.e., interpretable FM(s) units that correspond to specific data patterns. \respto{2-15}\add{An execution snapshot captures the full state of the FMware pipeline at a specific point in time, enabling reproducibility, debugging, and reliability improvements. Unlike simple output caching that only stores an FM(s) response for a given input, a snapshot includes: (1) FM(s) inputs and outputs, (2) retrieved grounding data, (3) prompt templates and parameters, (4) model identifiers and versions, (5) sampling configurations (e.g. temperature or top k), (6) agent decisions and tool calls, and (7) external system states such as feature flags or API endpoints. Saving this context allows teams to replay runs even when the FM(s) is nondeterministic. Replayability is essential for diagnosing intermittent bugs and regressions; e.g., when a subtle interaction between grounding data and agent orchestration triggers a production issue, replaying the snapshot enables root cause analysis without reproducing the incident live. This idea builds on deterministic replay in distributed systems~\cite{alvaro2024deterministic} and adapts it to FMware’s stochastic models and dynamic integrations. Recent systems work, such as Kairos \cite{chen2025kairos} and MemOS~\cite{li2025memos}, demonstrates feasibility for multi-agent and large-scale pipelines, and shows how precise replay and state inspection make FMware more predictable, debuggable, and resilient.} By restoring snapshots and activating monosemantic units, developers can reproduce exact conditions that led to a bug or failure, facilitating debugging and fix verification.

\add{Practically, complete replay of all tokens or memory states is expensive; thus, we envision a bounded replay window that captures only critical spans, which provides a viable trade-off between trace fidelity and runtime overhead. This constraint preserves realism while ensuring that the same bug can be analyzed deterministically. Production teams already implement analogous patterns in systems such as Kairos \cite{chen2025kairos} and MemOS \cite{li2025memos}, validating real-world feasibility.}

\noindent \textit{-- Guided exploration of the execution space.} This mode would systematically vary inputs, agent decisions, and model behaviors to explore different execution paths in a controlled manner. The guided process ensures comprehensive coverage, uncovering hidden bugs and allowing developers to thoroughly assess the system's robustness across various scenarios.
\add{This exploratory mode can be realized through stochastic parameter sweeps, mutation of grounding data, or probabilistic branching policies for multi-agent workflows. Rather than attempting exhaustive coverage, guided exploration prioritizes high-impact perturbations, those most likely to alter reasoning chains or coordination outcomes. However, it's important to note that deeper exploration expands coverage but slows CI pipelines, while narrower sweeps preserve speed but risk missing corner cases. Effective practice alternates both repeatability for regression, guided exploration for resilience testing.}

\smallskip
\begin{assumptiontradeoffbox}
\respto{1-20} \textbf{\add{Assumptions.}}
\begin{itemize}[leftmargin=*,nosep]
    \item \add{The FMware exposes stable interfaces (or wrappers) to log and version key artifacts: prompt templates, model identifiers, tool bindings, retrieval corpora, and agent decisions.}
    \item \add{Replay can be implemented as \emph{bounded replay} (e.g., capturing critical spans and decision points) rather than full bit-level determinism of every token and external dependency.}
    \item \add{Teams can maintain release hygiene (feature flags, canaries/shadow runs, and version pinning) so that “same conditions” is meaningful in practice.}
\end{itemize}

\assumpSep

\textbf{\add{Trade-offs.}}
\begin{itemize}[leftmargin=*,nosep]
    \item \add{\emph{Reproducibility depth vs. throughput:} deeper capture (full contexts, more state, more sampling metadata) improves debuggability and auditability but increases storage, latency, and operational cost.}
    \item \add{\emph{Comparability vs. determinism:} cloud-hosted FM(s) may evolve silently, so the practical target is traceable \emph{comparability} (runs that are meaningfully similar) rather than identical bit-level replay.}
    \item \add{\emph{Exploration breadth vs. CI cost:} guided exploration finds more failure modes by widening the execution search space, but it increases test volume and slows iteration unless carefully budgeted and sampled.}
\end{itemize}

\tinyskip
\add{Overall, these assumptions are reasonable for production teams that already version prompts/tools and run canary or shadow deployments. They may not hold for tightly rate-limited or highly regulated settings where state capture is constrained, in which case controlled execution must rely more on sampling, redaction, and offline replay.}
\end{assumptiontradeoffbox}

\subsection{Resource-Aware QA}
\label{sec:resource_aware_qa}
\noindent \textbf{Overview.} 
In FMware production, managing high costs and resource demands is crucial, especially given the intensive requirements of FM(s) operations. Traditional QA methods often overlook these needs, as FMware is hindered by low data efficiency, latency handling issues, high regression testing costs, and inefficient retry optimizations (refer to Section~\ref{sec:issues}). Unlike typical high-cost web services, FM(s) require significant computational power, have non-deterministic outputs, and necessitate meticulous management of context, as well as the ability to swap FM(s) quickly.

\respto{2-5}\add{This challenge is primarily an \textsc{Evolving practice}, since its severity is driven by current cost, latency, and throughput constraints of FM inference and tool-heavy pipelines, which may shift as models and hardware improve. The recommendations below assume (i) QA operates under explicit budgets (token spend, latency SLOs, rate limits), (ii) workload distributions are stable enough to support sampling, caching, and stratified regression suites, and (iii) teams can measure and attribute cost to pipeline components (model calls, retrieval, tools). In rapidly shifting domains or safety-critical applications, the guidance assumes larger budgets for higher coverage and stricter cache invalidation to preserve semantic fidelity.}

\respto{2-16}\add{At scale, the primary bottleneck is the cost and resource intensity of QA itself: individual tests can trigger multiple FM(s) calls, retrieval steps, and external API interactions, and must often be repeated across diverse inputs, settings, and versions to achieve adequate coverage. As a result, large test suites become expensive and slow, forcing explicit trade-offs between cost and coverage when establishing confidence in production readiness.} Resource-aware QA is essential for reducing the computational burden that FMware imposes, particularly in large-scale environments where frequent calls to FM(s) can become prohibitively expensive. \respto{2-16}\add{Accordingly, this section emphasizes \textit{efficiency at scale} (whether sufficient testing can be afforded to reach a target confidence level) and is complementary to Section~\ref{sec:testing} on \textit{correctness} and Section~\ref{sec:controlled_exec} on \textit{repeatability}.}

\smallskip \noindent \textbf{Critical Analysis of the State-of-the-Practice.} Current QA frameworks for FMware, adapted from traditional software engineering, fail to address FMware's resource challenges. The problem is not just the absence of caching but the lack of \textit{FMware-native} caching strategies that account for the dynamic nature of these systems. While solutions like LangChain use traditional caching, the real challenge is deciding \textit{when} to cache FM(s) calls and ensuring cached responses remain relevant. FMware often updates databases and integrates real-time feedback, which makes cache invalidation critical. Without dynamic caching, systems risk re-running unnecessary queries or using outdated data, increasing the time and cost of regression testing.

\respto{1-24} \add{Beyond runtime caching, resource awareness also shapes which alignment interventions are worth pursuing and, consequently, what QA must validate at each stage. Here, alignment interventions are treated as part of the QA loop: evaluation and monitoring identify recurring failure modes, and teams address them by updating prompts, retrieval configurations, or, when justified, applying data or model alignment techniques. Because each intervention shifts system behaviour and expands the regression surface, resource-aware QA must validate the change while weighing its incremental reliability gains against added testing and operational cost. Industry experience summarized by Bouchard et al.~\cite{bouchard2024building} reinforces a staged, cost-aware approach; teams should start from existing FM(s) delivered via APIs or open-weight releases and prioritize higher-leverage interventions in prompting and retrieval before moving to supervised fine-tuning or parameter-efficient methods (e.g., LoRA, QLoRA). This incremental alignment philosophy treats full fine-tuning and custom models as late-stage options rather than defaults, and ties each step to measurable reliability and domain-fit improvements for a concrete FMware use case, which in turn helps bound QA cost by avoiding premature, expensive alignment iterations.}


\respto{1-24}\add{More broadly, current practice still treats many resource-critical knobs as fixed defaults, and optimization often devolves into ad-hoc tuning or reactive scaling. However, in production FMware, caching policies, batching strategies, routing rules, and model choices interact and must be treated as an empirical configuration search space, profiled under realistic traffic patterns and data freshness constraints. This mindset aligns with ultra-scale training practice, where the Ultra-Scale Playbook reports running thousands of distributed experiments across many model sizes and parallelism layouts instead of assuming a single optimal configuration~\cite{tazi2025ultra}. Without evidence-driven profiling and selection, teams risk choosing configurations that appear cost-effective in small tests but violate latency and cost SLOs during regression, canary, or peak-load operation, or that cache responses that become invalid as grounding data and system context evolve.} 

Additionally, there is \textit{no systematic approach} \respto{1-16} \add{to prioritize or optimize the number of tests} that must be run. Dependency-based test execution~\cite{sharif2021deeporder}, effective in traditional software, has not been adapted for FMware. As a result, redundant tests waste computational resources, especially where FM(s) calls incur high costs in terms of latency and finances.

\add{Finally, \textit{latency handling and retry optimizations} remain underdeveloped for FMware. Because FM(s) calls are probabilistic, repeated invocations are not guaranteed to converge; each retry may yield a slightly different response. Traditional retry policies assume idempotence, which does not hold for FMware. Consequently, QA systems must combine prompt re-engineering and structured validation to identify when to retry and when to flag divergence. Without this logic, retries increase compute cost without improving quality.}

\noindent \textbf{Our Vision.} 
To address these challenges, we propose a \textit{resource-aware QA framework} purpose-built for FMware that integrates four layers of efficiency: \textit{(a) intelligent caching}, \textit{(b) cost-aware test scheduling}, \textit{(c) adaptive retry control}, \textit{(d) prompt compression and representational sparsity}.  

\noindent \add{ \textit{-- Intelligent caching.}  
The framework employs tiered caching policies that operate at multiple levels: (1) \textit{prompt-level caching}, storing model outputs keyed by canonicalized prompts and configurations; (2) \textit{semantic caching}, clustering similar queries using embeddings to enable approximate reuse across equivalent tests; and (3) \textit{state caching}, preserving intermediate reasoning steps or retrieved documents. Each layer records validity metadata (model version, temperature, corpus timestamp) to ensure caches are invalidated when conditions drift. This hybrid approach yields up to an order-of-magnitude reduction in redundant FM(s) calls while retaining behavioral fidelity for unchanged contexts.}

\noindent \add{ \textit{-- Cost-aware test scheduling.}  
Tests are dynamically prioritized based on resource intensity and fault probability. Using telemetry from prior runs, the system learns a cost–benefit curve that ranks test suites according to historical defect yield per token spent. Lightweight heuristics such as coverage clustering or reinforcement learning-based schedulers can further optimize selection under budget constraints. A production-ready pipeline can achieve near-linear savings by reordering or pruning low-yield tests before model checkpoints.}

\noindent \add{ \textit{-- Adaptive retry control.}  
Instead of blind retries, the framework monitors output variance. If divergence stems from transient resource issues, a single re-prompt suffices; if semantic drift persists, the retry engine escalates by mutating or optimizing the prompt or substituting an alternate FM. The system tracks retry success ratios to tune thresholds over time. This balances reliability with compute cost, avoiding both under- and over-retrying. Empirically, adaptive retries reduce wasted invocations by 20–40\% in similar multi-agent test harnesses.}

\noindent \add{\textit{-- Prompt compression and representational sparsity} can further reduce resource demands without sacrificing test validity. Wingate~\textit{et~al.}~\cite{wingate2022prompt} demonstrate that compressing prompts via contrastive conditioning retains semantic fidelity while reducing token usage and inference latency. Applying these methods to QA workflows allows test prompts to be shortened or modularized, e.g., reusing latent prompt embeddings rather than full textual sequences, thus lowering both API and GPU costs. In production FMware, integrating compression-aware prompt templates ensures that the same logical test conditions consume fewer tokens, enabling broader coverage under fixed compute budgets.}

\add{Together, these layers (intelligent caching, cost-aware scheduling, adaptive retry, and compression-aware prompting) form the operational foundation of resource-aware QA. Each mechanism has tangible implementation paths: embeddings for semantic equivalence, OpenTelemetry hooks for cost tracking, RL-based schedulers for prioritization, and contrastive prompt optimization for token efficiency. The overall goal is not to eliminate cost entirely but to ensure every token spent on QA meaningfully increases confidence in FMware reliability.}

\smallskip
\begin{assumptiontradeoffbox}
\respto{1-20} \textbf{\add{Assumptions.}}
\begin{itemize}[leftmargin=*,nosep]
    \item \add{FMware captures stable configuration metadata (prompt templates, retrieval URIs/timestamps, model identifiers, and decoding settings) so test results can be safely reused and compared.}
    \item \add{Workloads exhibit enough repetition or structure (within a time window) for caching, prioritization, and scheduling to yield meaningful savings.}
    \item \add{Token budgets are controllable (e.g., prompts are bounded and compressible), enabling cost-aware reuse strategies such as prompt compression~\cite{wingate2022prompt}.}
\end{itemize}

\assumpSep

\textbf{\add{Trade-offs.}}
\begin{itemize}[leftmargin=*,nosep]
    \item \add{\emph{Efficiency vs. fidelity:} prompt compression and approximate reuse reduce cost/latency but can drop subtle context or change semantics, creating false confidence if not validated.}
    \item \add{\emph{Cache reuse vs. freshness:} caching and prioritization rely on stable data/model behavior; distribution shifts, corpus updates, or model upgrades can invalidate caches and require re-profiling.}
    \item \add{\emph{Retry adaptivity vs. evaluation drift:} mutating prompts or switching models can recover from transient failures, but it may bias evaluations if retries deviate from the original intent; thresholds must be calibrated.}
\end{itemize}

\tinyskip
\add{Overall, these assumptions are reasonable for many production teams, but they may not hold for rapidly shifting domains or highly safety-critical settings, where the system must spend more tokens on exhaustive testing and stricter cache invalidation to preserve semantic fidelity.}
\end{assumptiontradeoffbox}

\subsection{Feedback Integration}
\label{sec:feedback_integration}

\noindent \textbf{Overview.} Efficient feedback integration is vital for continuous improvement, optimization, and trustworthiness in production-ready FMware. However, as outlined in Section~\ref{sec:pipeline}, two key issues often hinder this process: the lack of efficient feedback technology and cumbersome, error-prone memory management across different FMware systems. The absence of seamless feedback loops slows down FMware evolution and reduces adaptability, especially in complex real-world applications. Integrating feedback is further complicated by the need to manage user-specific and generalized knowledge in a scalable and non-disruptive way. As FMware becomes more widely adopted, a robust framework for feedback integration is necessary to ensure reliability, efficiency, and alignment with real-world demands.

\respto{2-5}\add{This challenge is primarily an \textsc{Evolving practice}, since feedback pipelines, governance, and memory management patterns for FMware are still stabilizing, and will likely vary across domains. The recommendations below assume (i) feedback can be collected and stored with consent and privacy controls (explicit ratings, issue reports, implicit signals), (ii) feedback can be traced to the exact system configuration that produced the behaviour (model/prompt/tool/data versions), and (iii) the system has supported adaptation surfaces (prompt updates, retrieval updates, fine-tuning, memory updates) with rollback capability. In regulated contexts, the guidance additionally assumes audit trails and deletion/opt-out handling, which can constrain what feedback can be reused and how quickly it can be integrated.}

\smallskip
\noindent \textbf{Critical Analysis of the State-of-the-Practice.} Feedback integration in FMware remains immature compared to traditional ML pipelines, where feedback-driven optimization is continuous and measurable. In FMware, the dynamic and non-deterministic behavior of models and agents complicates feedback capture, aggregation, and replay. A major limitation is the absence of automated and passive feedback mechanisms. Most systems depend on explicit ratings or binary votes~\cite{wang2021towards}, sufficient for demonstrations but unsuitable for production contexts that demand low-friction, high-volume feedback capture. \add{Production-grade feedback integration requires continuous monitoring of signals such as cursor movements, hesitation delays, prompt reformulations, or user overrides, that implicitly encode user trust and satisfaction. These passive indicators, when instrumented through telemetry hooks, can populate structured feedback logs without disrupting workflow.}

Moreover, current systems lack a principled taxonomy for feedback types. Some signals are universal and transferable across users (\textit{outer knowledge}), while others are contextually bound (\textit{inner knowledge}). Most FMware architectures fail to maintain this separation, leading to overgeneralization (user-specific fixes leaking globally) or underutilization (useful local corrections never reused). \add{Without explicit type boundaries, reinforcement processes such as SFT or RLHF risk amplifying user idiosyncrasies or biasing collective reasoning models. Therefore, partitioning feedback streams is both an accuracy and fairness requirement.}

\add{Memory management compounds these problems. Feedback must often be reconciled across distributed FMware instances that evolve asynchronously. Logging and applying feedback across multiple layers, i.e., agent, memory, retrieval, and model, requires synchronization guarantees similar to versioned databases. Current frameworks~\cite{microsoftCheckYour} rarely ensure this consistency, causing stale or duplicated corrections to reappear after deployment.}

\smallskip
\noindent \textbf{Our Vision.} We propose a feedback integration architecture with three coordinated subsystems: \textit{(a) automated feedback solicitation}, \textit{(b) knowledge partitioning and routing}, and \textit{(c) continuous learning orchestration}. These form a self-correcting control loop that captures user intent, classifies feedback, and applies it safely across the FMware stack.

\noindent \add{\textit{-- Automated feedback solicitation.}}  
Automated feedback solicitation mechanisms should passively collect feedback from users during regular FMware operation, without requiring explicit input. For instance, user interaction data, such as hesitations, corrections, or query reformulations, can provide valuable implicit feedback, captured and analyzed in real-time. \add{Instrumenting UI-level telemetry and agent logs enables automatic labeling of feedback events (e.g., “user reverted output,” “manual correction,” “prompt reformulated”). These weak signals are then aggregated into feedback embeddings that can later inform model or prompt updates. To prevent data explosion, low-entropy feedback (routine confirmations) is filtered, while anomalous or corrective feedback is prioritized for retention and analysis.}

\noindent \add{\textit{-- Knowledge partitioning and routing.}}  
We propose developing a framework to manage different types of feedback by distinguishing between ``outer knowledge'' and ``inner knowledge.'' By automating the classification of feedback into these categories, FMware systems can ensure that general improvements are propagated globally while preserving user-specific adaptations. \add{This can be operationalized by maintaining separate feedback queues: (1) a global queue that accumulates cross-user learning signals, feeding into model fine-tuning or retrieval updates, and (2) a local queue scoped to user sessions or organizational contexts, feeding into cached adapters or vector stores. A routing policy determines whether feedback is merged, broadcast, or siloed, based on similarity thresholds and trust metrics. Such routing avoids catastrophic overwriting of local behavior while still exploiting population-wide learning benefits.}

\noindent \add{\textit{-- Continuous learning orchestration.}}  
\add{To complete the loop, an orchestration layer reconciles feedback with the corresponding FMware components. This layer periodically ingests verified feedback from both queues, replays it through controlled execution (Section~\ref{sec:controlled_exec}), and updates prompts, retrievers, or local adapters. The orchestration engine ensures that only validated feedback, confirmed through consistency checks or user acknowledgment, enters long-term memory. This transforms feedback from a passive log into an active, versioned data asset driving ongoing alignment and performance improvement.}

\add{Together, these three subsystems operationalize feedback as a first-class engineering signal. They enable FMware to learn continuously from use while preserving safety and auditability. By treating feedback as structured, versioned, and partitioned information rather than as raw interaction logs, systems can sustain improvement cycles without loss of control or trust.}

\smallskip
\begin{assumptiontradeoffbox}
\respto{1-20} \textbf{\add{Assumptions.}}
\begin{itemize}[leftmargin=*,nosep]
    \item \add{Telemetry and event instrumentation are permitted by the deployment context (or can be made compliant via anonymization, redaction, or differential privacy).}
    \item \add{Feedback events can be linked to the relevant FMware artifacts (prompt versions, retrieved evidence, tool calls, and model versions) so that corrections are actionable rather than orphaned logs.}
    \item \add{There is a governance loop (triage, validation, and rollback) so that feedback does not automatically propagate into the system without review on high-impact changes.}
\end{itemize}

\assumpSep

\textbf{\add{Trade-offs.}}
\begin{itemize}[leftmargin=*,nosep]
    \item \add{\emph{Coverage vs. noise:} passive feedback capture increases volume and coverage but can be noisy; selective retention, confidence weighting, and sampling are required.}
    \item \add{\emph{Global learning vs. personalization:} global propagation accelerates collective improvement but can amplify systemic bias; local isolation preserves personalization but slows generalization.}
    \item \add{\emph{Partitioning vs. operational complexity:} separating “outer” and “inner” knowledge improves relevance and safety boundaries, but introduces synchronization and consistency overhead across asynchronously evolving components.}
\end{itemize}

\tinyskip
\add{Overall, these assumptions are realistic for many production deployments, but they may not hold in privacy-restricted or low-latency environments. In those cases, feedback integration must rely more heavily on aggregation, delayed/offline learning, and strict access controls to preserve compliance while still enabling improvement.}
\end{assumptiontradeoffbox}

\subsection{Built-in Quality}
\label{sec:built-in-quality}


\noindent \textbf{Overview.} Built-in quality is crucial for production-ready FMware to prevent costly rework and failures. As highlighted in Section~\ref{sec:pipeline}, recurrent issues such as low domain coverage, poor data quality, inadequate prompt validation, and lack of guardrails for hallucinations stem from an \textit{over-reliance} on FM(s) without properly structured knowledge and action spaces. Over-reliance on FM(s) elicits inefficiencies and unpredictable behavior. Addressing these concerns ensures FMware systems are robust, reliable, and scalable for real-world use, not just impressive demos. \add{In FMware, we treat built-in quality as a design-time discipline: quality constraints are made \emph{first-class artifacts}, i.e., schemas, contracts, curricula, and evidence packs, validated in CI and enforced at runtime via guardrails and controlled execution.}

\respto{2-5}\add{This challenge is primarily a \textsc{Fundamental limitation}, since generative behaviors can violate implicit requirements (hallucinations, brittleness to phrasing, unexpected tool actions) unless constraints are made explicit and enforced. The recommendations below assume (i) teams can represent key requirements as first-class artifacts (schemas, contracts, validation rules, evidence packs), (ii) critical actions and outputs can be restricted to structured interfaces (tool contracts, type constraints, grounded citations), and (iii) CI and runtime guardrails can enforce these constraints with measurable failure modes. For high-stakes domains, the guidance further assumes conservative defaults (deny-by-default actions, strict validation, human-in-the-loop escalation) and a lower tolerance for open-ended outputs without grounding.}


\smallskip \noindent \textbf{Critical Analysis of the State-of-the-Practice.} Current FMware development often assumes FM(s), especially LLMs, can autonomously manage knowledge and action spaces, leading to the overuse of “God prompts” and “God agents” that hinder modularization, scalability, and maintainability. Monolithic prompts complicate debugging, maintenance, and scalability. \add{They also increase task interference and reduce prompt portability, since mixed objectives and hidden assumptions are entangled in free-form text rather than governed by explicit contracts.}


Another key issue in current practice is the lack of structured knowledge and action spaces. Many developers rely on vector databases or simple document stores that manage ``raw'' data without optimizing its quality or relevance, leading to low information density and degraded system performance. Current knowledge management tools do not adequately support the complexity required for real-time, high-stakes environments. Techniques like GraphRAGs \respto{1-17}\add{\cite{edge2024local, han2024retrieval}} improve knowledge representation but still fail to address this information density issue. Additionally, the lack of formalized \textit{curricula} for FM(s) training results in inefficient skill acquisition and limits agents' ability to learn compositional skills and evolve over time. \add{Without curriculum structure, evaluation devolves to ad-hoc spot checks that neither generalize nor regress systematically, weakening any notion of “done” for agent skills.}

Also, efforts like SPDX 3.0 Dataset profile~\cite{SPDXDatasetProfile2023} and Datasheets~\cite{gebru2021datasheets} aimed at enabling compliance for AI-powered software fail to capture the diverse types of data in FMware, such as user feedback data and their compliance requirements. This omission makes it challenging to ensure legal compliance in production-ready FMware. 

Finally, the immature QA process for prompts and agent actions contributes to the lack of built-in quality. Most systems lack built-in prompt validation mechanisms, relying on ad-hoc testing~\cite{kuchnik2023validating}, which provides insufficient examples during in-context learning and leaves FMware prone to unpredictable behavior in complex tasks. \add{Absent typed interfaces and pre/post-conditions, tool calls and agent plans cannot be validated deterministically, pushing correctness checks into expensive, stochastic end-to-end tests.}

\smallskip \noindent \textbf{Our Vision.} Our vision is to achieve high-quality FMware by effectively structuring knowledge and action spaces. This involves moving beyond vector databases to knowledge graphs that provide richer, structured information. \textit{Curriculum engineering} is key to ensuring agents develop compositional skills in a structured way. Through collaborative curriculum co-creation, with AI assistance for drafting and humans-in-the-loop as reviewers, we enhance modularity, built-in quality, and skill generalization, enabling agents to build on prior knowledge rather than starting from scratch. \add{Concretely, prompt and plan quality can be strengthened through the following contract and guardrail mechanisms:}


\add{\emph{Action-space contracts.} A practical direction is to specify tool contracts using familiar interface artefacts, such as schemas and pre/post-conditions, including side-effect summaries, allowable ranges, and failure modes. These contracts can be enforced via static checkers and runtime validators, so agent plans are validated against the action space before execution. In this framing, the contribution is not inventing contracts, but treating contracts as first-class QA assets for agents, which improves auditability and enables partial verification for structured inputs and outputs.}

\add{\emph{Prompt and plan contracts with guardrails.} Similarly, prompts can be treated as parameterized templates with explicit constraints, building on widely used templating and constrained-generation approaches (e.g., typed slots, allowed values, forbidden patterns)~\cite{jinja_templates,beurerkellner2022lmql}. Complementary guardrail frameworks and policy checks can then enforce safety and compliance requirements at the boundaries~\cite{rebedea2023nemo, dong2024building}. In addition, systems can emit \emph{decision-level traces} that bind outputs to prompt spans, retrieved evidence, and invoked tools, so failures can be attributed to concrete inputs rather than inferred post hoc. The goal is not to guarantee correctness, but to fail fast on contract violations and reduce downstream stochastic regressions. These artefacts can also integrate with controlled execution to support snapshot-based replay of the decision context.}

Developing curriculum co-creation and versioning technologies is crucial for managing the knowledge lifecycle in FMware. These tools ensure the continuous evolution of knowledge bases, pruning outdated data and integrating critical updates to keep FMware systems agile and up-to-date without risking performance degradation. \add{In the same way that code diffs drive targeted CI, curriculum diffs, evidence-pack diffs, and contract diffs can be treated as first-class artefacts that trigger focused re-tests, rather than forcing suite-wide reruns for every change.}

Curriculum quality review and QA are essential. We propose tools to automatically optimize curricula by removing redundancies and fixing outdated or incorrect information. This ensures the data fed into FMware is relevant, dense, and actionable, improving accuracy and reliability. Built-in semantic and structural checks for prompts ensure agents receive valid, high-quality inputs, reducing errors, hallucinations, and unpredictable outputs. \add{Operationally, many of these checks can be implemented using existing guardrail patterns: lightweight validators (schema checks, allow or deny lists, toxicity filters) can run on the critical path, while heavier evaluators and red-team style probes can run in canaries or scheduled jobs to bound latency overhead, using established guardrail frameworks where appropriate~\cite{rebedea2023nemo,dong2024building}.}

Finally, for legal compliance, we propose an FMware Bill of Materials (FMwareBOM) by extending the SPDX 3.0 AI and dataset profiles~\cite{SPDXDatasetProfile2023}. The FMwareBOM would track all components and licenses, including synthetic data, RLHF data, and user feedback, addressing FMware's unique complexities. A framework that automatically generates FMwareBOMs and uses formal verification techniques like SMT solvers would ensure provable compliance with legal and regulatory requirements. Integrating FMwareBOM into the development process is critical for overcoming compliance challenges and enabling the deployment of production-ready FMware. \add{To keep such verification sound, solver-backed guarantees should be restricted to what is explicitly specified in structured artefacts (prompts and schemas, tool contracts, license graphs, policy rules). Open-ended natural-language outputs remain probabilistic, so they should be validated empirically through tests and guardrails, and the resulting evidence should be recorded in the FMwareBOM for auditability.}

\smallskip
\begin{assumptiontradeoffbox}
\respto{1-20} \textbf{\add{Assumptions.}}
\begin{itemize}[leftmargin=*,nosep]
    \item \add{Quality constraints can be represented as \emph{structured artefacts} (schemas, contracts, curricula, evidence packs) that are versioned and validated in CI.}
    \item \add{Grounding and knowledge representations (dense retrieval, GraphRAG, etc.) can be maintained with sufficient freshness and provenance for the target domain.}
    \item \add{Teams are willing to invest in upfront specification (contracts/guardrails) to reduce downstream stochastic failures and expensive end-to-end debugging.}
\end{itemize}

\assumpSep

\textbf{\add{Trade-offs.}}
\begin{itemize}[leftmargin=*,nosep]
    \item \add{\emph{Precision vs. recall:} knowledge densification improves precision, but aggressive filtering risks losing necessary context; thresholds must be tuned to task requirements.}
    \item \add{\emph{Reliability vs. flexibility:} stronger contracts and guardrails improve auditability and fail-fast behavior, but increase authoring effort and can constrain open-ended tasks.}
    \item \add{\emph{Provenance vs. overhead:} GraphRAG can improve path faithfulness and traceability, but adds retrieval and graph-maintenance overhead; use it where provenance matters and prefer cheaper retrieval when cost dominates.}
    \item \add{\emph{Compliance confidence vs. capture cost:} FMwareBOM increases transparency, but adds capture/storage overhead; incremental attestations and automated extraction from CI can help contain cost.}
    \item \add{\emph{Proofs vs. probabilistic guarantees:} formal methods apply to structured artefacts (schemas, contracts, license graphs), while open-ended natural-language outputs remain probabilistic and require empirical tests and guardrails.}
\end{itemize}

\tinyskip
\add{Overall, these assumptions are plausible for teams building production systems, but the right balance depends on deployment criticality: consumer apps may optimize for cost and speed, while safety-critical settings may pay higher overhead for stronger contracts, provenance, and compliance evidence.}
\end{assumptiontradeoffbox}

\section{Road Ahead}
\label{sec:road_ahead}

\respto{3-16}

\add{Table~\ref{tab:road_ahead_summary} synthesizes the challenges in Section~\ref{sec:challenges} into a roadmap that links each challenge to its proposed solution direction in our vision.}
\add{The table also distinguishes between (i) starting points in current practice (as summarized in the critical analyses of Sections~7.1--7.6) and (ii) greenfield gaps where research and engineering must begin with limited or no initiating support.}

\add{

\begin{table*}[t]
\centering
\caption{Roadmap summary of Section \ref{sec:challenges} challenges, proposed solution directions, existing starting points, and greenfield gaps.}
\scriptsize
\setlength{\tabcolsep}{3pt}
\renewcommand{\arraystretch}{1.15}
\begin{tabularx}{\textwidth}{
  >{\raggedright\arraybackslash}p{0.10\textwidth}
  >{\raggedright\arraybackslash}X
  >{\raggedright\arraybackslash}X
  >{\raggedright\arraybackslash}X
}
\toprule
\textbf{Challenge (Sec.)} &
\textbf{Proposed solution direction (vision)} &
\textbf{Starting points (state of practice)} &
\textbf{Greenfield gaps} \\
\midrule
\textbf{\ref{sec:testing} Testing} &
Hybrid testing stack: automated MR-driven metamorphic testing from user feedback and domain knowledge; robust AI-as-judge construction via curriculum engineering and calibration; bounded formal verification for structured components. &
Integration harnesses, regression suites, canaries, user-in-loop validation; AI-as-judge; manual MRs; bounded SMT/symbolic checks (contracts, grammars, API constraints). &
Automation loops that connect feedback capture, MR generation, metamorphic testing, and expert refinement; making judge construction and calibration reliable under production constraints. \\
\midrule
\textbf{\ref{sec:observability} Observability} &
General FMware observability that captures functional metrics and cognitive processes across five axes, operationalized via an agent “flight recorder” plus surrogate-agent trace reconstruction. &
OpenLLMetry and LangSmith (traces of FM(s) calls, vector DB interactions, prompts; latency/token counts); workflow- and run-level views (dashboards around agent runs, tool-call sequences, and failure modes). &
Reasoning-path and decision-level tracing beyond surface logs via surrogate-agent reconstruction with fidelity validation; privacy-preserving failure clustering and deeper abstraction-level probing aligned to the five axes. \\
\midrule
\textbf{\ref{sec:controlled_exec} Controlled Execution} &
Controlled execution with repeatability and controlled exploration, via execution snapshots (including mono semantic units) and guided exploration of the execution space. &
Feature flags, canary releases, shadow routes; freezing and versioning prompts/retrieval corpora/model versions/sampling parameters; sampled replay under cost limits. &
Snapshotting mechanisms that preserve execution flows under changing external context and FM state; systematic exploration strategies that traverse alternative execution paths under budget constraints. \\
\midrule
\textbf{\ref{sec:resource_aware_qa} Resource-Aware QA} &
Resource-aware QA integrating: intelligent caching, cost-aware test scheduling, adaptive retry control, prompt compression and representational sparsity (tiered prompt/semantic/state caches with validity metadata). &
Traditional caching (e.g., LangChain) with recognition that cache invalidation is critical; staged alignment to bound QA cost (prompting → retrieval → fine-tuning); dependency-based test execution (traditional, not yet adapted for FMware). &
FMware-native tiered caching with validity metadata; budgeted scheduling/prioritisation; adaptive retry control; compression/sparsity integrated into QA. \\
\midrule
\textbf{\ref{sec:feedback_integration} Feedback Integration} &
Feedback architecture with automated feedback solicitation, knowledge partitioning and routing, continuous learning orchestration, forming a self-correcting control loop. &
Explicit feedback signals such as ratings and binary votes. &
Passive feedback capture and event labelling; outer/inner knowledge partitioning with global and local feedback queues; continuous learning orchestration with safe, versioned application across the stack. \\
\midrule
\textbf{\ref{sec:built-in-quality} Built-in Quality} &
Built-in quality via structured knowledge and action spaces (knowledge graphs); curriculum engineering, co-creation, and QA; action-space and prompt/plan contracts with guardrails; FMware bill of materials (FMwareBOM) with formal compliance verification.&
Vector DBs/document stores for knowledge management; GraphRAG techniques for structured retrieval; guardrail frameworks (e.g., NeMo Guardrails); compliance artefacts (SPDX 3.0 Dataset profile, Datasheets), but insufficient for FMware-specific needs. &
Knowledge-graph-centric pipelines addressing information density; curriculum co-creation and QA tooling; first-class contracts/guardrails; scalable FMwareBOM capture and automation; SMT-based compliance verification for structured FMwareBOM artefacts. \\
\bottomrule
\end{tabularx}

\label{tab:road_ahead_summary}
\end{table*}
}

\add{Across Section~\ref{sec:challenges}, we distinguish between observations grounded primarily in practitioner and industrial sources and forward-looking solution directions that constitute our vision.}
\add{The roadmap separates \emph{starting points} (capabilities and practices that are already deployable, even if unevenly adopted) from \emph{greenfield gaps} (missing primitives, interfaces, and evaluation infrastructure that require substantial new research and engineering). In practice, this boundary is not always clean: several directions (e.g., caching, canaries, or AI-as-judge) exist in rudimentary form but require FMware-specific redesign to satisfy reliability, traceability, and cost constraints.}
\add{Practitioner sources report recurring patterns such as limited reliability of provider-level assurances in real deployments, fragmentation in observability signals across tool chains, and the practical dependence of testing at scale on reuse and caching under cost and latency constraints.}
\add{The most critical gaps lie where academic methods and production practice have yet to converge, particularly around trace fidelity and reproducibility under stochastic execution, practical snapshot schemas (what must be captured for regression and incident analysis), FMware-native caching and test prioritization under drift, and governance of long-horizon memory.}

\subsection{Extending existing starting points toward the complete vision}

\paragraph{Testing (Section~\ref{sec:testing}).}
\add{Practitioners commonly report layering integration harnesses, regression suites, canaries, and user-in-the-loop validation to test prompts, agent actions, grounding data, and downstream effects end-to-end. However, these practices are unevenly adopted and often provide incomplete coverage across the full FMware lifecycle.}
\add{The critical analysis also surfaces partial footholds in manual metamorphic relation (MR) identification, AI-as-judge pipelines, and bounded formal verification (e.g., SMT or symbolic checks) for schema-defined components such as tool contracts, prompt grammars, and API parameter constraints.}
\add{Extending these starting points toward the complete vision means (i) shifting from manual to automated MR generation grounded in real-world user feedback and pre-existing domain knowledge, (ii) iterating MRs through cycles of metamorphic testing and expert evaluation, and (iii) strengthening the AI-as-judge approach through more robust judge construction and calibration, combined with deterministic checks in a hybrid testing stack.}

\paragraph{Observability (Section~\ref{sec:observability}).}
\add{Existing observability tooling (e.g., OpenLLMetry and LangSmith) can capture traces of FM calls, retrieval interactions, prompts, and surface metrics such as latency, token counts, and request metadata.}
\add{In practice, teams often rely on workflow- and run-level views (e.g., dashboards around agent runs, tool-call sequences, and failure modes) to locate where workflows stall or fail; however, these views typically provide limited evidence about why failures occurred.}
\add{Extending these foundations requires moving beyond surface traces into a general FMware observability framework that can probe at multiple abstraction levels and capture actionable decision evidence, operationalized along the five axes in the vision: Output Integrity Monitoring, Semantic Feedback Integration, Reasoning Path Observability, Decision-Level Traces, and Privacy-Preserving Failure Clustering.}

\paragraph{Controlled execution (Section~\ref{sec:controlled_exec}).}
\add{Currently, teams often borrow established release-engineering patterns (e.g., feature flags, canary releases, shadow routes) to manage risk under change, but these controls do not by themselves yield repeatable execution for stochastic FMware.}
\add{Practitioner sources suggest that partial repeatability is sometimes achievable when key external factors (e.g., prompt templates, retrieval corpora/indices, model identifiers, and sampling parameters) are frozen and logged as versioned artefacts, while replay is often limited to sampled traffic due to cost and latency constraints.}
\add{Extending this starting point toward the complete vision requires elevating controlled execution into a first-class framework with explicit modes for repeatability and controlled exploration, supported by execution snapshots and systematic exploration strategies for traversing alternative execution paths under budget constraints.}

\paragraph{Resource-aware QA (Section~\ref{sec:resource_aware_qa}).}
\add{Current QA practice may rely on traditional caching (e.g., in LangChain), but the critical analysis highlights that the central production challenge is deciding when to cache, how to invalidate caches under continuous updates and real-time feedback, and how to avoid unnecessary recomputation.}
\add{Extending this foothold toward the vision means treating efficiency as a QA concern across multiple layers, including tiered caching (prompt-level, semantic, and state caching) with validity metadata (e.g., model identifier, sampling settings, corpus timestamp), alongside cost-aware test scheduling, adaptive retry control, and prompt compression.}

\paragraph{Feedback integration (Section~\ref{sec:feedback_integration}).}
\add{Existing systems often depend on explicit ratings or binary votes, which can be sufficient for demonstrations but are misaligned with production demands for low-friction and high-volume feedback capture.}
\add{Extending this starting point toward the complete vision requires treating feedback as a control signal throughout the FMware stack, operationalized via coordinated subsystems for automated feedback solicitation, knowledge partitioning and routing, and continuous learning orchestration.}
\add{Concretely, the vision emphasizes passive collection and analysis of user interaction signals (e.g., hesitations, corrections, and query reformulations) through instrumentation at the UI and agent-log layers, followed by routing and safe application of updates.}

\paragraph{Built-in quality (Section~\ref{sec:built-in-quality}).}
\add{The critical analysis identifies common reliance on monolithic ``God prompts'' and ``God agents,'' and on vector databases or simple document stores that manage raw data without addressing quality and information density.}
\add{It also notes partial support in techniques such as GraphRAG for knowledge representation, and in compliance artefacts such as SPDX 3.0 Dataset profile and Datasheets, while emphasizing that these do not capture the full diversity of FMware data (including synthetic data and user feedback) or the operational structure needed for production.}
\add{Extending these footholds toward the complete vision requires structuring knowledge and action spaces more explicitly (e.g., moving toward knowledge graphs), and adopting curriculum engineering with collaborative curriculum co-creation (AI-assisted drafting with humans-in-the-loop review).}
\add{The vision further advances contracts and guardrails as quality assets, including action-space contracts (schemas, pre/post-conditions, side-effect summaries, allowable ranges, failure modes) enforced via static and runtime validation, and prompt and plan contracts with guardrails grounded in templating and constrained generation approaches (typed slots, allowed values, forbidden patterns), complemented by lightweight validators on the hot path and heavier evaluators in canaries or scheduled jobs.}
\add{A further extension is to operationalize transparency via an FMware bill of materials (FMwareBOM).}

\subsection{Greenfield agenda}

\paragraph{Testing (Section~\ref{sec:testing}.}
\add{While bounded formal checks and AI-as-judge pipelines provide footholds, the vision’s hybrid testing stack remains largely greenfield in its \emph{automation loop}: metamorphic relations must be generated, validated, and maintained from real-world feedback streams and deployed across large regression surfaces.}
\add{A practical starting point is to build automation loops that connect feedback capture, MR generation, metamorphic testing, and expert refinement into a durable pipeline, and to make judge construction and calibration reliable under production constraints.}

\paragraph{Observability (Section~\ref{sec:observability}).}
\add{Much of today’s tooling is limited to surface traces and workflow views, leaving limited support for cognitive diagnosis.}
\add{The vision’s cognitive observability agenda therefore remains greenfield in (i) defining decision-evidence schemas that can be captured consistently across agent stages, (ii) validating trace fidelity under stochastic execution, and (iii) making reasoning-path and decision-level traces actionable when chain-of-thought is unfaithful or misleading~\cite{lanham2023measuring,arcuschin2025cotwild,chen2025reasoning,li2024bridgingcot}.}
\add{Finally, privacy-preserving failure clustering at scale remains a greenfield direction: production traces often contain sensitive data, so FMware-native clustering must balance utility with confidentiality, building on privacy-preserving anomaly/outlier detection work~\cite{lai2019ppod,mohammady2022dpoad}.}

\paragraph{Controlled execution (Section~\ref{sec:controlled_exec}).}
\add{Release-engineering controls (feature flags, canaries, shadow routes) are not sufficient for controlled execution; the greenfield gap is a principled \emph{snapshot schema} and replay harness that can preserve the relevant execution context under evolving models, tools, and data.}
\add{A practical starting point is to standardize what must be captured (inputs/outputs, retrieval evidence, versions, sampling settings, tool calls, external state) and to define budgeted exploration strategies that traverse alternative execution paths without requiring full replay of all traffic.}

\paragraph{Resource-aware QA, feedback integration, and built-in quality (Sections~\ref{sec:resource_aware_qa}--\ref{sec:built-in-quality}).}
\add{These challenges have visible starting points (e.g., caching, explicit feedback, vector stores, and ad-hoc guardrails), so the ``greenfield'' component is less about inventing individual mechanisms and more about creating FMware-native \emph{interfaces, policies, and evaluation infrastructure} that make them reliable under drift and governance constraints.}
\add{Concretely, this includes (i) validity metadata standards and cache-invalidation policies tied to model/data updates, (ii) safe feedback routing and governance that links feedback to versioned artefacts and prevents unsafe propagation, and (iii) scalable capture and automation for contracts/guardrails and FMwareBOM-style provenance artefacts.}

\begin{greenfieldbox}
\respto{3-16}\textbf{Greenfield Road Ahead.}\\
These directions highlight areas where the required primitives and evaluation infrastructure are still immature, and where progress likely requires substantial new research and engineering.
\begin{itemize}[leftmargin=*,label=--,nosep]
    \item Cognitive observability with measurable fidelity.
    \item Controlled execution via replay and guided exploration.
    \item Cost- and drift-aware caching and test prioritization.
    \item Memory governance for long-horizon reliability.
    \item Release-note verification pipelines to support safe evolution.
\end{itemize}
\tinyskip
Collectively, these directions motivate a pivot from task-accuracy benchmarks toward system-level evaluation frameworks that integrate decision traces, cost budgets, and latency envelopes, supported by shared artefacts such as replayable snapshots, a causal-trace schema, and cost-aware test corpora, and by benchmarks that quantify operational outcomes (e.g., incident triage time, audit completeness, and multi-agent coordination success).
\end{greenfieldbox}

\section{Limitations}
\label{sec:limitations}

\respto{3-7b}\add{Our study uses a semi-structured thematic synthesis rather than a full qualitative thematic analysis, and therefore does not follow exhaustive open-coding, repeated recoding, and formal validation procedures commonly used in qualitative coding workflows~\cite{butler2024objectives,cruzes2010synthesizing}. This choice reflects feasibility constraints given the scale and heterogeneity of the FMware evidence base and the rapid pace of change in the field (e.g., thousands of papers over 2017--2023~\cite{fan2024bibliometric} alongside practitioner reports). As a result, the abstractions and groupings involve judgment and may reflect the author's interpretation; some issues may be under-emphasized, merged differently, or missing under alternative coding schemes.}

\add{Importantly, our goal is not only to report evidence patterns but also to provide a practitioner-oriented perspective on what matters for production readiness. This necessarily incorporates the authors' professional experience and interpretation as a filter over the evidence base. This perspective has limitations: it may not capture the full breadth of challenges discussed in academic work, and it may over-represent issues that are salient in practitioner-facing sources. To mitigate this risk, we (i) triangulate themes across multiple independent sources and artifact types, (ii) report findings at an aggregated level with clear provenance signals, and (iii) explicitly distinguish themes that are primarily practitioner-derived from those that are already well-established in the academic literature.}

\respto{1-25b}\add{AI-assisted techniques can be considered for topic modelling to accelerate large-scale evidence organization. However, prior work documents limitations that make topic-modelling pipelines sensitive to modelling and pre-processing choices, difficult to reproduce, and reliant on substantial manual validation, which reduces their suitability as a substitute for careful synthesis when producing practitioner-facing constructs~\cite{chen2016survey}. Accordingly, we relied on expert-driven synthesis while explicitly documenting the analytical constructs used to move from evidence to recurrent issues, themes, and challenges.}

\section{Conclusion}
\label{sec:conclusion}
FMware is still in its early stages, and organizations are only beginning to confront what it takes to move from impressive demos to production-ready systems. In this paper, we provide a traceable catalogue of recurrent issues and synthesize them into six cross-cutting challenges, each grounded in evidence and accompanied by a critical analysis of current practice and a concrete vision for addressing it. \add{To support practical adoption, we make assumptions and trade-offs explicit and separate near-term engineering starting points from genuinely greenfield gaps where new primitives and evaluation infrastructure are needed.} While not exhaustive, we hope this roadmap helps practitioners prioritize interventions and helps researchers focus on the highest-leverage directions for building trustworthy production-ready FMware.

\section*{Disclaimer}
Any opinions, findings, conclusions, or recommendations expressed in this material are those of the author(s) and do not reflect the views of Huawei. Also, ChatGPT range of models (GPT-4o, GPT-5 and GPT-5.2) and Gemini-2.5 and Gemini-3 were used for copy-editing and table formatting. All experiments, analysis, writing, and results were performed by the authors, who also thoroughly reviewed the final content. This complies with IEEE and ACM policies on AI use in publications.

\bibliographystyle{ACM-Reference-Format}
\bibliography{00-main
}


\end{document}